\documentclass[final,3p,times,twocolumn]{elsarticle}
\def\astrobj#1{#1}

\def\arcs{\rlap{.}$^{\prime\prime}$}
\def\arc{$^{\prime\prime}$}
\usepackage{amssymb}
\usepackage{graphics}

\journal{New Astronomy}

\begin{document}

\begin{frontmatter}

\title{Galaxy Secular Mass Flow Rate Determination Using the Potential-Density Phase Shift Approach: Application to Six Nearby Spiral Galaxies}

\author{Xiaolei Zhang}

\address{
Department of Physics and Astronomy, George Mason University, 
4400 University Drive, Fairfax, VA 22030, USA
\newline E-mail: xzhang5@gmu.edu}

\author{Ronald J. Buta}

\address{Department of Physics and Astronomy, University of Alabama, 
514 University Blvd E, Box 870324, Tuscaloosa, AL 35487, USA
\newline E-mail: buta@sarah.astr.ua.edu}

\begin{abstract}
Using the potential-density phase shift approach developed by the present authors in earlier publications, we estimate the magnitude of radial mass accretion/excretion rates across the disks of six nearby spiral galaxies (\astrobj{NGC 628}, \astrobj{NGC 3351}, \astrobj{NGC 3627}, \astrobj{NGC 4321}, \astrobj{NGC 4736}, and \astrobj{NGC 5194}) having a range of Hubble types.  Our goal is to examine these rates in the context of bulge building and secular morphological evolution along the Hubble sequence.  Stellar surface density maps of the sample galaxies are derived from SINGS 3.6$\mu$m and SDSS $i$-band images using colors as an indicator of mass-to-light ratios. Corresponding molecular and atomic gas surface densities are derived from published CO(1-0) and HI interferometric observations of the BIMA SONG, THINGS, and VIVA surveys. The mass flow rate calculations utilize a volume-type torque integral to calculate the angular momentum exchange rate between the basic state disk matter and what we assume to be density wave {\it modes} in the observed galaxies. This volume-type integral contains the contributions from both the gravitational surface torque couple and the advective surface torque couple at the nonlinear, quasi-steady state of the wave modes, in sharp contrast to its behavior in the linear regime, where it contains only the contribution from the gravitational surface torque couple used by Lynden-Bell \& Kalnajs in 1972. The potential-density phase shift approach yields angular momentum transport rates several times higher than those estimated using the Lynden-Bell and Kalnajs approach. And unlike Lynden-Bell and Kalnajs, whose approach predicts zero mass redistribution across the majority of the disk surface (apart from the isolated locations of wave-particle resonances) for quasi-steady waves, the current approach leads to predictions of significant mass redistribution induced by the quasi-steady density wave modes, enough for the morphological types of disks to evolve substantially within its lifetime.  This difference with the earlier conclusions of Lynden-Bell and Kalnajs reflects the dominant role played by collisionless shocks in the secular evolution of galaxies containing extremely non-linear, quasi-steady density wave modes, thus enabling significant morphological transformation along the Hubble sequence during a Hubble time. We show for the first time also, using observational data, that {\em stellar} mass accretion/excretion is just as important, and oftentimes much more important, than the corresponding accretion/excretion processes in the {\em gaseous} component, with the latter being what had been emphasized in most of the previous secular evolution studies.  
\end{abstract}

\begin{keyword}
galaxies: kinematics and dynamics; galaxies: structure;
galaxies: evolution; galaxies: spiral 
\end{keyword}

\end{frontmatter}

\section{INTRODUCTION}

The potential-density phase shift (PDPS), referring to a characteristic
radial distribution of the azimuthal offset between the gravitational potential 
minimum and the density maximum of a skewed internal gravitational perturbation 
(such as a skewed bar or an open spiral), has been shown to be both the 
driver as well as the result of a collective dissipation process occurring 
in any gravitational $N$-body disks having self-organized, quasi-steady 
nonlinear density wave {\it modes}, (Zhang 1996, 1998, 1999.  Hereafter Z96, 
Z98, Z99).  The PDPS can be determined from observations of galaxies, and 
its magnitude and distribution shed important light on the secular evolution 
of the stellar and gaseous distributions in disk galaxies of different Hubble 
types.

Zhang \& Buta (2007=ZB07) made the first demonstration that the PDPS approach 
can be used on observed galaxy images to determine two quantities of dynamical 
interest, for those galaxies containing self-organized density wave modes: 
the locations of the {\it corotation resonance (CR) radii} of individual 
modes, and the rate of radial mass accretion/excretion as a function of radius 
as a result of the interaction between these modes and the basic state of
the galactic disks.  Subsequently, Buta \& Zhang (2009=BZ09) applied the PDPS 
approach to locate the CR radii in 153 galaxies using near-infrared $H$-band 
images from the Ohio State University Bright Spiral Galaxy Survey 
(OSUBSGS, Eskridge et al.  2002), and brought attention to a range of 
characteristic bar-spiral combinations, as well as their associated 
characteristic PDPS distributions, among normal galaxies (some examples
of these characteristic bar-spiral combinations
will be further discussed in Appendix B of the current paper 
in the context of secular morphological evolution of galaxies).  
Among the original findings of ZB07 and BZ09 is the so-called ``super-fast" 
bar, where the bar corotation radius lies inside the ends of the bar, 
in contradiction to the previous conclusion of Contopoulos (1980), obtained
using passive orbit analysis ignoring the collective interactions
of stars in a self-organized bar {\em mode}, that a bar cannot extend 
beyond its own CR. This finding alone shows that the PDPS approach can lead 
to new insights into galaxy dynamical states that had not been considered 
previously.

In this paper, our focus is on observational estimates of mass flow
rates in a small sample of galaxies over a range of Hubble types.
Our goals are two-fold: to outline how to use the PDPS
method to estimate galaxy mass flow rates, and to examine the implications
of these rates with regard to angular momentum transport, bulge-building,
and long-term secular morphological evolution of galaxies. 

\subsection{Application of the PDPS Approach}

Observationally, measuring phase shift distributions, mass flow rates,
and torque couplings in galaxies using the PDPS approach requires
mainly a two-dimensional map of the surface mass density distribution,
$\Sigma (x,y)$, in units of $M_{\odot} pc^{-2}$. Reasonable
approximations to {\it stellar} surface mass densities can be inferred
from calibrated images using mass-to-light ratio estimates, which are
often based on the colors of the light. The preferred passbands for
this kind of calculation have been the near- and mid-infrared, where
the starlight more reliably traces the mass of the old stellar
population, and where extinction effects are much lower than in optical
bands (Meidt et al.  2012). An important assumption is that we can
reasonably deproject galaxy images by assuming outer isophotes are
intrinsically circular. To minimize deprojection uncertainties, we
limit our analysis in the current study
only to relatively face-on galaxies. In contrast to
the analysis in ZB07 and BZ09, which considered only the stellar
mass surface densities, in the current work we include also the
contributions from atomic and molecular gas, obtained from archival
radio interferometric observations, to the determination
of the total surface mass density of the galaxy disks.

In calculating the two-dimensional gravitational potential ${\cal V} (x,y)$
used in the PDPS approach, we applied the standard Poisson
integral approach to the total surface density map $\Sigma (x,y)$, 
employing the fast Fourier transform technique similar
to that used in Quillen et al. (1994).
We have experimented with various forms of the vertical density distribution
in ZB07 and BZ09, and found that these different choices of
profiles (including exponential, sech, or sech-squared functions) 
led to moderate differences in the magnitude of the potential obtained, but 
negligible difference in the locations of resonances determined through the PDPS
approach.

Our main {\it working hypothesis} is that the features we see in disk
galaxies are {\it quasi-stationary wave modes}, meaning they are
long-lived. This is at the moment still not a universally
accepted assumption.  For example, Sellwood (2011)
argues that short-lived transient patterns are the rule in simulations.
On the other hand, detection of azimuthal color-age gradients across 
spiral arms, predicted by density wave theory, favors that the patterns 
are long-lived (Gonz\'alez \& Graham 1996; Mart\'inez-Garc\'ia et al. 
2009a,b; Mart\'inez-Garc\'ia \& Gonz\'alez-L\'opezlira 2011). 
The presence of well-organized nested resonance patterns in grand-design
galaxies such as NGC 4321 (one of our sample galaxies) also supports
the quasi-steady modal origin of these patterns.  The success
of the PDPS approach itself in accurately predicting the
location of CR radii\footnote{Haan et al. (2009) had independently
accessed several CR radii determination methods and stated
that ``For our galaxies the phase-shift method appears to be
the most precise method with uncertainties of (5-10)\% ..."
(Haan et al. 2009, ApJ, 692, 1623),
Their above statement appears in page 1641, left column and second full
paragraph.}, is also an indirect confirmation of the
modal origin of the density wave patterns in grand-design galaxies,
because such an agreement should not be expected in the transient
wave picture.
{\it Thus we adopt as our working hypothesis that the density wave
patterns in grand design disk galaxies are quasi-steady modes,
and then examine if the consequences derived from this assumption
contradict with the observed properties of galaxies.}  Note that for
the dynamical mechanism described in Z96, Z98, Z99 to work in
predicting the secular mass flow rate, the
modes only need to be quasi-steady on a local dynamical timescale,
i.e, a few galactic rotation periods at a characteristic radius
of the galaxy, rather than the more stringent stability requirement
on a Hubble time, which would be unrealistic to expect given
the significant basic state evolution produced by these nonlinear
modes.  The evolving basic state of the galactic disk will 
gradually change the modal set and modal morphology compatible with it.

Our application of the PDPS approach will be based on several equations.
The first is the radial distribution of the phase shift itself (Z96,
Z98), which will be used to locate CR radii in each galaxy:

\begin{equation}
\phi_0(r) = {1 \over m} \sin^{-1}
\left ( {1 \over m}
{{\int_0^{2  \pi}
\Sigma_1
{ {\partial {{{\cal{V}}_1}}} \over {\partial \phi}}
d \phi}
\over
{ \sqrt{\int_0^{2  \pi}
{\cal{V}}_1^2
d \phi}}
\sqrt{\int_0^{2  \pi}
{\Sigma_1}^2
d \phi}  }
\right )
,
\end{equation}

\noindent
where $m$ is the azimuthal order of the perturbation, and $\phi_0$ is
defined to be positive if the potential lags the density in the
azimuthal direction, in the sense of galactic rotation, and negative if
it leads the density. For most galaxies the sense of galactic rotation
can be determined by the assumption that the density wave pattern is
trailing.  In the above expression, $\Sigma_1(r,\phi)$ and ${\cal
V}_1(r,\phi)$ are the nonaxisymmetric parts of the surface mass density
and gravitational potential, respectively.  From considerations of
the angular momentum exchange between the axisymmetric basic state
and the spontaneously-formed mode, it can be inferred that
the radial phase shift distribution of a single quasi-steady mode has one
positive hump followed by one negative hump (Z98), with the zero
crossing of the PDPS curve between these two humps occurring at CR. 
Multiple modes in the same galaxy would be manifested as a sequence of 
(positive, negative) hump pairs.  Each positive-to-negative (P/N) zero crossing
of the phase-shift-versus-radius curve would correspond to the
CR radius of a single mode -- if the underlying modal pattern is indeed
in a quasi-steady state.
Note that for the purpose of locating CR radii, 
no knowledge of the distance of the galaxy is required since a constant
mass-to-light ratio can lead already to fairly accurate CR
determination (ZB07).

The validity of using the P/N zero-crossings of the PDPS
distribution for locating CR radii does {\it not} rely on the
spiral/bar modes being exactly steady, only that they are modes (i.e.,
the approach is equally valid if the mode is uniformly growing and
uniformly decaying, while maintaining the modal shape -- this has been
seen in many N-body simulations of spiral modes, such as Donner \&
Thomasson 1994; Z98). The validity of the PDPS approach in determining
CRs is reduced if there is spurious non-modal content in the pattern
surface density that is in the process of being filtered out by the
galactic resonant cavity (i.e., if the mode is in the very early stages
of growing out of primordial inhomogeneous clumps, or else is due to tidal
excitation). Another issue regarding equation 1 is that P/N crossings
are not affected by the assumed value of $m$.

The second equation we will use is the mass flow rate as a function of
radius (Z96, Z98, ZB07, note that the minus sign present in front
of the integral sign for this equation in ZB07,
which was a typo, is now taken out.  The sign convention is for
$dM/dt$ to be greater than zero for accretion inside R, and
less than zero for excretion) 

\begin{equation}
{dM (R) \over dt} = 
{R \over {V_c}} \int_0^{2 \pi} 
\Sigma_1  {{\partial {\cal V}_1} \over {\partial \phi}} d \phi
\label{eq:eq2}
\end{equation}
where $R$ is the radius, $V_c$ is the circular velocity, and $\Sigma_1$
and ${{\cal V}_1}$ denote the perturbed surface mass density and
potential, respectively.  $\Sigma_1$ in particular can denote the perturbative
surface density of a given mass component (i.e., stars or gas),
but ${{\cal V}_1}$ used needs to be the total potential --
this is because the accretion mass cannot separate the forcing field
component, and responds only to the {\em total} forcing
potential\footnote{Dark matter is not included in our analysis because
its skewness is assumed to be small -- and if this is not the case,
the derived accretion rate will be higher than what we obtained in this
study.}.  In actual calculations,
one can in fact employ the sum total of perturbative and non-perturbation mass 
components, since the axisymmetric parts will naturally drop out
through the differentiation and integration process.

Equations 1 and 2 both show that if $\Sigma_1$ and ${\cal V}_1$ are
exactly in phase with each other, which can result for either an
extremely tightly-wound spiral or a perfectly straight bar (Z96,Z98),
then there would be no angular momentum exchange between the wave and
the disk matter in an annulus and no radial mass flow.  For open (skewed)
spirals or bars, there will always be a phase shift as a result of the Poisson
integral (this in fact is a result of earlier potential theory studies
by mathematicians of the 19th century, though no kinematic information
was included in such studies and thus no corotation determination was
considered).  For spontaneously-formed modes, its basic state density
distribution will be such that the phase shift changes sign at corotation,
leading to wave-basic-state angular momentum exchange at the quasi-steady
state of the modes and the secular mass redistribution of the basic state,
specifically, mass inflow inside CR and mass outflow outside CR.  
The exact validity of equation 2 for calculating the
basic state mass flow rate depends on the quasi-steady state
assumption (Z96, Z98, Z99, once again note that ``quasi-steady'' here
refers to modal pattern maintaining its shape on the order of
local dynamical time ), but even for the non-steady situation,
equation 2 can be used to estimate an approximate mass flow rate. In
addition to the $\Sigma$ and $\cal V$ maps, application of equation 2
requires knowledge of the galaxy rotation curve and the distance.

Other quantities used in the following include
gravitational and advective (surface) torque couplings (Lynden-Bell
\& Kalnajs 1972 = LBK72), in order to examine the secular angular momentum 
redistribution rate.  Employing the notation used by
Binney \& Tremaine (2008=BT08), the gravitational torque coupling
integral $C_g(R)$ can be obtained from

\begin{equation}
C_g(R) = {R \over {4 \pi G}} \int_{- \infty}^{\infty} \int_0^{2 \pi}
{{\partial {\cal V}} \over {\partial \phi}}
{{\partial {\cal V}} \over {\partial R}}  
d \phi dz,
\end{equation} 
and the advective torque coupling integral $C_a(R)$ can be obtained from

\begin{equation}
C_a(R) = R^2 \int_0^{2 \pi} \Sigma_0 V_R V_{\phi} d \phi ,
\end{equation}
where $V_R$ and $V_{\phi}$ are the radial and azimuthal velocity
perturbations relative to the circular velocity, respectively. 
The gravitational torque couple is due to the torquing of the inner
disk material on the outer disk material, or vice versa, across
the relevant surface under consideration, through gravitational interaction;
whereas the advective torque couple accounts for the angular momentum exchange 
between the inner and outer disk carried by the matter
crossing the same surface during the different phases of the
galactic rotation with respect to the phases of the density
wave (i.e. the so-called lorry transport effect, or Reynolds stress, 
see LKB72 or BT08).

The mass flow rate formula involves a volume-type of torque integral
that can be related to the above two surface-type torque couplings. The
integral,

\begin{equation}
T_1(R) \equiv R \int_{0}^{2 \pi}
\Sigma_1  {{\partial {\cal V}_1} \over {\partial \phi}} d\phi,
\label{eq:eq3}
\end{equation}
was introduced in Z96 and Z98 in the context of the self-torquing
of the disk matter in a unit-width annulus at $R$ by the potential of
the associated spontaneously-formed density wave mode (effectively
the torque on the matter in the annulus is applied by matter
both in the inner disk and in the outer disk away from the
annulus). The volume torque can be shown to be equal to the time rate 
of angular momentum exchange between
the density wave and the disk matter in a unit-width annulus at $R$,
for wave modes in approximate quasi-steady state. For such modes, Z99
showed that $T_1(R) = d(C_a+C_g)/dR$, and thus accounts for both types
of surface torque couple contributions. 

Past calculations of the secular angular momentum redistribution rate
[i.e. Gnedin et al. (1995), Foyle et al. (2010)] considered only 
the gravitational torque couple and ignored the advective torque
couple (which cannot be directly estimated using the observational
data, except by estimating the total volume torque $T_1(R)$, performing
a radial integration, and then
subtracting from it the contribution of the gravitational torque). The
results in section 2.3 below show that for the typical density wave amplitudes
usually encountered in observed galaxies, the advective contribution to
the total torque couple is several times larger than the contribution from the
gravitational torque, and also is of the same sense of angular momentum
transport as the gravitational torque couple -- another characteristic
unique to the nonlinear mode. Past calculations of gas mass accretion
near the central region of galaxies (e.g. Haan et al. 2009) are likely
to have significantly underestimated the gas mass flow rate for the
same reason.

\section{ANALYSES OF INDIVIDUAL GALAXIES}

In this section, we begin by analyzing the morphology and kinematics of six
bright, nearby galaxies, to set the stage for further evolutionary
studies in the latter part of this section. The parameters of these
galaxies are given in Table 1. For five of the galaxies, two stellar
mass maps were derived, one based on a SINGS (Kennicutt et al.  2003)
3.6$\mu$m IRAC image, and the other based on a SDSS $i$-band image.
HI and H$_2$ gas maps from archival radio interferometric
observations were then added onto these stellar surface
density maps to obtain the total disk surface density maps. The
procedure for calculating all maps is described in Appendix A. We note
that hot dust corrections had to be made to the 3.6$\mu$m maps, and
that no bulge/disk decompositions in this instance
were made to allow for bulge shapes.

\begin{figure}
\vspace{250pt}
\includegraphics{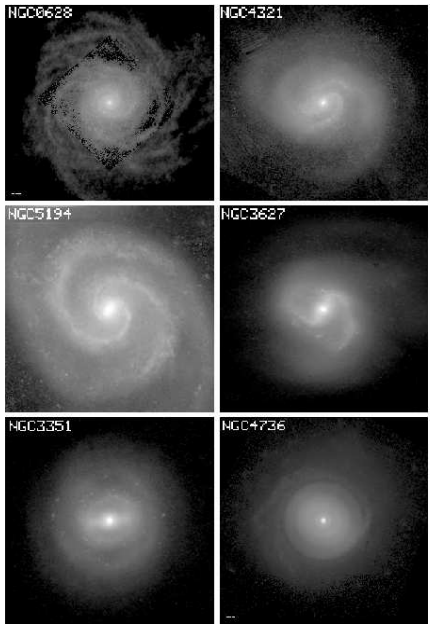}
\caption{Total mass maps of the six sample galaxies.}
\label{fg:Fig1}
\end{figure}

The total-mass maps derived are shown in Figure~\ref{fg:Fig1}. Since
parts of the discussions in this section concern CR radii, Appendix B
gives schematic representations of the different kinds of bar-spiral
configurations described by BZ09, including slow, fast, and super-fast
types.

\begin{table*}
\caption{Adopted parameters for sample galaxies.$^a$}
\halign{%
\rm#\hfil&
\qquad\rm\hfil#&
\qquad\rm\hfil#&
\qquad\rm\hfil#&
\qquad\rm\hfil#&
\qquad\rm\hfil#&
\rm#\hfil\cr
Galaxy & $i$ & $\phi_n$ & $h_R$ & $h_z$ & Distance & ~~~~~References for \cr
& (degrees) & (degrees) & (arcsec) & (arcsec) & (Mpc) & ~~~~~($i$, $\phi_n$) \cr
\noalign{\vskip 10pt}

NGC 628  & 6 & 25 & 64 (3.6) & 7.1 & 8.2 & ~~~~~Shostak et al. (1983)\cr
NGC 3351 & 40.6 & 13.3 & 44 (3.6) & 8.9 & 10.1 & ~~~~~3.6$\mu$m isophotes \cr
NGC 3627 & 60 & 173 & 66 (3.6) & 13.3 & 10.1 & ~~~~~RC2, Zhang et al. (1993) \cr
NGC 4321 & 31.7 & 153 & 63 (3.6) & 12.6 & 16.1 & ~~~~~Hernandez et al. (2005) \cr
NGC 4736 & 30 & 116 & 135 ($i$) & 27.1 & 5.0 & ~~~~~3.6$\mu$m,$g$,$i$ isophotes; Buta (1988) \cr
NGC 5194 & 20$\pm$5 & 170$\pm$3 & 50 ($K_s$) & 10.0 & 7.7 & ~~~~~Tully (1974)\cr
}
$^a$Col. 1: galaxy name; col. 2: adopted inclination; col. 3: adopted
line of nodes position angle; col. 4: adopted radial scale length, not based on
decomposition but from slope of azimuthally-averaged surface brightness
profile (filter used in parentheses); col. 5: adopted vertical scale
height derived as $h_z = h_R/5$ except for NGC 628 where $h_z = h_R/9$; col. 6: mean redshift-independent distance from NED; typical uncertainty
$\pm$1-2Mpc; col. 7: source of adopted orientation parameters.
\end{table*}

\subsection{Phase Shifts and Corotation Resonances}

\subsubsection{NGC 4321 (M100)}

\begin{figure}
\vspace{160pt}
\includegraphics{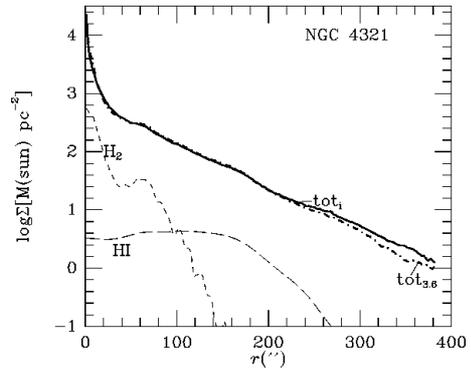}
\caption{Azimuthally-averaged surface mass density profiles of NGC
4321, based on the atomic (HI), molecular (H$_2$), and total (3.6$\mu$m
+ HI + H$_2$ and $i$ + HI + H$_2$) mass maps. The profile based on the
3.6$\mu$m image is called tot$_{3.6}$ while that based on the $i$-band
is called tot$_i$.}
\label{fg:Fig2}
\end{figure}

NGC 4321 (M100) is the bright Virgo spiral, of mid-infrared (MIR) RC3
type SAB(rs,nr)bc (Buta et al. 2010).  It lies at a distance of
approximately 16 Mpc. The surface densities of the gaseous mass
components HI and H$_2$, derived from VIVA (Chung et al.  2009) and
BIMA SONG (Helfer et al. 2003) observations, respectively, and the
total mass profiles using the IRAC 3.6$\mu$m image and the SDSS
$i$-band image (``tot$_{3.6}$" and ``tot$_i$", respectively), are
illustrated in Figure \ref{fg:Fig2} for this galaxy.  This shows that
the 3.6$\mu$m and $i$-band total mass maps give very similar average
surface density distributions for M100.

\begin{figure}
\vspace{300pt}
 \includegraphics{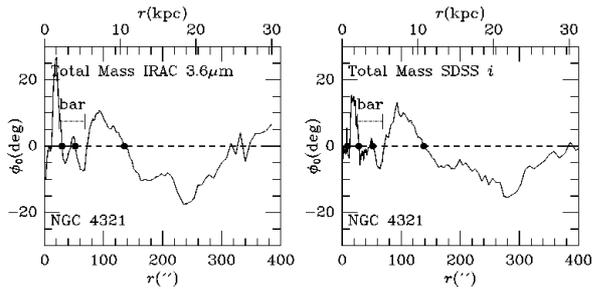}
 \includegraphics{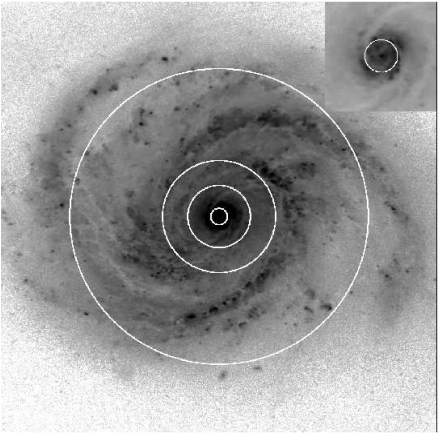}
\caption{{\it Top:} Potential-density phase shift versus galactic radius
for NGC 4321 derived using the total mass map with IRAC and SDSS data.
CR radii are indicated by the filled circles. The radius range of the
bar from ZB07 is indicated.
{\it Bottom:} Corotation circles overlaid on a deprojected
SDSS $g$-band image of NGC 4321.
The main frame covers an area 6\rlap{.}$^{\prime}$4 square, while the inset is
0\rlap{.}$^{\prime}$8 square. The units are mag arcsec$^{-2}$.}
\label{fg:Fig3}
\end{figure}

ZB07 presented a phase shift analysis for M100 using the
3.6$\mu$m SINGS image, and assumed a constant mass to light ratio
($M/L$) without gas. (Our current analysis allows for color-dependent
stellar $M/L$ corrections.) Four corotation resonances (CRs) were
found, and were shown to correspond to well-defined morphological
features. Figure~\ref{fg:Fig3}, top, shows the radial-dependence of
potential-density phase shift derived using the {\em total} (stellar
plus gaseous) maps for the IRAC 3.6 $\mu$m band (left frame) and the
SDSS i-band (right frame).  We can see that these two maps give similar
CR predictions [in terms of the positive-to-negative (P/N) zero
crossings of the phase shift curves], except near the very center of
the galaxy where the factor of 2 better spatial resolution of the SDSS
map allows the possible detection of a new inner nuclear pattern.  We
have also conducted tests by calculating phase shift distributions
using images from the different optical and NIR wave bands that we have
access to, assuming constant M/L, and found that the phase shifts
derived from the above total mass maps (calibrated with radial
dependent M/L) show the best coherence, most likely due to the best
kinematical and dynamical mutual consistency between the potential and
density pair used for the total mass analysis, as is the case for
physical galaxies. Table 2 lists the (unweighted) average CR radii from
the 3.6$\mu$m and $i$-band maps.

Figure~\ref{fg:Fig3}, bottom shows the four CR circles for M100
superposed on the $g$-band SDSS image. The inset shows only the inner
region, which in the $g$-band includes a nuclear pseudoring and in the
longer wavelength bands includes a nuclear bar. The CR$_1$ radius
determined from the new maps, (8\arcs 3$\pm$0\arcs 7) is smaller than
what we obtained previously (13\arc) in ZB07, but turns out to better
correspond to the features of the resolved inner bar. The nuclear bar
seems to extend either a little beyond or just up to its CR and
terminates in a broad, swept-up spiral section (a nuclear pseudoring) at
the location indicated by the next N/P crossing. The next CR, CR$_2$,
appears to be related to an inner spiral that breaks from near the
nuclear pseudoring. The curved dust lanes in this spiral are on the
leading sides of the weak bar. CR$_3$ appears to lie completely within
the main spiral arms and could be the CR of the bar itself. The radius
of CR$_3$, 52\arc, is slightly less than that determined by ZB07,
59\arc, and may lie a little inside the ends of the primary bar.
CR$_4$, at 137\arc, lies within the outer arms. These values should be
compared with r(CR) = 97\arc $\pm$15\arc\ obtained by Hernandez et al.
(2005) using the Tremaine-Weinberg method applied to an H$\alpha$
velocity field.  Their value is close to the average of our CR$_3$ and
CR$_4$ radii. We believe our two-outer-CR result is more reasonable
than the single outer CR result from the TW method for this galaxy,
because on the $g$-band overlay image one can clearly discern the dust
lanes moving from the inner/leading edge of the spiral to
outer/trailing edge across CR$_4$, indicating the location where the
angular speeds of disk matter and density wave switch their relative
magnitude. CR$_3$ on the other hand is tied to the main bar rather
than the spiral.

It appears that the patterns surrounding CR$_1$ and CR$_3$ are best
described as super-fast bars (ZB07, BZ09) of a dumb-bell shape with
their thick ends created by the different pattern speeds interacting in
the region of their encounter (see also the classification in Appendix
B). As we noted in section 1, super-fast bars are an unexpected
finding that has come out of the application of the phase shift method, 
because they contradict the
results of passive orbit analyses that imply that a bar cannot extend
beyond its own CR due to a lack of regular orbital support (Contopoulos 1980).

\subsubsection{NGC 3351 (M95)}

\begin{figure}
\vspace{160pt}
\includegraphics{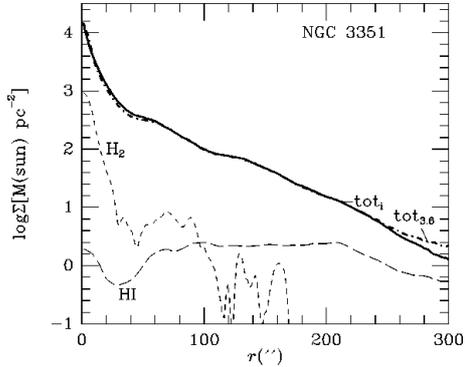}
\caption{Azimuthally-averaged surface mass density profiles of NGC
3351. The layout is the same as in Figure 2.}
\label{fg:Fig4}
\end{figure}

NGC 3351 (M95) is a barred spiral galaxy of MIR RC3 type (R')SB(r,nr)a
(Buta et al. 2010). It lies at a distance of approximately 10 Mpc. In
spite of the early-type classification, the galaxy has a very small
bulge. Its actual bulge, in the form of a bright nuclear ring, is
regarded as a pseudobulge by Kormendy (2012).  

\begin{figure}
\vspace{300pt}
 \includegraphics{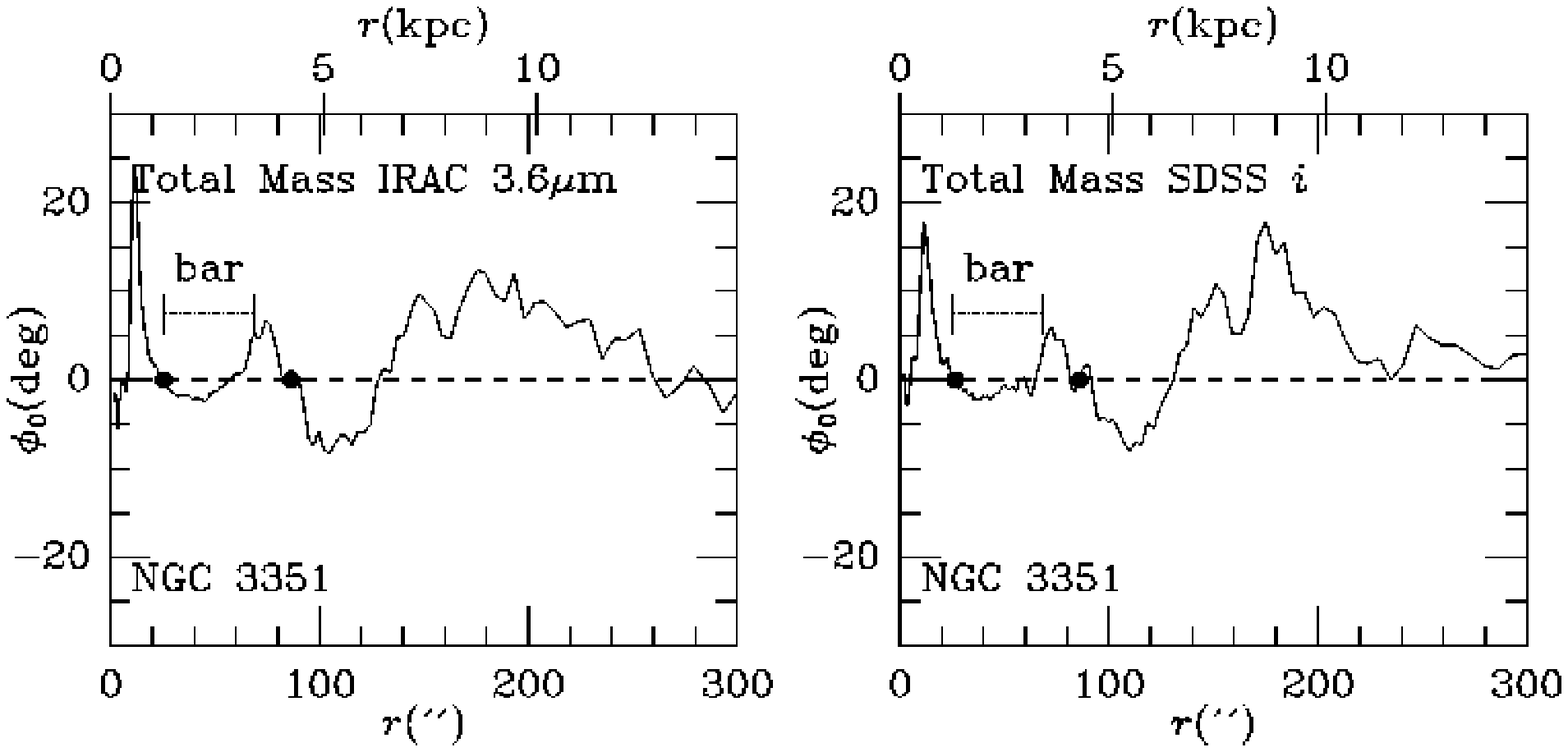}
 \includegraphics{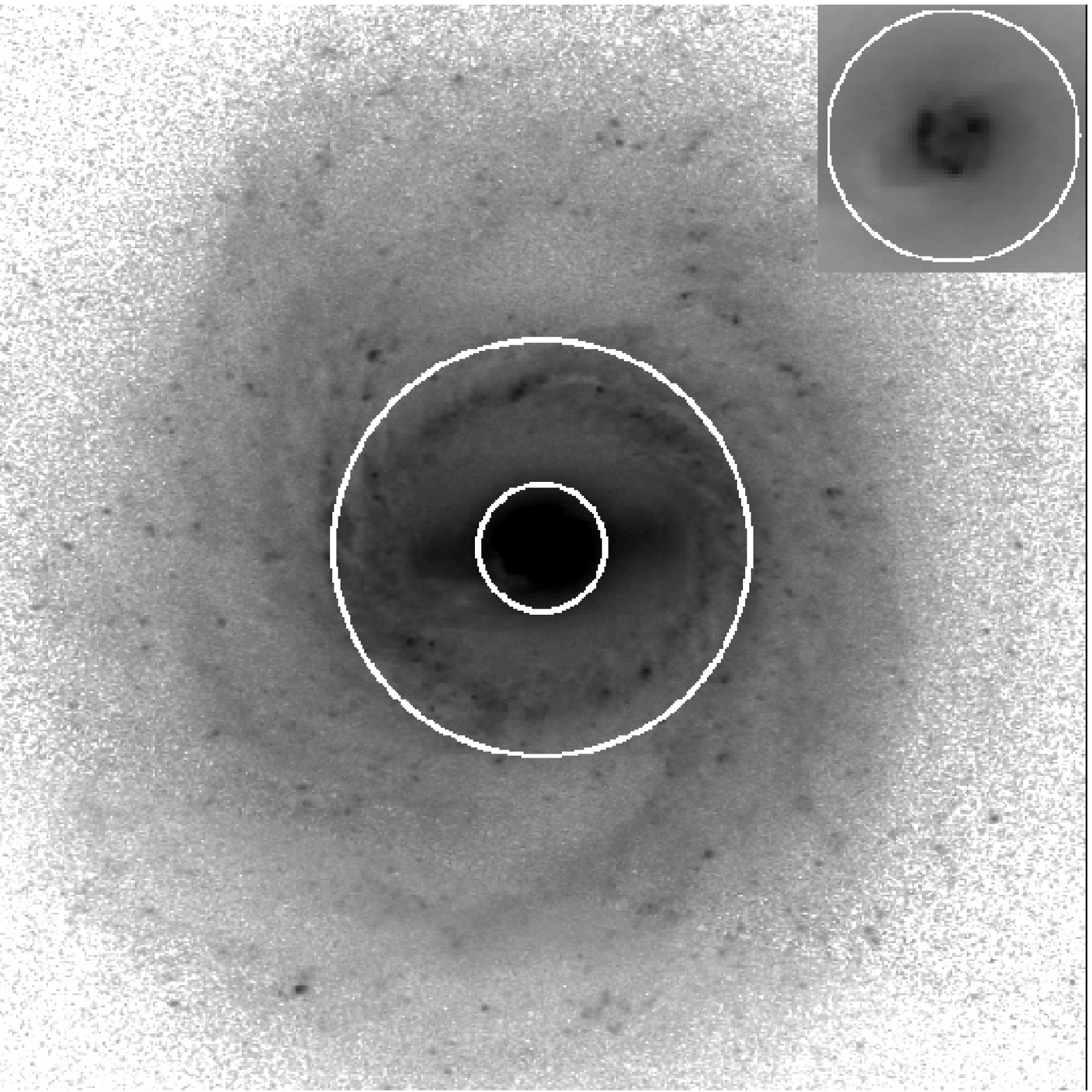}
\caption{{\it Top:} Potential-density phase shift versus galactic
radius for galaxy NGC 3351 derived using the total mass map with IRAC
and SDSS data. {\it Bottom:} Deprojected $g$-band image of NGC 3351 in
units of mag arcsec$^{-2}$, with major phase shift crossings superposed
as solid circles. The main frame covers an area 7\rlap{.}$^{\prime}$5
square, while the inset is 0\rlap{.}$^{\prime}$93 square.}
\label{fg:Fig5}
\end{figure}

The surface densities of the different mass components for this galaxy
derived using IRAC, SDSS, BIMA SONG and THINGS data are shown in Figure
\ref{fg:Fig4}.  Phase shift analyses for NGC 3351 were carried out
using the IRAC and SDSS mass surface density images based on the
3.6$\mu$m SINGS and $i$-band SDSS images, plus the gas maps
(Figure~\ref{fg:Fig5}, top).  The phase shift distributions for both
the IRAC and SDSS total mass maps show a major CR near $r$ = 86\arc,
and also another CR at $r$=26\arc.

Figure~\ref{fg:Fig5}, bottom shows the two CR radii as solid circles
superposed on the $g$-band image.  Two prominent N/P crossings follow
these CR radii at 65\arc\ and 128\arc. For comparison, the inner ring
of NGC 3351 has dimensions of 71\arcs\ 0$\times$67\arcs\ 8 and lies
close to the first N/P crossing, implying that the ring/spiral is a
separate mode whose CR is CR$_2$ at 86\arc. In Figure~\ref{fg:Fig5},
top also, the extent of the main part of the bar is indicated, and both
mass maps show the same thing: the phase shifts are negative across the
main part of the bar, implying (as discussed in ZB07 and BZ09) that the
corotation radius of the bar is CR$_1$, not CR$_2$. Thus, NGC 3351 is
also a case of a super-fast bar.

The inner pseudoring at the location of the first major N/P crossing
would, in passive orbit analysis in a galaxy with a bar and
non-self-gravitating clouds, be considered an inner 4:1 resonance ring
as identified by Schwarz (1984) and Buta \& Combes (1996). This ring
would be directly related to the bar and have the same pattern speed as
the bar. Our phase shift analysis, however, suggests that the apparent
ring/pseudoring could be due to a ``snow plough'' effect of two sets of
inner/outer patterns having different pattern speeds, and which thus
accumulate mass at that radius.  The strong and symmetric arms which
end a little within the second CR circle might be related to what
Contopoulos and collaborators had advocated: that symmetric patterns
sometimes end at the inner 4:1 resonance (counter-examples are
discussed by ZB07 and BZ09). The spiral patterns outside the second CR
circle in this case are more fragmented.  It is not yet clear what
different dynamics would make some spirals extend all the way to OLR
and others end mostly within CR or inner 4:1.  Empirical evidence so
far shows that those galaxies that have CRs intersecting the arms (such
as NGC 5247 analyzed in ZB07) tend to be Sb, Sc or later types, such
that the arms are going through the transitional phase from a skewed
long inner bar to either a bar-driven spiral or an inner organized
spiral plus outer diffused arms.  Therefore, the ones where the spiral
ends at the inner 4:1 resonance or CR (i.e. the NGC 3351 type) should
tend to be the more mature and more steady types, i.e., in earlier-type
disks compared to those that have arms being crossed by the CR circle.

\subsubsection{NGC 5194 (M51)}

\begin{figure}
\vspace{160pt}
\includegraphics{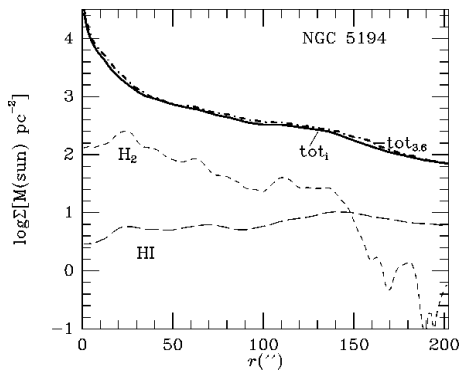}
\caption{Azimuthally-averaged surface mass density profiles of NGC
5194. The layout is the same as in Figure 2.}
\label{fg:Fig6}
\end{figure}

NGC 5194 (M51) is a well-known interacting grand design spiral, of
MIR RC3 type SAB(rs,nr)bc (Buta et al. 2010).  It lies at a distance of
approximately 8.4 Mpc.  The surface densities of the different mass
components for this galaxy derived using IRAC, SDSS, BIMA SONG and
THINGS data are shown in Figure \ref{fg:Fig6}.  It can be seen that the
hot dust-corrected total 3.6$\mu$m surface density profile and that from
the $i$-band data are very similar.

\begin{figure}
\vspace{300pt}
 \includegraphics{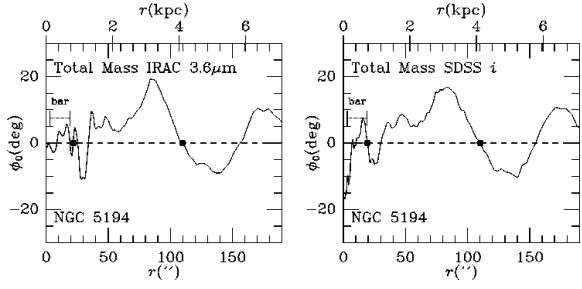}
 \includegraphics{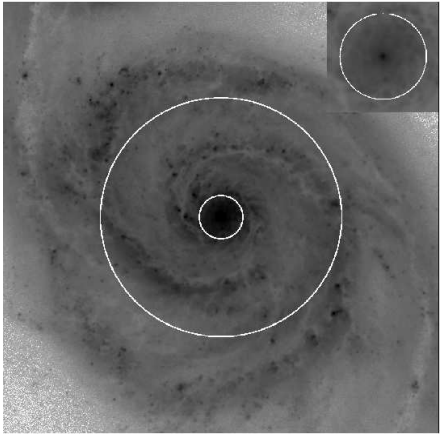}
\caption{{\it Top}: Calculated phase shift vs galaxy radii for NGC 5194
(M51) using the IRAC data as well as SDSS data.  
Two corotation radii (filled circles) are indicated.
{\it Bottom}: CR circles overlaid on the deprojected $g$-band image
of NGC 5194 (M51).
The main frame covers an area 6\rlap{.}$^{\prime}$6 square, while the inset is
0\rlap{.}$^{\prime}$93 square. The units are mag arcsec$^{-2}$.} 
\label{fg:Fig7}
\end{figure}

The potential-density phase shift method was used to derive the main CR
radii for this galaxy (Figure~\ref{fg:Fig7}). By focusing on the area
that just excludes the small companion NGC 5195, which is likely to lie
outside of the M51 galactic plane and thus have minor influence on the
internal dynamics of M51 at the epoch of observation (a conjecture
which was subsequently confirmed), the phase shift analysis gives two
major CR radii (P/N crossings on the phase shift plot, represented by
solid circles on the overlay image) followed by two
negative-to-positive (N/P) crossing radii (not shown on the overlay).
The latter are believed to be where the inner modes decouple from the
outer modes. The CR radii, at
21$^{\prime\prime}$$\pm$3$^{\prime\prime}$ and 110$^{\prime\prime}$
(Table 2), match very well the galaxy morphological features (i.e., the
inner CR circle lies near the end of an inner bar/oval, and the first
N/P crossing circle [$r$=30\arc] is where the two modes decouple).
Also, for the outer mode, the CR circle seems to just bisect the
regions where the star-formation clumps are either concentrated on the
inner edge of the arm, or on the outer edge of the arm, respectively --
a strong indication that this second CR is located very close to where
the pattern speed of the wave and the angular speed of the stars match
each other (this can be compared to a similar transition of arm
morphology across the CR in a simulated galaxy image of Z96, Fig. 3).
This supports the hypothesis that the spiral patterns in this galaxy
are intrinsic modes rather than being tidal and transient.  Tidal
perturbations in this case serve to enhance the {\em prominence} of the
intrinsic mode, but do not alter its modal shape.

\subsubsection{NGC 3627 (M66)}

\begin{figure}
\vspace{160pt}
\includegraphics{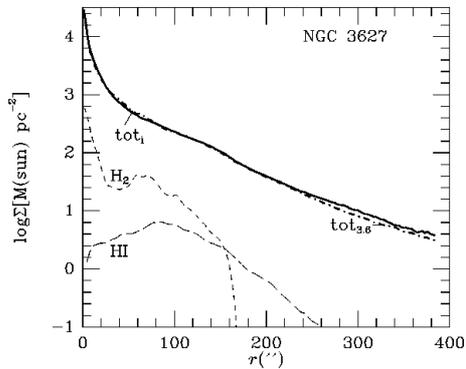}
\caption{Azimuthally-averaged surface mass density profiles of NGC
3627. The layout is the same as in Figure 2.}
\label{fg:Fig8}
\end{figure}

The intermediate-type barred spiral galaxy NGC 3627, of MIR RC3 type
SB(s)b pec (Buta et al. 2010), is a member of the interacting group the
Leo Triplet (the other two members being NGC 3628 and NGC 3623).  It
lies at a distance of approximately 10 Mpc.  The surface densities of
the different mass components for this galaxy derived using IRAC, SDSS,
BIMA SONG and THINGS data are shown in Figure \ref{fg:Fig8}.  As for NGC
3351, 4321, and 5194, the total 3.6$\mu$m mass profile and the total
$i$-band mass profile are in good agreement over a wide range of
radii.

Zhang, Wright, \& Alexander (1993) observed 
this galaxy in high-resolution CO(1-0) and
HI, and found that the choice of an outer CR location at
220\arc\ coinciding with the outer HI cutoff could account for many
observed morphological features.  Subsequent observations, e.g. those
of Chemin et al. (2003) in the H$\alpha$ line, determined that an
inner CR enclosing the central bar is a more reasonable choice. Our new
surface density map allows us to re-evaluate the question of CR
determination for this galaxy, bearing in mind that the PDPS method
gives the most reliable CR determination only for potential-density
pairs that have achieved dynamical equilibrium, a condition which is
likely to be violated for this strongly-interacting galaxy that has
suffered serious damage in its outer disk (e.g., Fig. 1 of Zhang et al.
1993).

\begin{figure}
\vspace{300pt}
\includegraphics{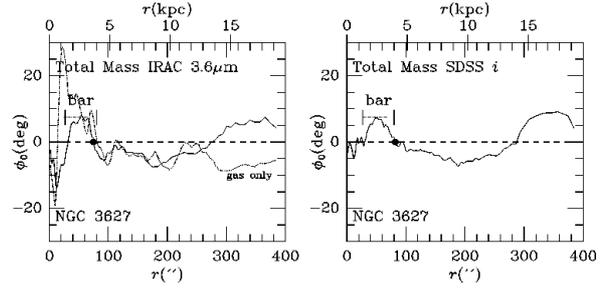}
\includegraphics{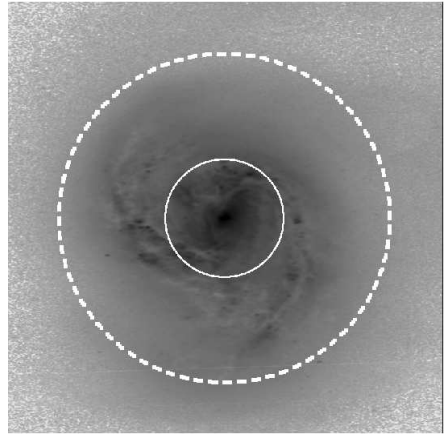}
\caption{{\it Top:} Calculated phase shifts between the stellar mass
and total potential, and the HI gas mass and the total potential,
for NGC 3627. {\it Bottom:}
Overlay of CR circles on the $g$-band image for NGC 3627.
The main frame covers an area 9\rlap{.}$^{\prime}$6 square.
The units are mag arcsec$^{-2}$.}
\label{fg:Fig9}
\end{figure}

In Figure~\ref{fg:Fig9}, top, we show the phase shifts with respect to the
total potential of the total mass distribution, as well as for the total
gas distribution. The total mass phase shifts display only one inner CR
at 78\arcs 6 $\pm$ 3\arcs 5, which can be compared to the CR radius
obtained by Chemin et al. (2003) of $\sim $70\arc, yet the large-radius
phase distribution shows clear evidence of the disturbance caused by
the interaction, and thus (by implication) a lack of dynamical equilibrium 
for this galaxy. The main CR appears to encircle the bar ends in NGC 3627
(Figure \ref{fg:Fig9}, bottom), which would make the galaxy a 
``fast bar" case according to BZ09 (see also Appendix B).

The total gas phase shift shows the presence of a second possible CR
between 220\arc-260\arc, close to the one adopted in Zhang et al.
(1993) of 220\arc.  The radius of this possible CR is ill-defined since
the phase shift barely reaches zero and does not become a clear P/N
crossing.  Figure~\ref{fg:Fig9}, bottom, shows the $g$-band image
overlaid with the two possible CR circles, selecting 220\arc\ for the
outer CR as the minimum likely value. The total mass map shows two
faint outer arms that extend towards the outer CR circle -- in fact these
outer spiral arms extend much further than the impression given in the
$g$-band image, as can be seen in the non-deprojected HI
surface density profile (Zhang et al. 1993, Figure 7; as well as a
similar HI image on the THINGS website).

According to the one-CR view of Chemin et al. (2003), the main spiral
would be bar-driven. However, this view would be contradicted by the
two strong disconnected bow-shocks at the CR circle, indicating the
interaction of two pattern speeds (these disconnected bow shock
segments in fact show up more clearly in the gas surface density map).
Truly bar-driven spirals do not contain these disconnected bow-shock
segments, and the spirals would appear as the further continuation of a
skewed bar (see Appendix B).  The outer spiral arms in the case of NGC
3627 clearly do not connect smoothly with the inner bar -- in fact,
there is evidence of two sets of spiral patterns, the inner set with
short arms appears bar-driven whereas the outer set is offset in phase
in the azimuthal direction from the inner one.

With this additional evidence, we propose that prior to
the encounter with NGC 3628, NGC 3627 originally had a two-pattern
structure similar to the CR$_3$ and CR$_4$ regions of NGC 4321.
The interaction tore out a large part of the outer surface density
from NGC 3627, including a segment of the outer spiral arm (Zhang et
al. 1993, Figure 5, clump L).  Therefore the galaxy is evolving
towards a new dynamical equilibrium, with the possibility
of losing the coherent outer spiral pattern eventually.

\subsubsection{NGC 628}

\begin{figure}
\vspace{160pt}
\includegraphics{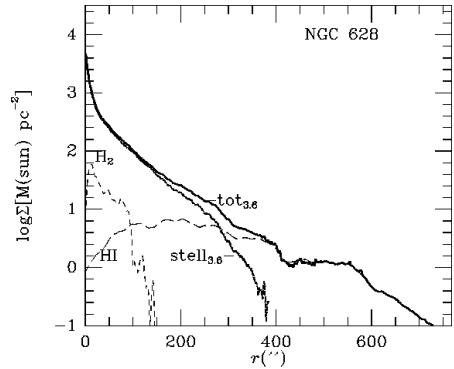}
\caption{Azimuthally-averaged surface mass density profiles of NGC
628. The HI and H$_2$ profiles are shown, and the stellar and
total mass profiles based on a 3.6$\mu$m image.}
\label{fg:Fig10}
\end{figure}

NGC 628 is a late-type spiral of MIR RC3 type SA(s)c (Buta et al.
2010).  It lies at a distance of 8.2 Mpc.  The surface densities of the
different mass components for this galaxy derived using IRAC, BIMA SONG
and THINGS data are shown in Figure \ref{fg:Fig10}.  For this galaxy,
no SDSS data are available.  Unlike the other galaxies in our sample,
NGC 628 (due partly to its late morphological type) has considerable HI
gas at large radii, and this accounts for the significant departure of
the total mass profile from the stellar density profile alone.

\begin{figure}
\vspace{300pt}
\includegraphics{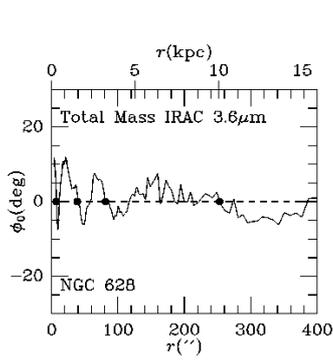}
\includegraphics{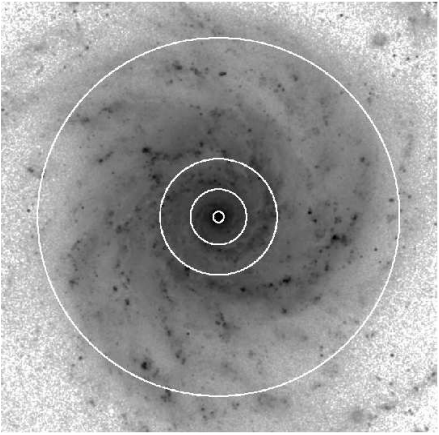}
\caption{{\it Top:} Phase shift between the total mass surface density
and total potential for NGC 628. {\it Bottom:} Deprojected
$B$-band image of NGC 628 overlaid with CR circles.
The main frame covers an area 10\rlap{.}$^{\prime}$1 square.
The units are mag arcsec$^{-2}$.}
\label{fg:Fig11}
\end{figure}

The phase shift plot (Figure~\ref{fg:Fig11}, top) shows
noisier organization compared to all the other
galaxies in this sample, which is consistent with
this galaxy being of very late type (Buta \& Zhang 2009). In the
secular evolution picture this corresponds to a
young pattern still in the process of settling down towards
a dynamical equilibrium state.  Nevertheless, there are 
indications of four P/N crossings which are represented
by the filled circles in the figure.  Table 2 lists the radii
of these crossings.  Figure~\ref{fg:Fig11}, bottom, shows
the overlay of CR circles on a deprojected $B$-band
image of NGC 628.

\subsubsection{NGC 4736}

\begin{figure}
\vspace{160pt}
\includegraphics{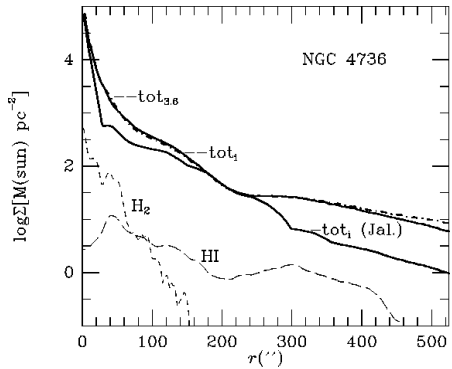}
\caption{Azimuthally-averaged surface mass density profiles of NGC
4736. The layout is the same as in Figure 2, except that graph includes
the profile scaled according to the mass-to-light ratio profile
used by Jalocha et al. (2008).}
\label{fg:Fig12}
\end{figure}

NGC 4736 is an early-type spiral of MIR RC3 type (R)SAB(rl,nr',nl,nb)a
(Buta et al. 2010), implying a galaxy with many distinct features.  It
lies at a distance of about 5 Mpc. In addition to the usual maps with
color-dependent mass-to-light ratio scalings, we have also derived a
map using the SDSS $i$-band image and the mass-to-light profiles
obtained through a fitting procedure by Jalocha et al. (2008). This is
because the usual color-based approach gives a stellar surface density
distribution that leads to a predicted rotation curve much higher than
the observed rotation curve in the central region of the galaxy. We
have not made further correction to account for the fact that Jalocha
et al. used the Cousins $I$-band whereas we are using the SDSS
$i$-band, since we expect the uncertainties in the $M/L$
determination would be larger than these differences. The surface mass
density curves for this galaxy are shown in Figure~\ref{fg:Fig12}. The
tot$_{3.6}$ and tot$_i$ curves were derived in the same manner as for
the other galaxies. The tot$_i$ (Jal.) profile is based on the purely
radius-dependent $M/L$ from Jalocha et al. (2008).

\begin{figure}
\vspace{300pt}
\includegraphics{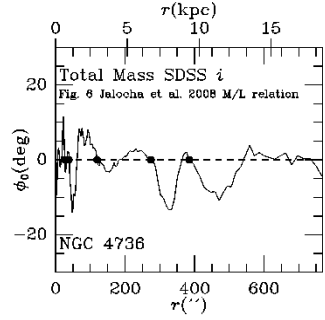}
\includegraphics{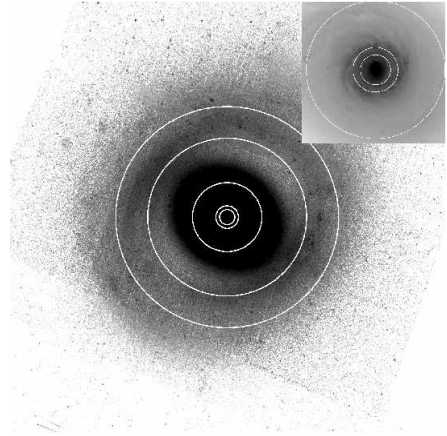}
\caption{{\it Top:} Phase shift between the total mass surface density
and total potential for NGC 4736, derived using the SDSS data.
{\it Bottom:} SDSS $g$-band image of NGC 4736 with the CR circle overlay.
The main frame covers an area 25\rlap{.}$^{\prime}$4 square, while the
inset has an area 4\rlap{.}$^{\prime}$25 square.
The units are mag arcsec$^{-2}$.}
\label{fg:Fig13}
\end{figure}

The phase shift derived using the total mass map and the overlay of CR
circles on the image is given in Figure~\ref{fg:Fig13}. The phase
shift plot (Figure~\ref{fg:Fig13}, top) shows several well-delineated
P/N crossings (Table 2) which appear to correspond well to the resonant
structures in the image (Figure~\ref{fg:Fig13}, bottom). CR$_1$ could
be associated with an inner nuclear bar, while CR$_2$ is a mode
associated with the bright spiral inner pseudoring.  CR$_3$ could be
associated with an intermediate spiral pattern outside the prominent
inner ring.  The fact that CR$_4$ passes through the gap between the
inner and outer rings suggests that it is the actual CR of the massive
oval.  CR$_5$ may be associated with the outer ring pattern. 

\begin{table*}
\caption{Corotation radii from potential-density phase shifts}
\halign{%
\rm#\hfil&
\qquad\rm\hfil#&
\qquad\rm\hfil#&
\qquad\rm\hfil#&
\qquad\rm\hfil#&
\qquad\rm\hfil#&
\rm#\hfil\cr
Galaxy & CR$_1$ & CR$_2$ & CR$_3$ & CR$_4$ & CR$_5$ & ~~~~~Filters \cr
& (arcsec) & (arcsec) & (arcsec) & (arcsec) & (arcsec) & ~~~~~$$ \cr
\noalign{\vskip 10pt} 

NGC 628  &  7.4 & 39.2 & 81.5 & 253.0\rlap{:} & .... & ~~~~~3.6 \cr
NGC 3351 & 26.2$\pm$0.4 & 86.5$\pm$5.1 & .... & .... & .... & ~~~~~3.6,$i$\cr
NGC 3627 & 78.6$\pm$3.5 & ... & .... & .... & .... & ~~~~~3.6,$i$\cr
NGC 4321 & 8.3$\pm$0.7 & 28.8$\pm$0.6 & 51.9$\pm$0.0 & 136.9$\pm$1.6 & .... & ~~~~~3.6,$i$\cr
NGC 4736 & 27 & 39 & 120 & 275 & 385 & ~~~~~ $i$  \cr
NGC 5194 & 21.3$\pm$3.1 & 110.1$\pm$0.1 & .... & .... & .... & ~~~~~3.6,$i$\cr
}
\end{table*}

\subsection{Rotation Curves and Secular Mass Flow Rates} 

\begin{figure}
\vspace{370pt}
\includegraphics{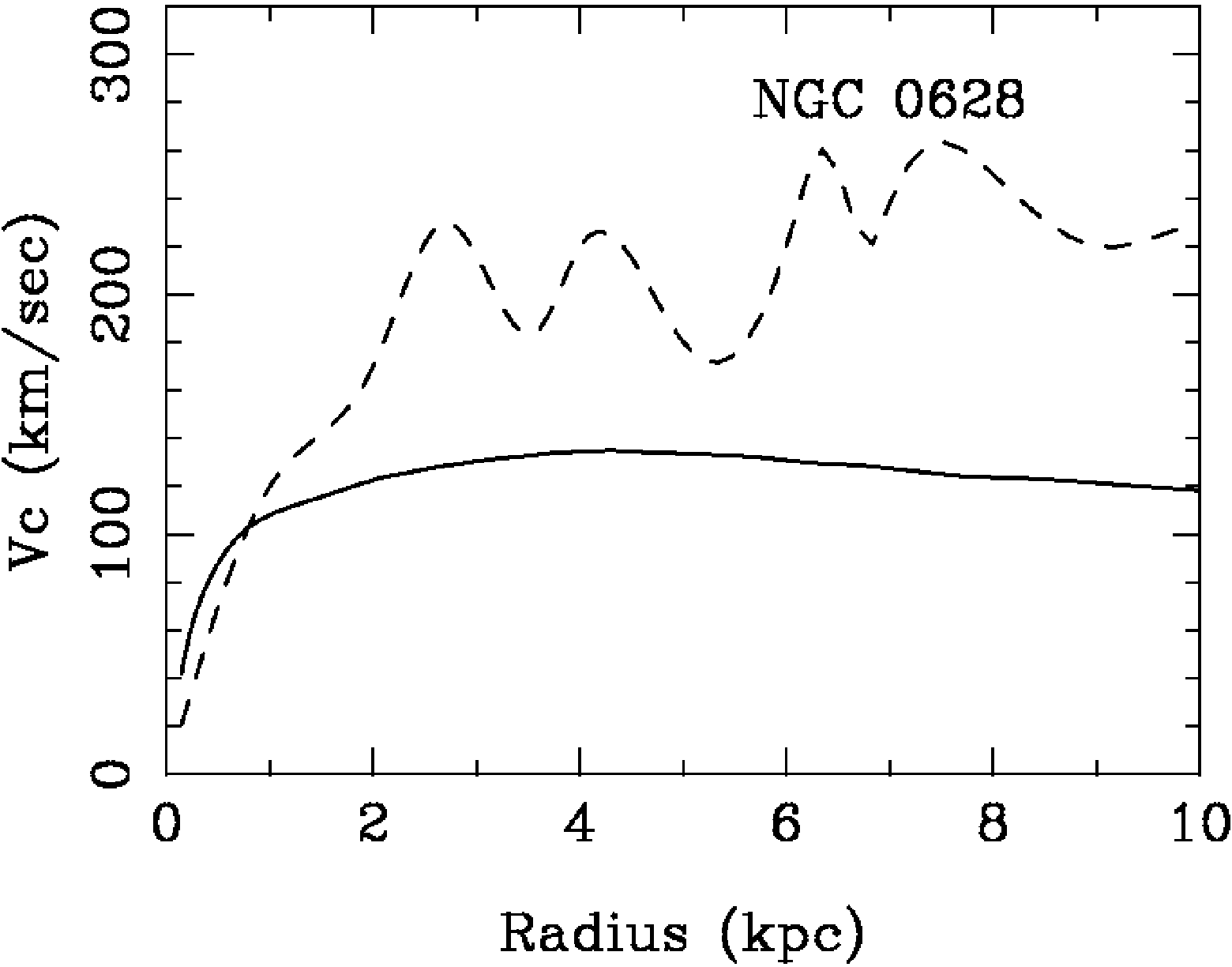}
\includegraphics{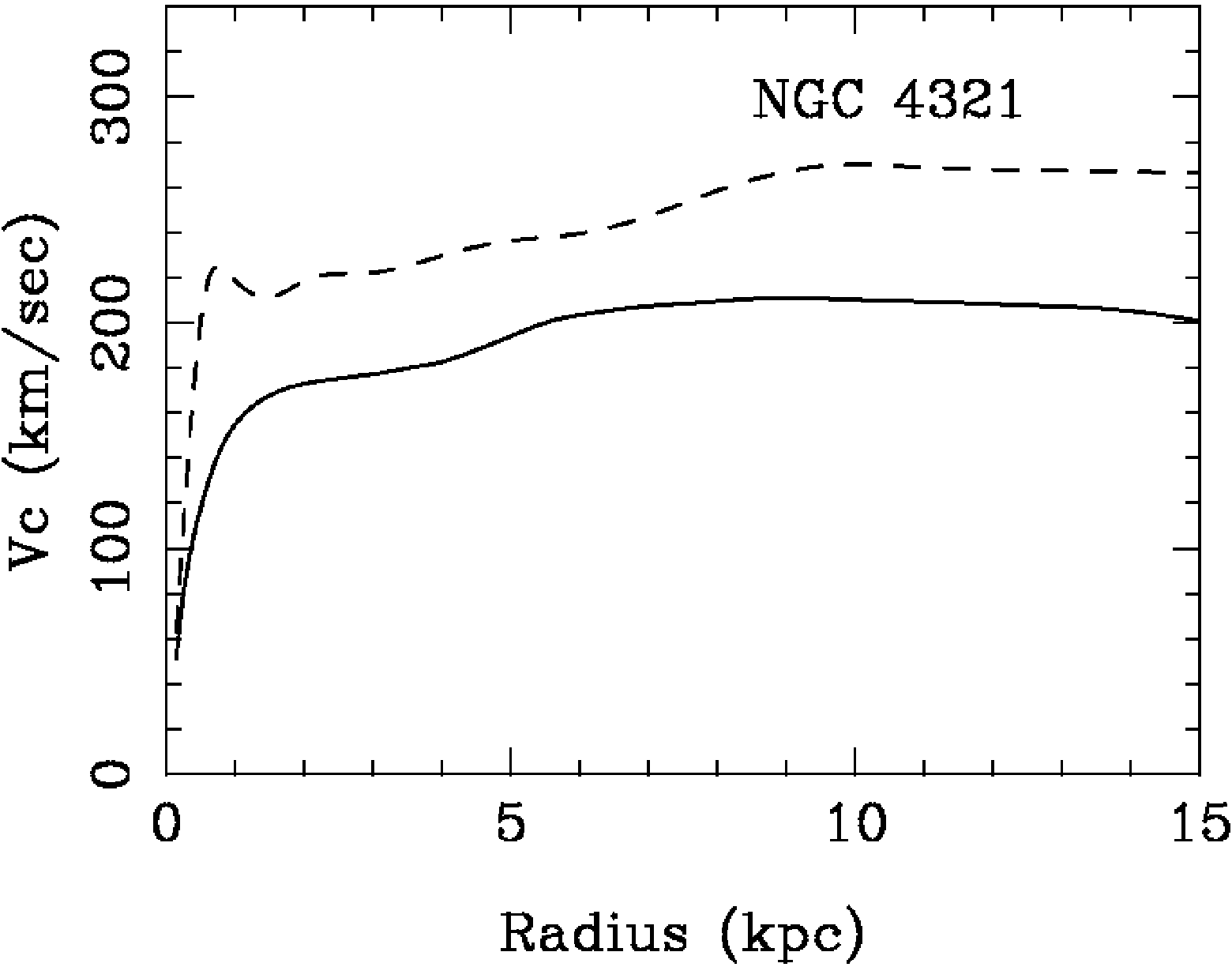}
\includegraphics{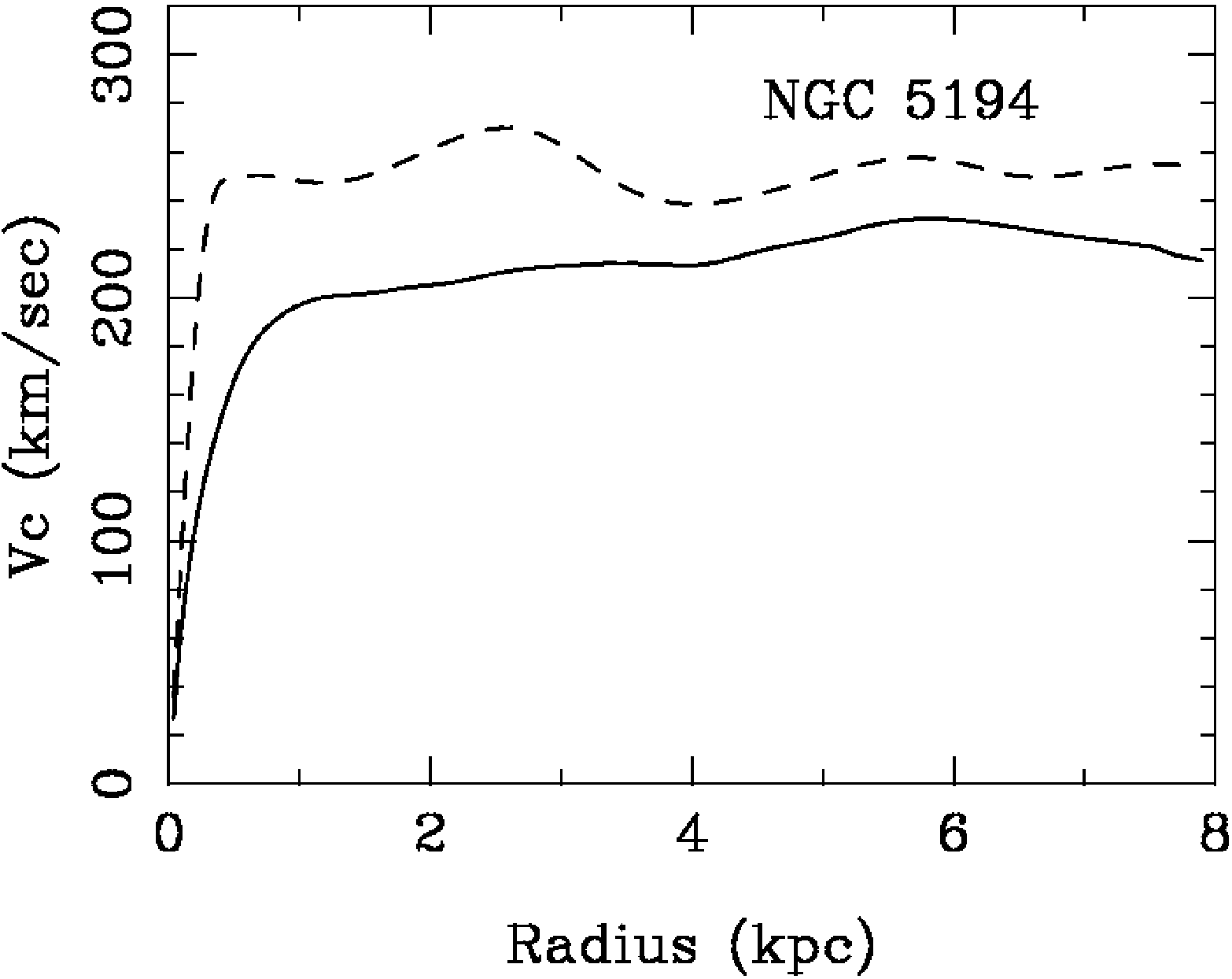}
\includegraphics{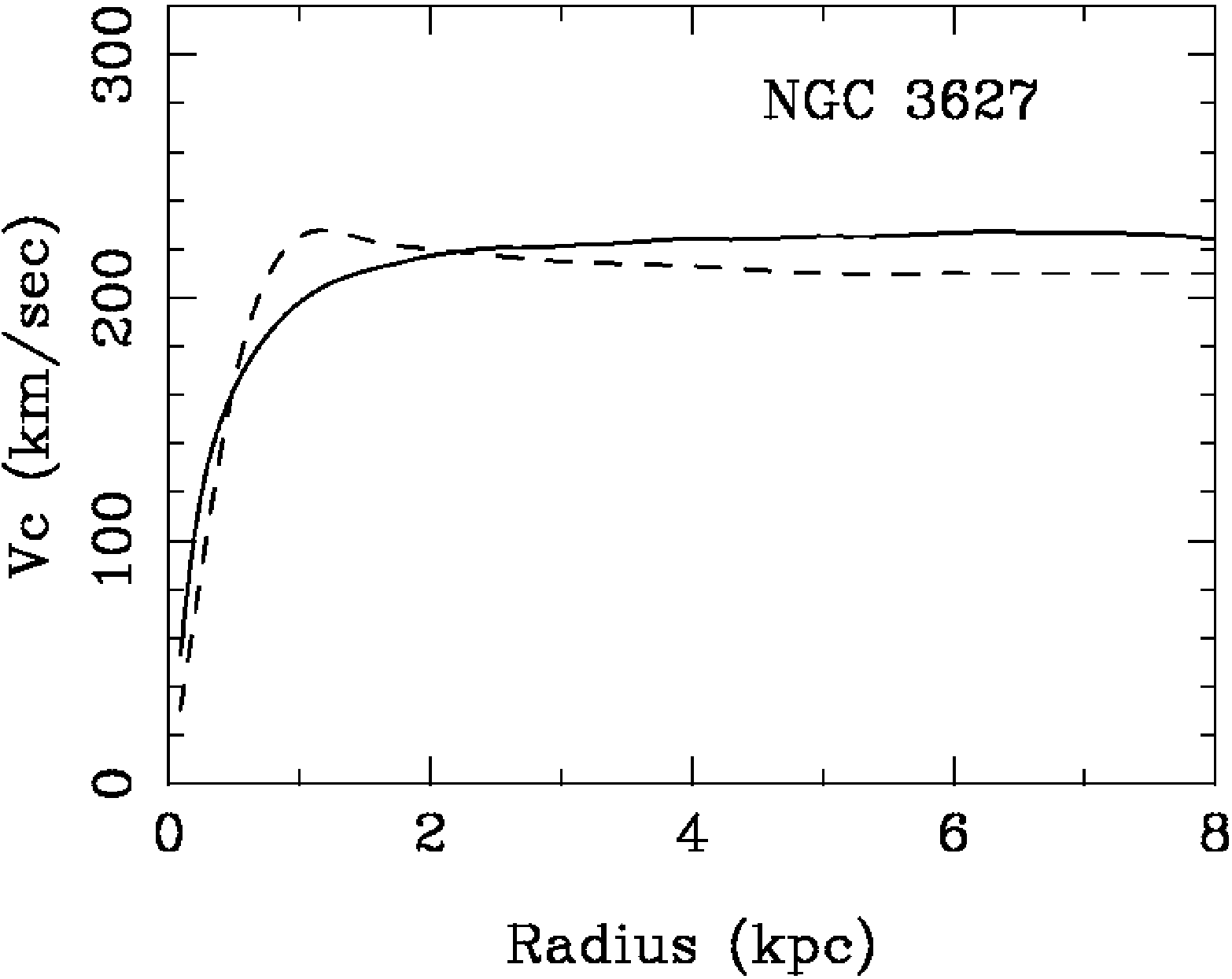}
\includegraphics{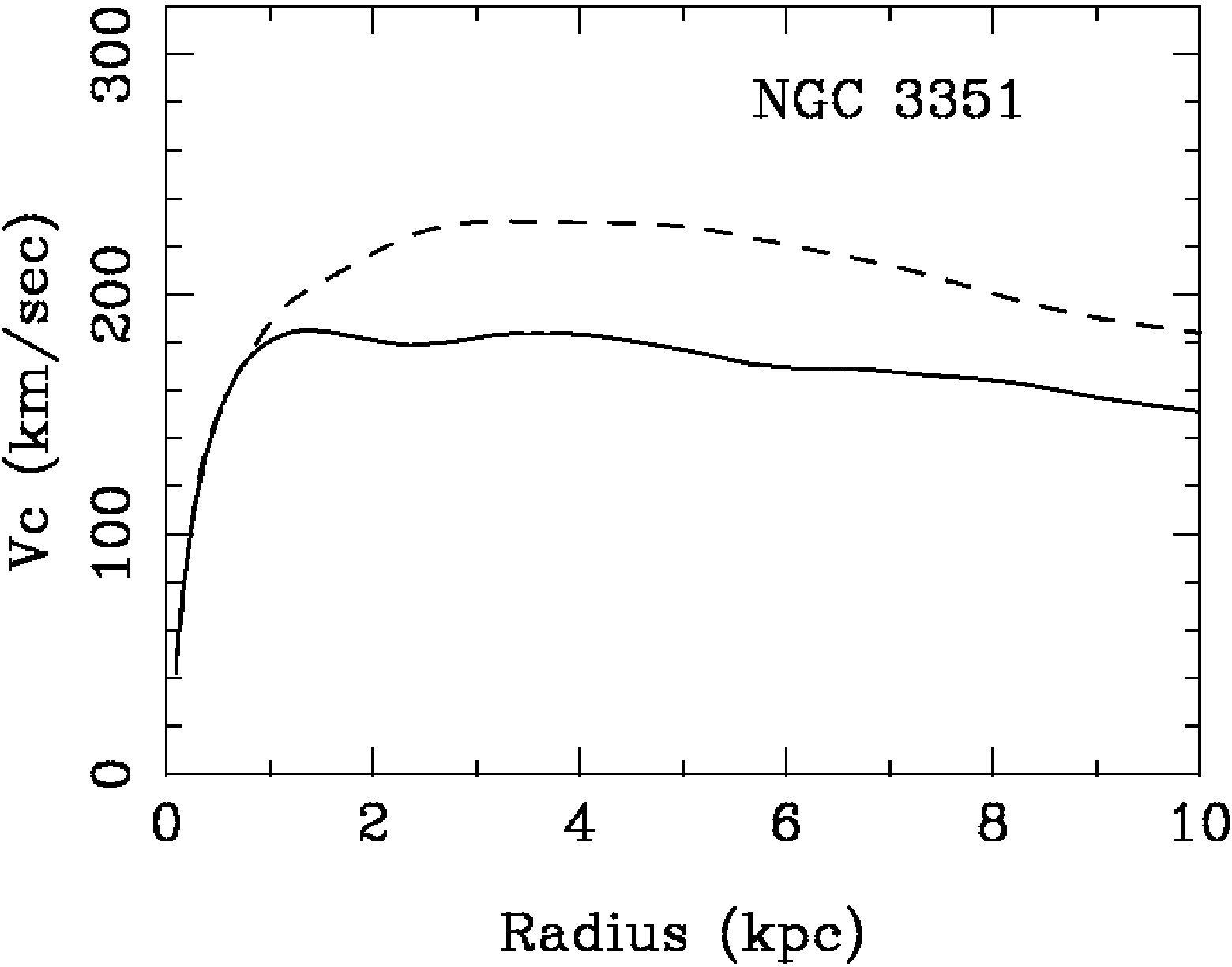}
\includegraphics{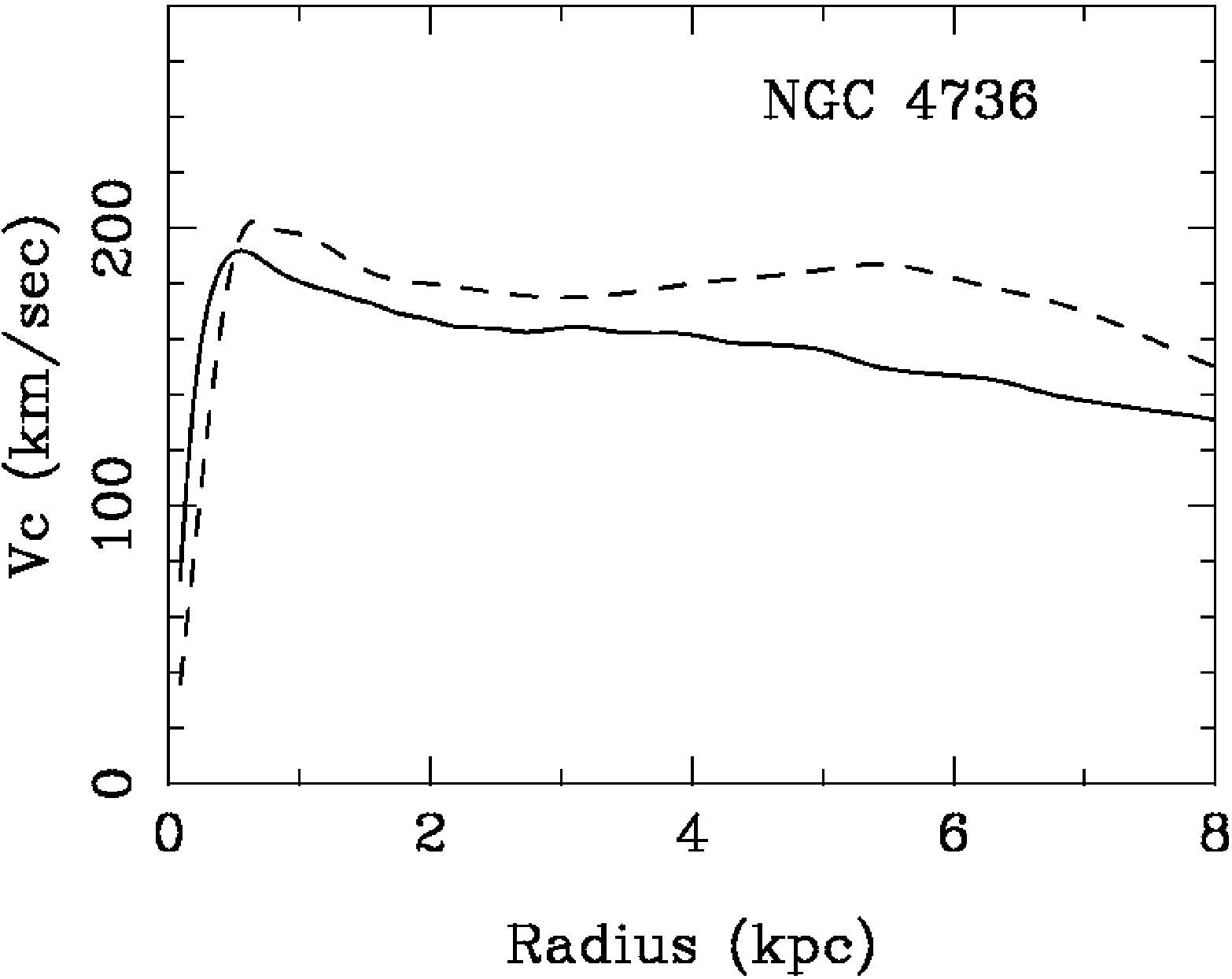}
\caption{Comparison of rotation curves for the six sample galaxies.
{\it Solid lines:} Disk rotation curves inferred from total disk surface density
(stellar plus atomic and molecular hydrogen mass. A very small 
correction to account for helium and the metal abundance in the gas mass has not
been made).  The inferred disk rotation curves
were derived using IRAC 3.6 $\mu$m data for NGC 628, an average
of IRAC 3.6 $\mu$m and SDSS i-band data for NGC 4321, 3351, 3627, 5194, and
using SDSS i-band data for NGC 4736, plus the atomic and molecular gas
contribution from the VIVA, THINGS and BIMA SONG observations.
{\it Dashed lines:} Observed rotation curves: 
NGC 628 - Nicol (2006);
NGC 3351 - Inner part from Devereux, Kenney, \& Young (1992)'s CO 1-0 observation, outer part from Buta (1988) H-alpha observations;
NGC 3627 - Inner part from Zhang et al. (1993)'s CO 1-0 observation (rescaled to the current distance value, and changed to use terminal velocity instead of peak velocity of the molecular contours), outer part from Chemin et al. (2003)'s H-alpha observation;
NGC 4321 - Sofue et al. (1999);
NGC 4736 - From Jalocha et al. (2008), adapted from the compilation of Sofue et al. (1999);
NGC 5194 (Sofue 1996)
}
\label{fg:Fig14} 
\end{figure}

\begin{figure}
\vspace{370pt}
\includegraphics{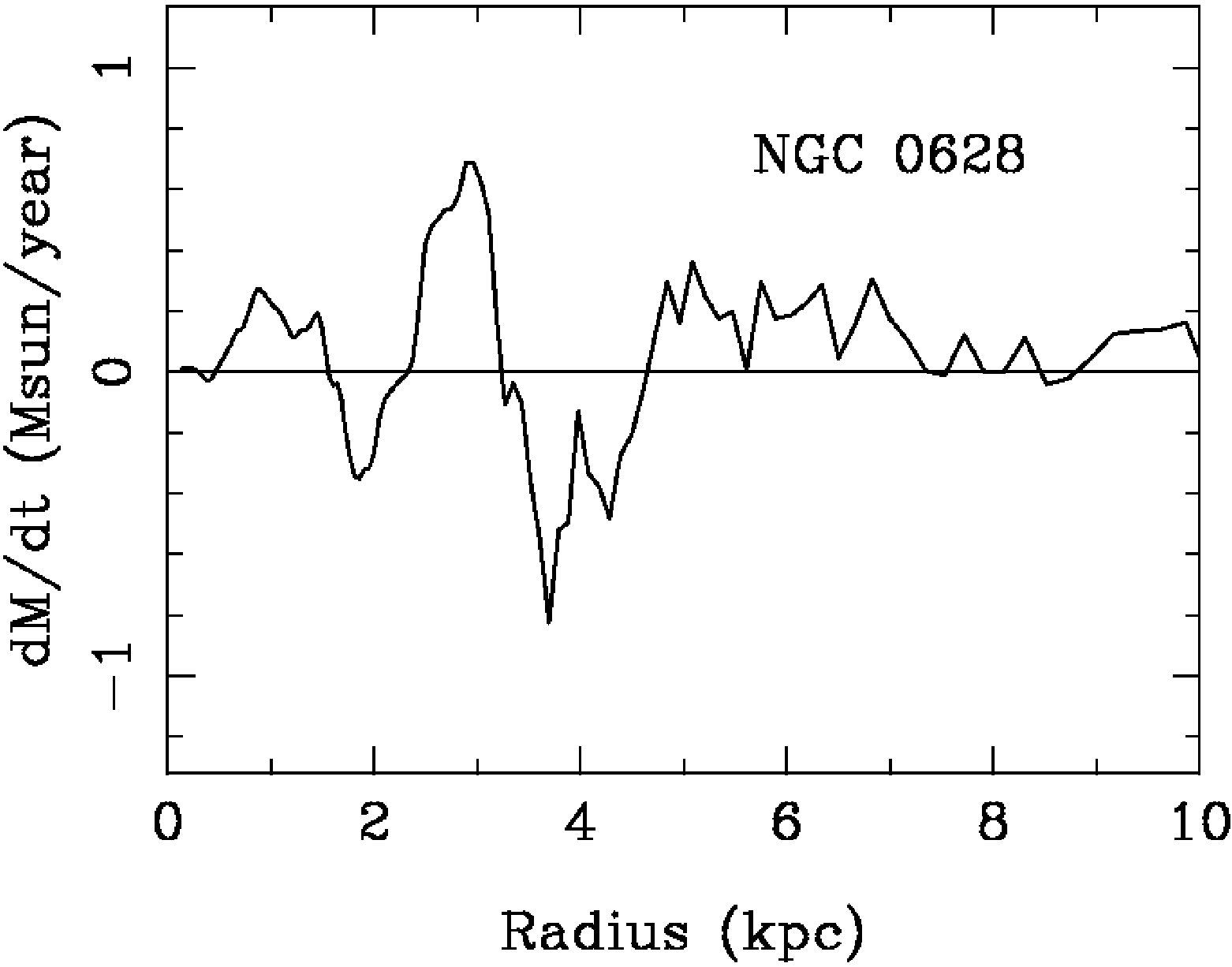}
\includegraphics{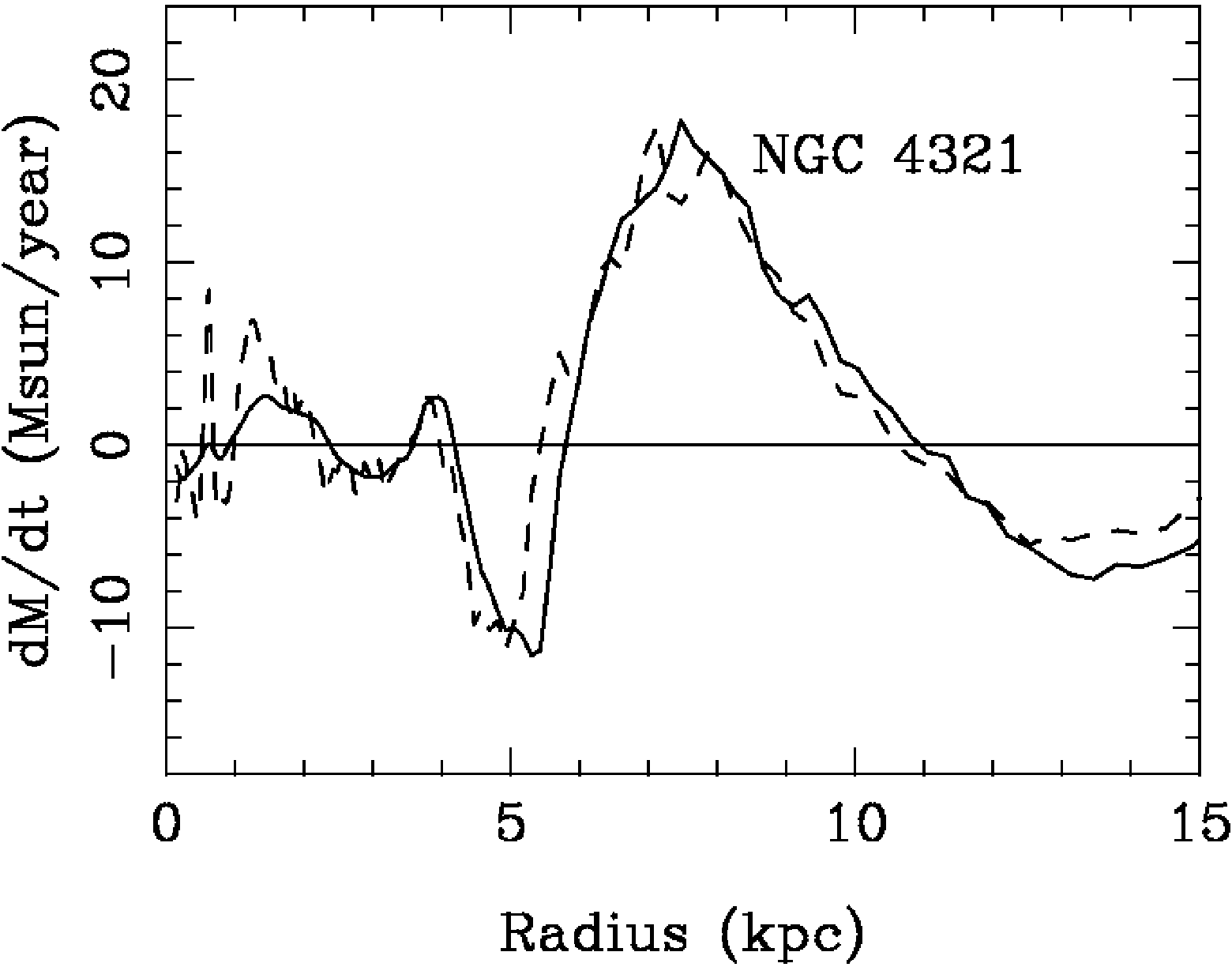}
\includegraphics{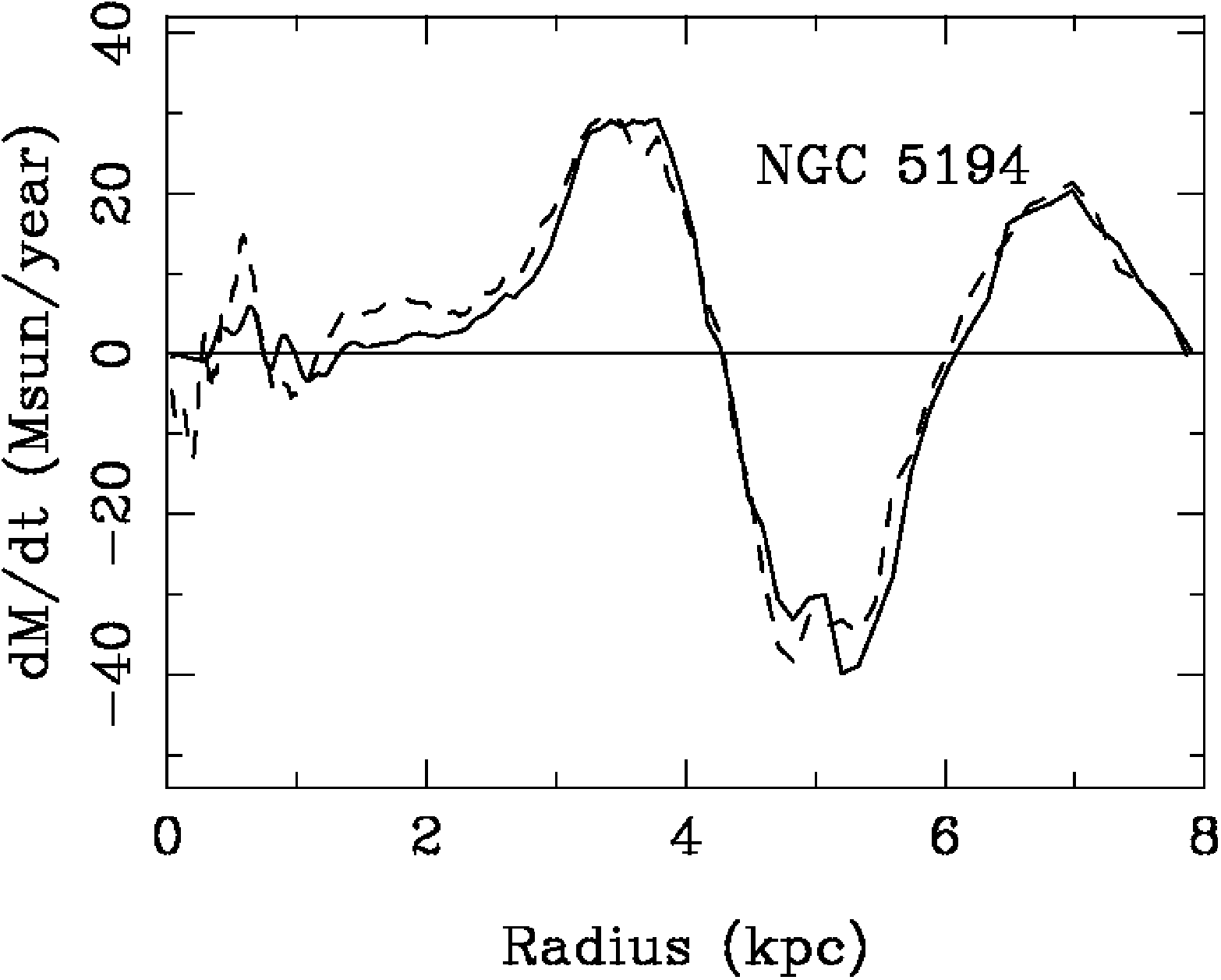}
\includegraphics{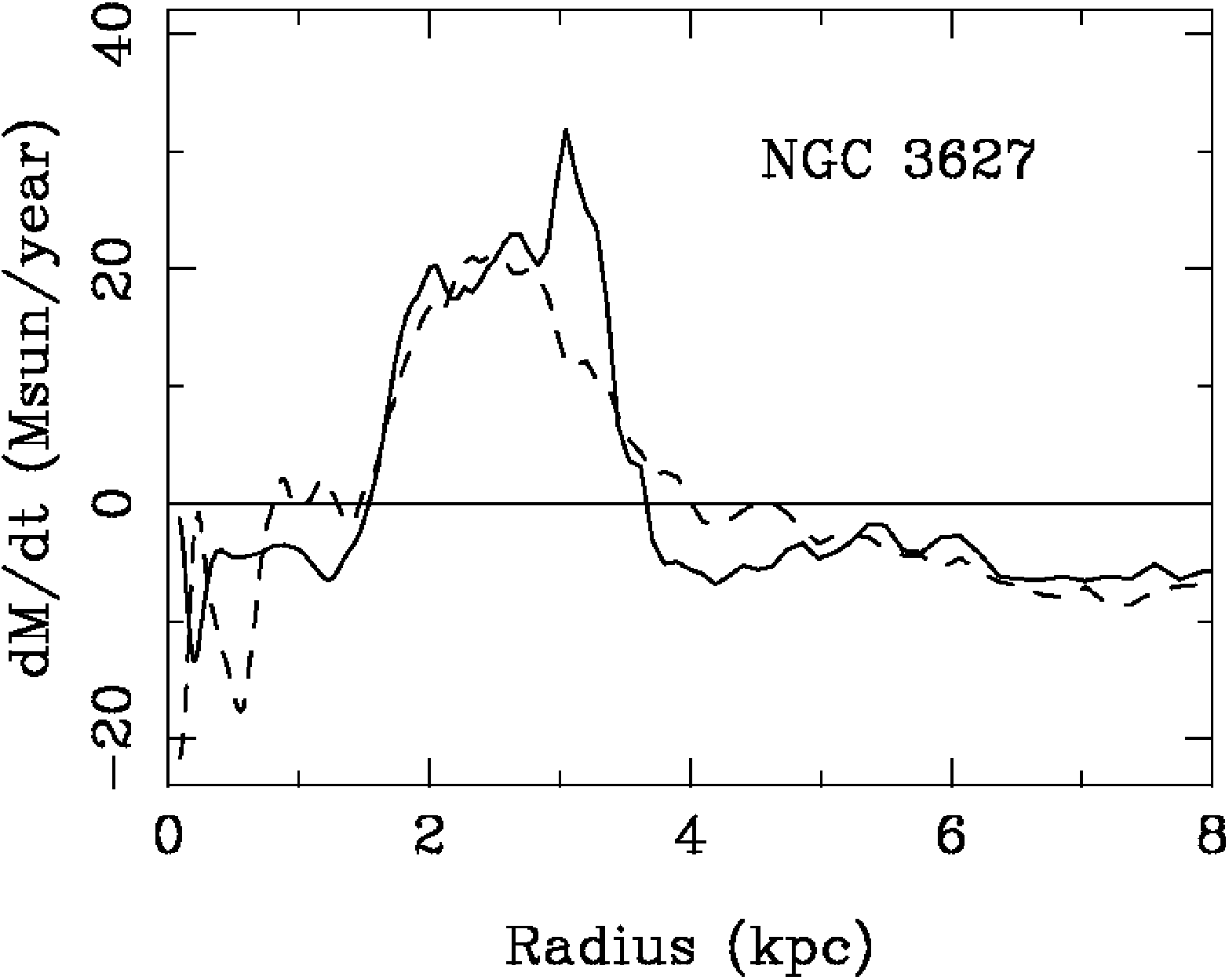}
\includegraphics{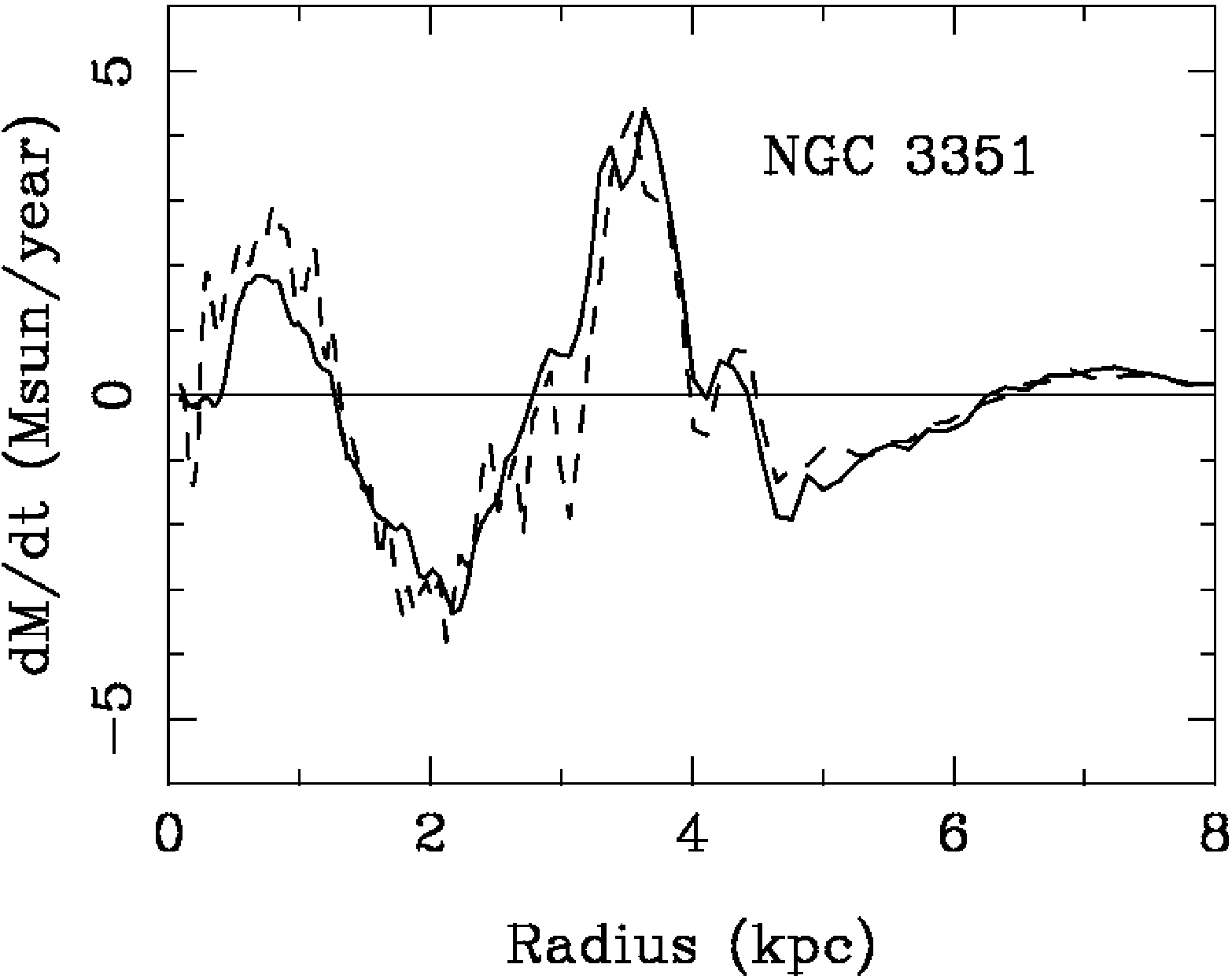}
\includegraphics{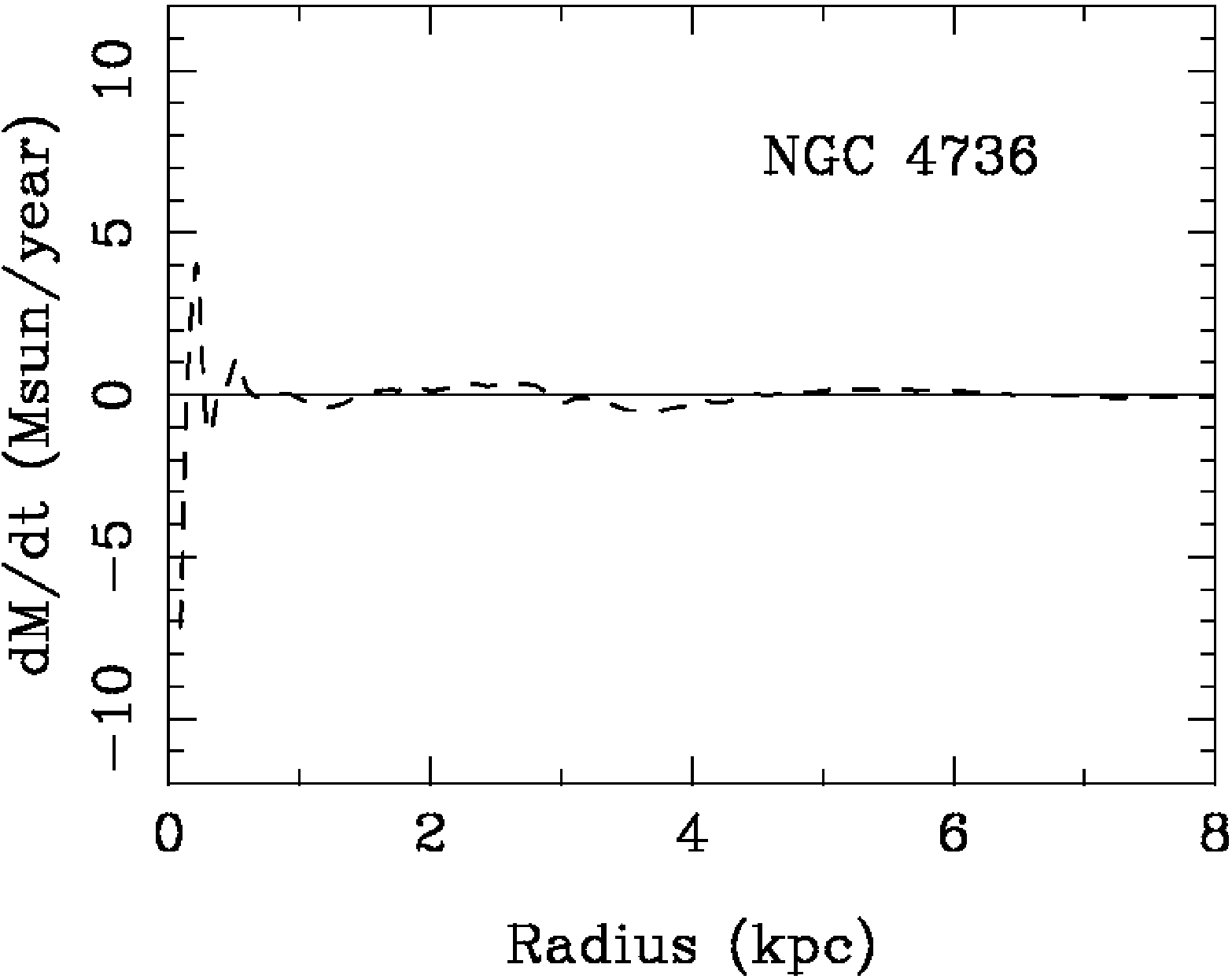}
\caption{Total radial mass flow rate for the six sample galaxies. Positive
portion of the curves indicate inflow, and negative portion of the
curves indicate outflow. The solid lines have stellar mass derived from
IRAC 3.6$\mu$m data, and the dashed lines have stellar mass derived
from SDSS i-band data.  Atomic and molecular gas mass from VIVA, THINGS
and BIMA SONG were added to obtain the total mass maps which were used
to derive these flow rates.}
\label{fg:Fig15}
\end{figure}

In the next group of figures (Figure~\ref{fg:Fig14} and
Figure~\ref{fg:Fig15}), we present the rotation curves (observed
versus disk-total-mass-derived) as well as the total mass flow rates
for our sample galaxies. The arrangement of the different
frames, NGC 628, NGC 4321, NGC 5194, NGC 3627, NGC 3351, NGC 4736, from
left to right, then top to bottom, is chosen to be roughly along the
Hubble sequence from late to early types, in order to reveal any
systematic trends along this sequence. Because two of the
intermediate types (NGC 5194 and NGC 3627) are strongly interacting
galaxies, they might deviate from the quiescent evolution trends.
Since there is usually a limited radial range for the availability of
observed rotation curves, our subsequent plots in this section (Figures
\ref{fg:Fig14}-\ref{fg:Fig20}) will be displayed with smaller ranges
than were used for the corresponding phase shift plots.

From Figure~\ref{fg:Fig14}, we observe that the contribution of the
disk baryonic matter (excluding the contribution from helium and heavy
elements) to the total rotation curve increases as the galaxy's Hubble
type changes from late to early along the Hubble sequence.  For
early-type galaxies such as NGC 4736, the entire rotation curve may be
accounted for by disk baryonic matter (Jalocha et al. 2008).  Note that
the close match between the observed and disk-inferred rotation curves
for NGC 3627, which is unusual for its intermediate Hubble type, may
be a result of the close encounter with NGC 3628 which is likely to have
stripped a large portion of its halo.

The mass flow rates obtained from equation 2 are not sensitively
dependent on the choice of scale height in the potential calculation
(i.e. a factor of 3.3 change of vertical scale height from 12.6" 
to 3.8" in the case of NGC 4321 changed the scale of mass flow 
by less than 10\% for the entire radial range considered).  So the
scale heights we used from Table 1 for each individual galaxies
should give a fairly good estimate of the mass flow rates.
From Figure~\ref{fg:Fig15}, the mass
flow rates for the various Hubble types range from a few solar masses
per year to about a few tens of solar masses a year, except for the
very late-type galaxy NGC 628 which is still in the process of settling
into a significant galaxy-wide mass flow pattern.  The
intermediate-type galaxies appear to have the largest mass flow rates
(e.g., NGC 1530, studied in ZB07, has mass flow rates on the order of
100 solar masses a year and is also an intermediate-type galaxy), whereas
for the early-type galaxies the mass flow is more concentrated to the
central region.

The averaged radial flow velocities $v_R(R)$ corresponding to the
averaged mass flow rates $dM/dt(R)$ displayed in Figure~\ref{fg:Fig15}
can be obtained by dividing the mass flow rate $dM/dt$ by $2 \pi R
\Sigma_0$, where $\Sigma_0$ is the local averaged surface density. This
gives a radial velocity typically on the order of a few km/sec.
Since the non-circular velocities in a galaxy possessing highly
nonlinear density wave patterns can be significantly larger than these
radial flow velocities, a direct measurement of the radial flow
velocities may not be as reliable a way of estimating the mass flow
rate compared to the torque/phase shift approach described here.

Although both inflow and outflow of mass are present at a given epoch
across a galaxy's radial range, there is a general trend for the mass
to concentrate into the inner disk with time, together with the
build-up of an extended outer envelope, consistent with the direction
of entropy evolution. This is confirmed in Figure~\ref{fg:Fig16}, which
shows that for late and intermediate-type galaxies the gravitational
torque couple (as well as the advective torque couple which has similar
shape) has mostly positive values in the inner disk, indicating that
the angular momentum is consistently being channeled out of the
galaxy.  The mass flow and angular momentum channeling progressively
moves from the outer disk to the inner disk as a galaxy evolves from
late to the early Hubble types.

The radial mass flow rates shown in these curves are the lower bounds
on the actual rates in physical galaxies, since the nonaxisymmetric
instabilities are expected to result in skewed distributions in parts
of the thick-disk and halo as well, and thus result in the radial
redistribution of the halo dark matter along with the luminous disk
matter.  Besides the skewed-pattern induced collective radial
mass inflow/outflow, we also expect a small amount of additional inflow due
to differential rotation and dynamical friction. 

\subsection{Relative Contributions from Gravitational and Advective
Torque Couples}

\begin{figure}
\vspace{370pt}
\includegraphics{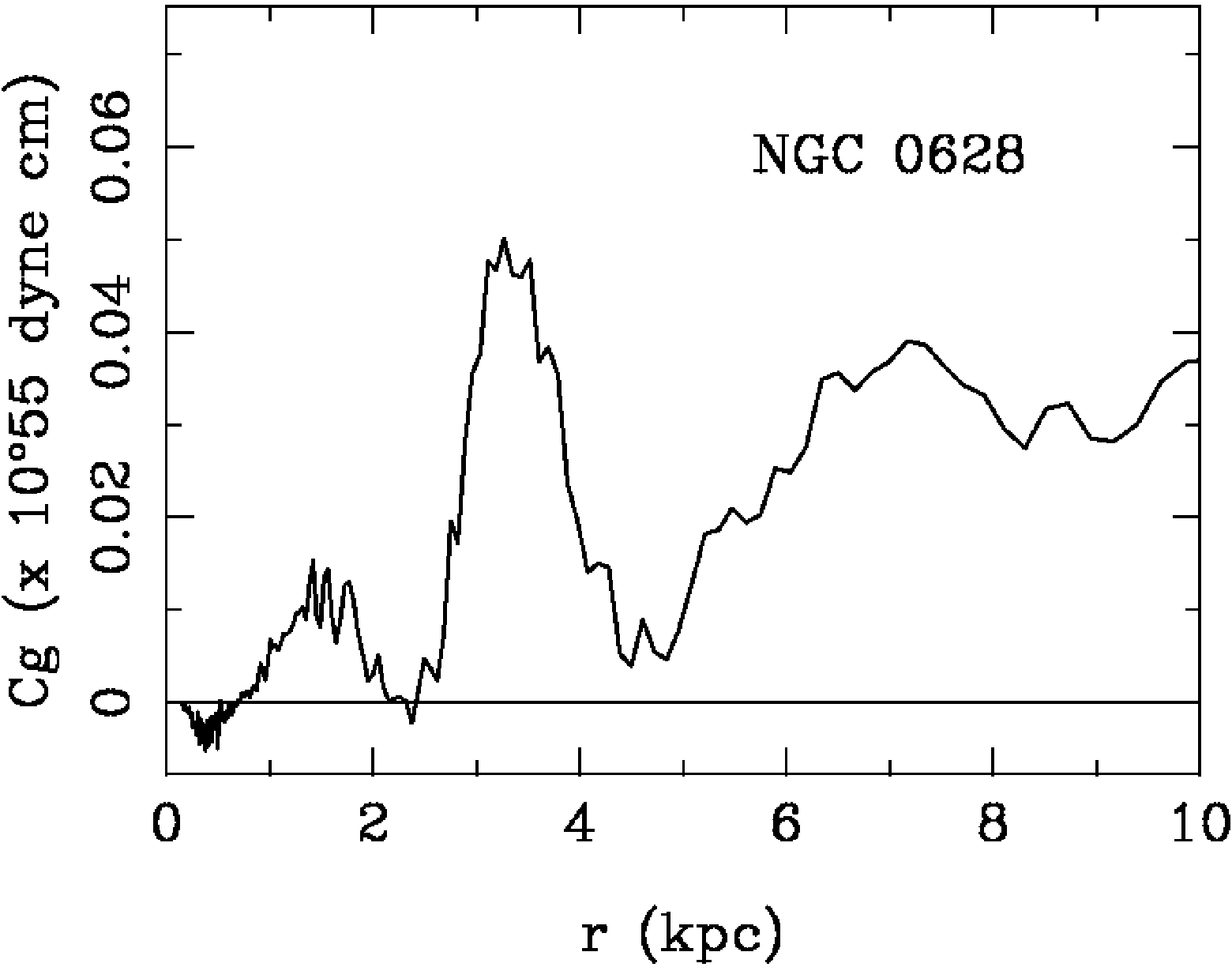}
\includegraphics{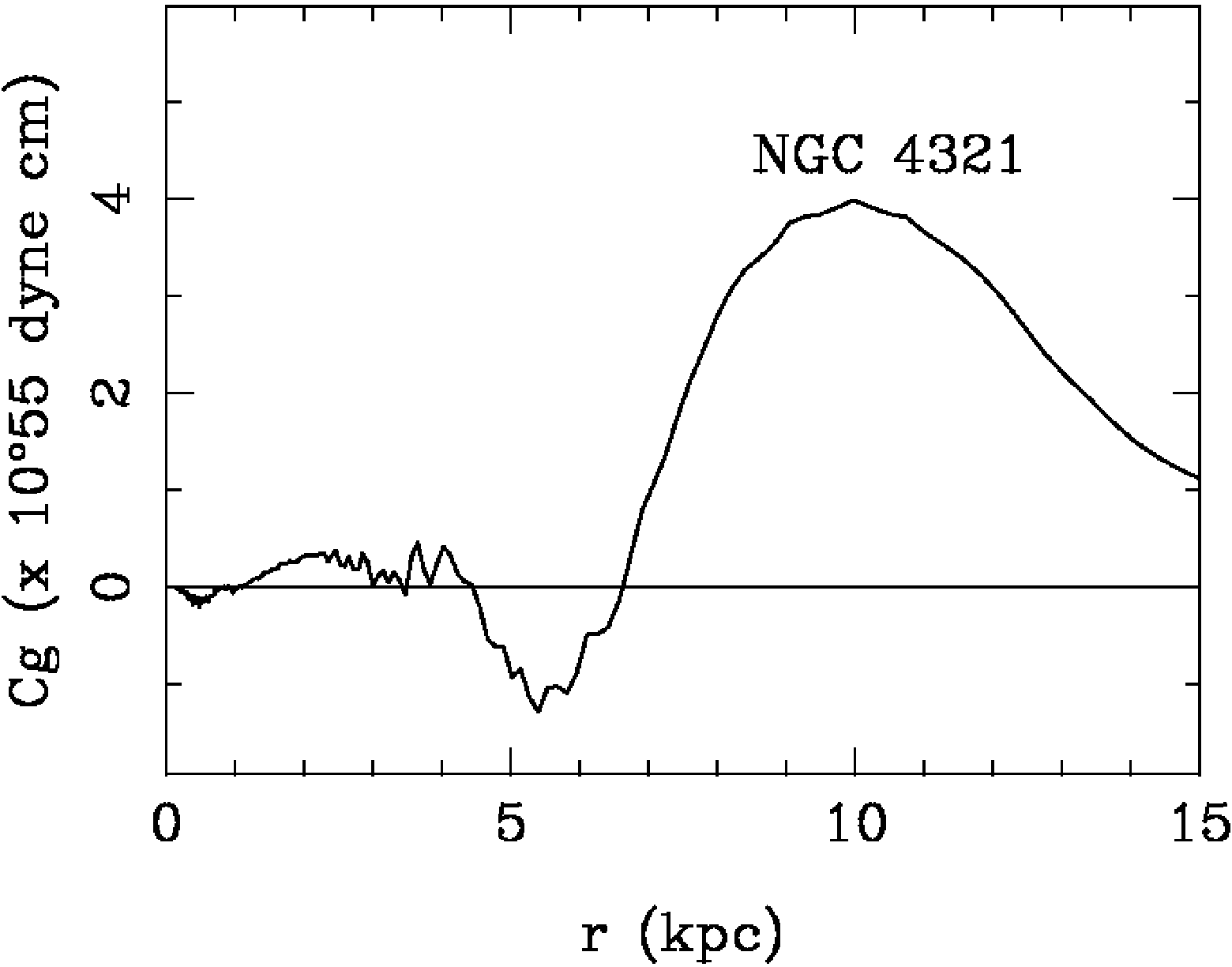}
\includegraphics{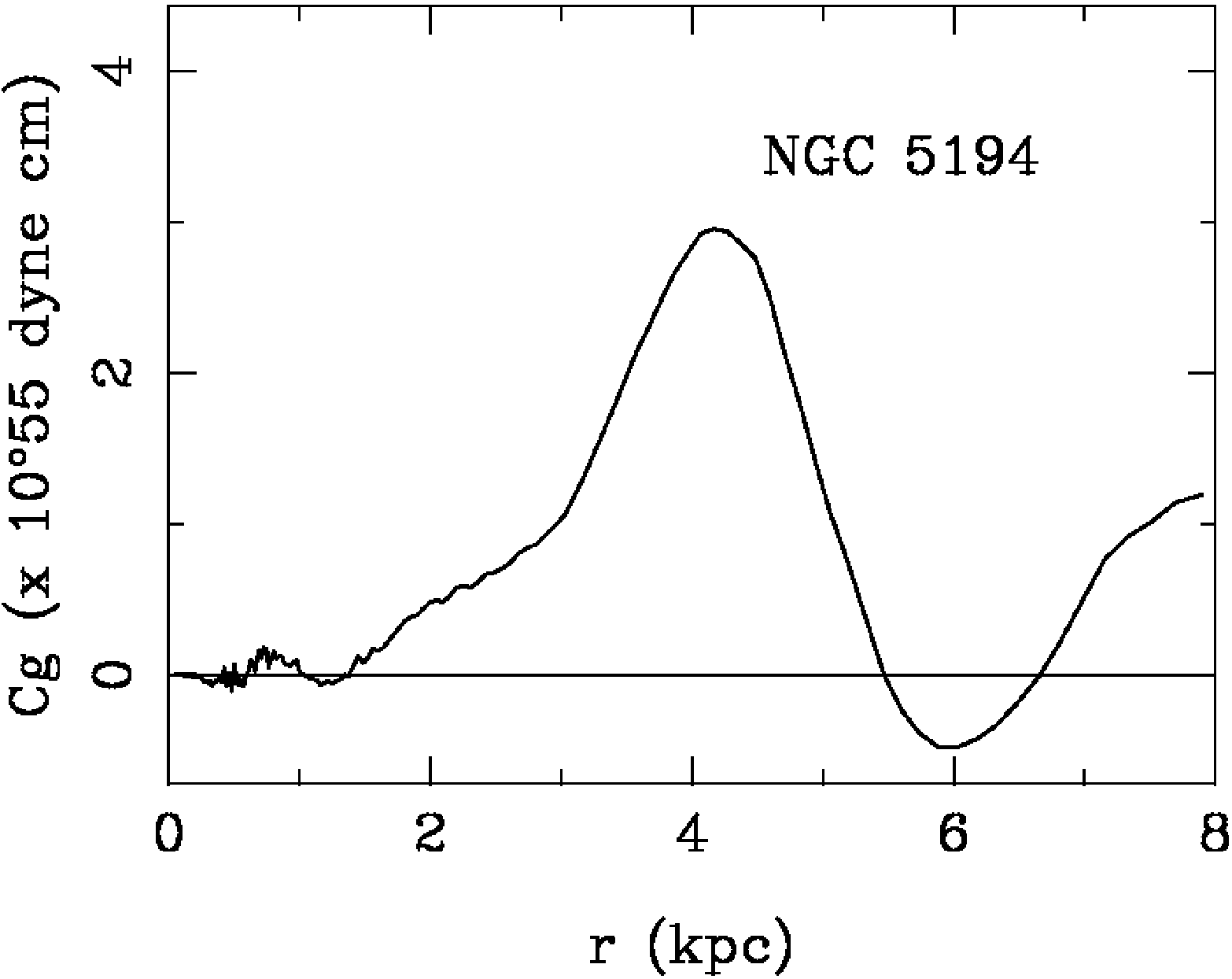}
\includegraphics{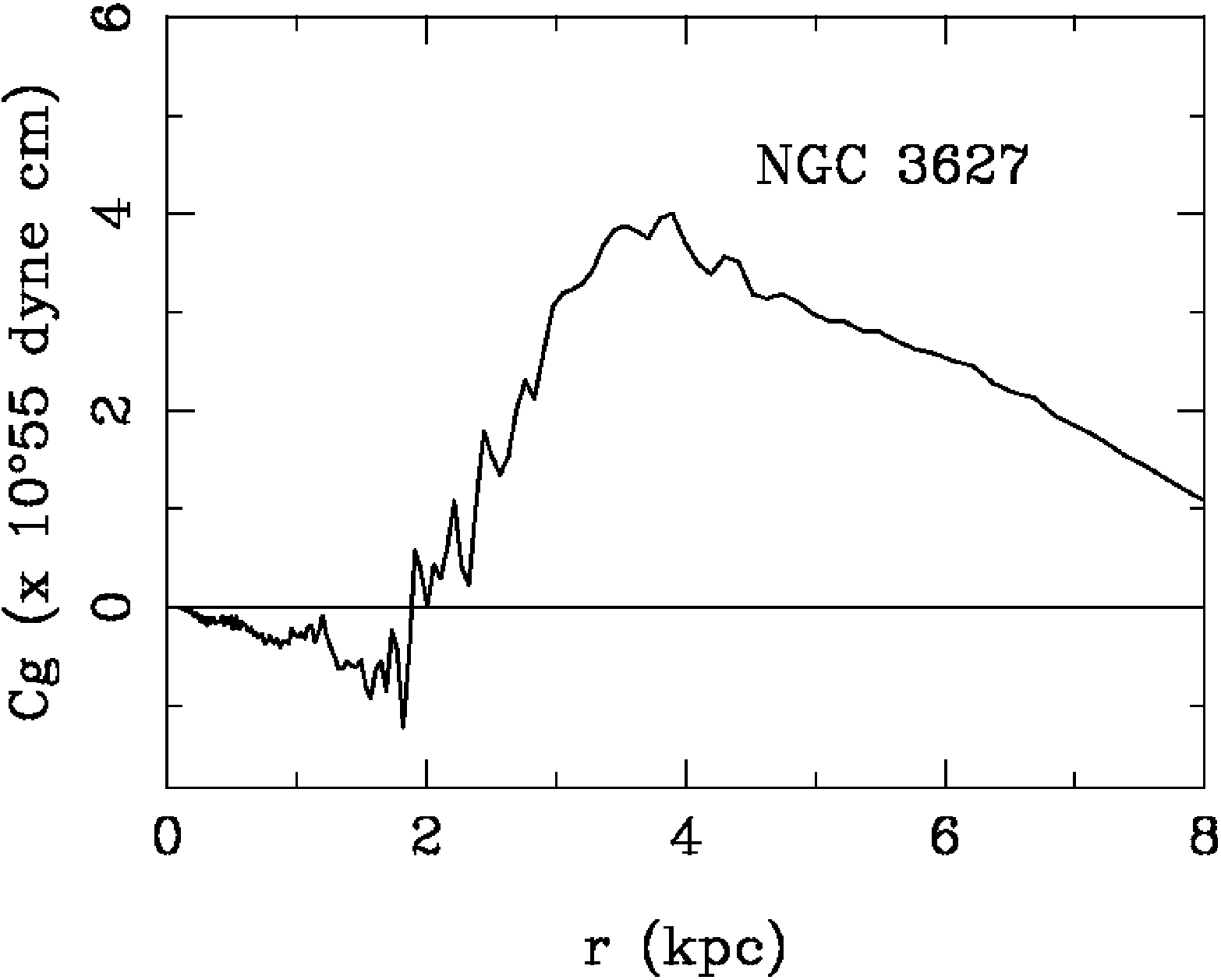}
\includegraphics{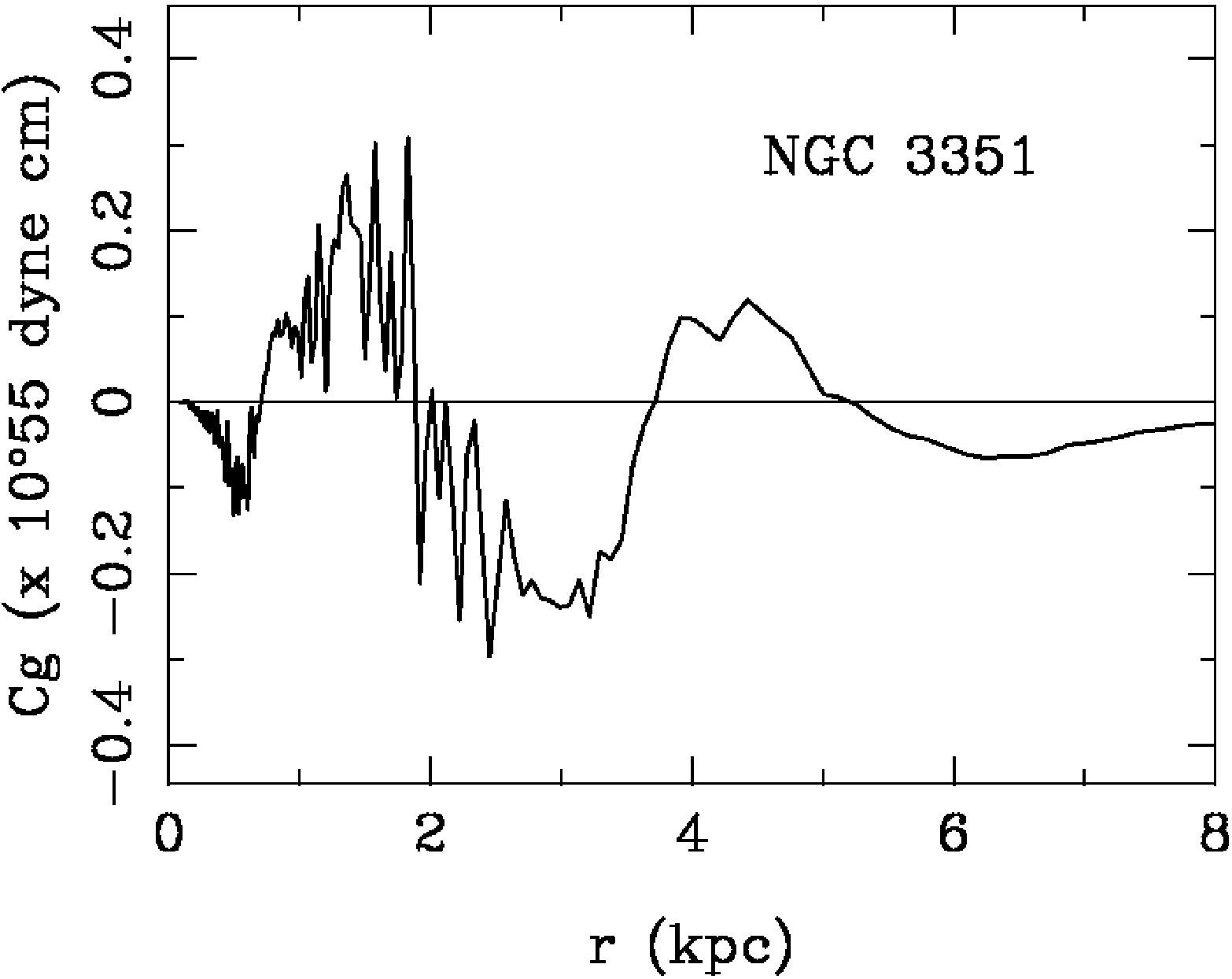}
\includegraphics{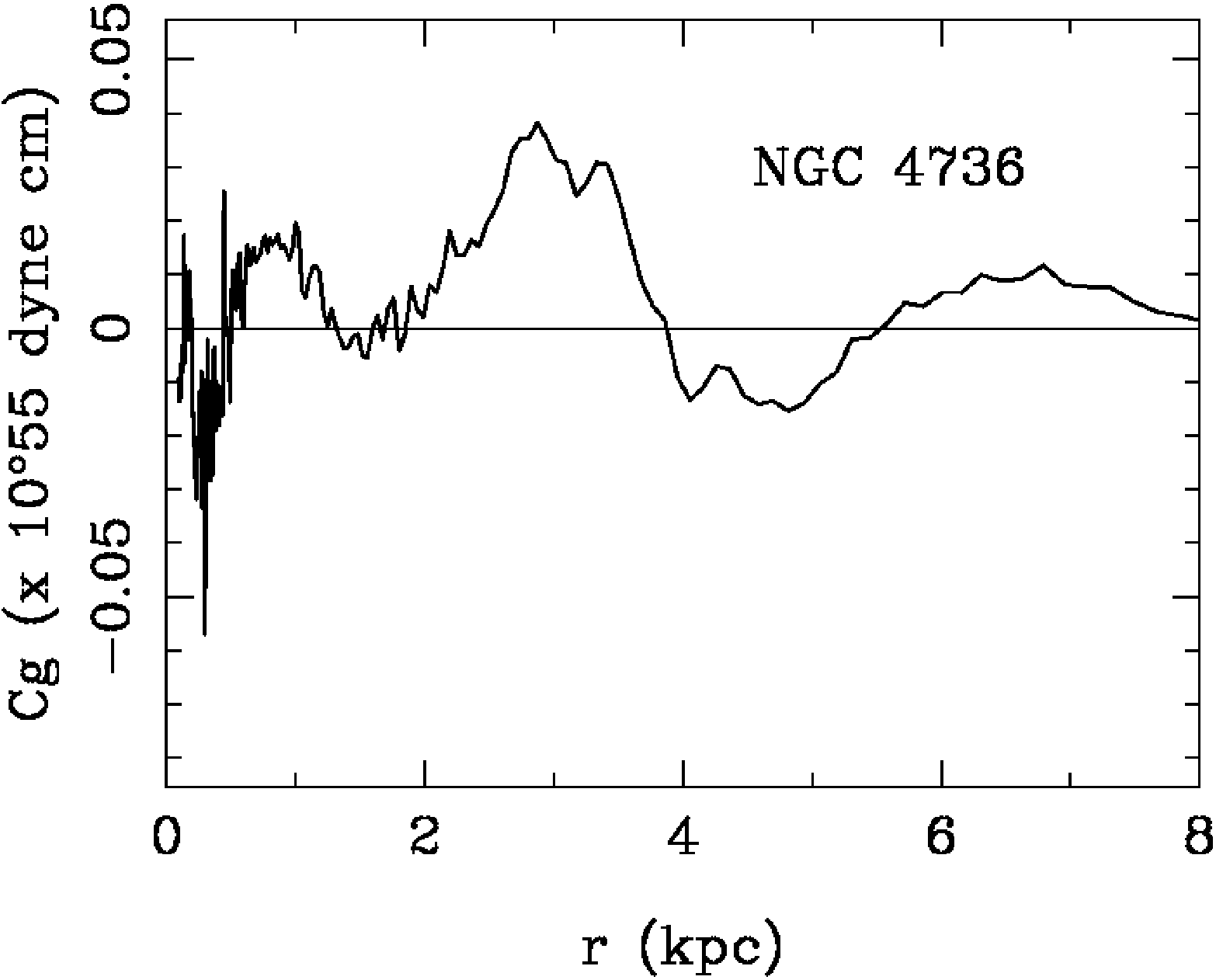}
\caption{Calculated gravitational torque coupling for the six sample
galaxies, using IRAC 3.6 $\mu$m data for NGC 628,
an average of IRAC 3.6 $\mu$m and SDSS i-band data for NGC 4321, 3351,
5194, 3627, and SDSS i-band data for NGC 4736, to these the
usual gas maps were added.}
\label{fg:Fig16}
\end{figure}

In Figure~\ref{fg:Fig16}, we plot the calculated gravitational torque
couplings for the six sample galaxies. We note these characteristics:

\noindent
1. The torque curve for M100 is very similar in shape to the one
calculated for the same galaxy by Gnedin et al. (1995), although the
scale factor is 10 times smaller than obtained in their paper. Part of
the difference can be accounted for from the difference in galaxy
images used (i.e. wave bands) and in the galaxy parameters used between
these two studies. We have used an $R$-band image as in Gnedin et al.,
and rescaled the galaxy parameters to be in agreement with what they
had used. However, as was also found by Foyle et al. (2010), after
these adjustments the resulting scale still differs by a factor of
$\sim 5$ from that in Gnedin et al. (1995).

\noindent
2. For most of the galaxies in the sample, $C_g$ has a primary peak
which is centered in the middle of the disk, suggesting that the mass
inflow/outflow should be focused near the peak location of the main
bell curve. This location is expected to change with time for
a given galaxy as secular evolution proceeds because the basic state
configuration and the resulting modal configuration will both evolve
due to the redistribution of the disk mass. By examining this
small sample of six galaxies, we can already see a trend that the main
peak of the $C_g$ curve moves from the outer region of the galaxy (as
for NGC 628) to the mid-disk (as for NGC 4321, 5194 and 3627), and then
to the central region (NGC 3351 and 4736).

\noindent
3. The existence of multiple humps in the gravitational torque couple
\footnote{as well as in the advective and total torque couple, since
these latter two are found to be of similar shape to the gravitational
torque couple for self-sustained modes} in a single galaxy, and by
implication multiple nested modes of differing pattern speeds, shows
that the secular mass flow process is {\it not a one-way street}.
During the lifetime of a galaxy a given radial range will experience
inflows and outflows of different magnitudes, but there is an overall
trend towards a gradual increase of central concentration together with
the build-up of an extended envelope, consistent with the entropy
evolution direction of self-gravitating systems. Galaxies accomplish
this in a well-organized fashion, by first employing a dominant mode
across the galaxy's outer disk, and then with more of the modal
activity moving towards the central region of a galaxy, by bifurcating
the single dominant mode into multiple nested modes.

\begin{figure}
\vspace{370pt}
\includegraphics{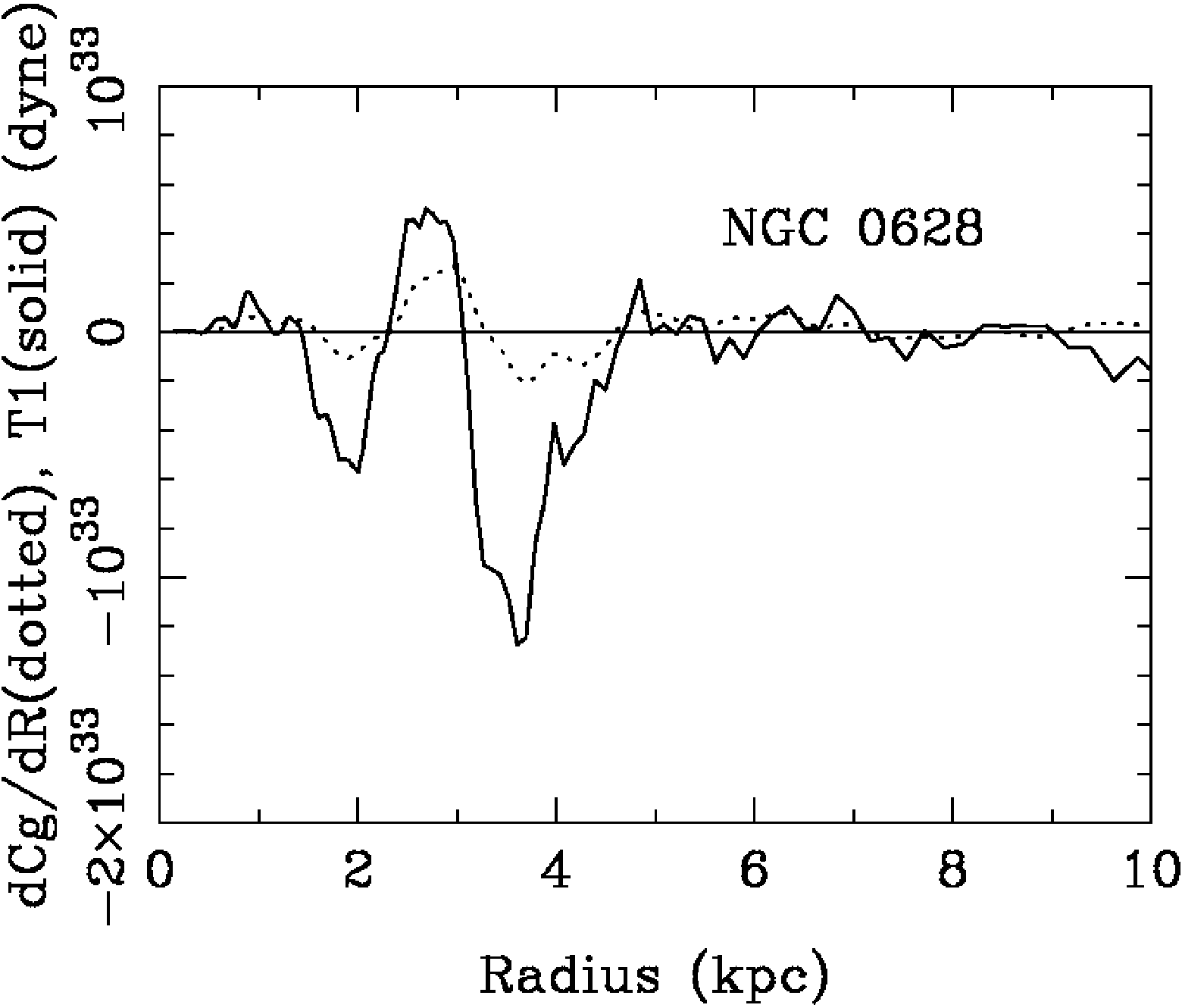}
\includegraphics{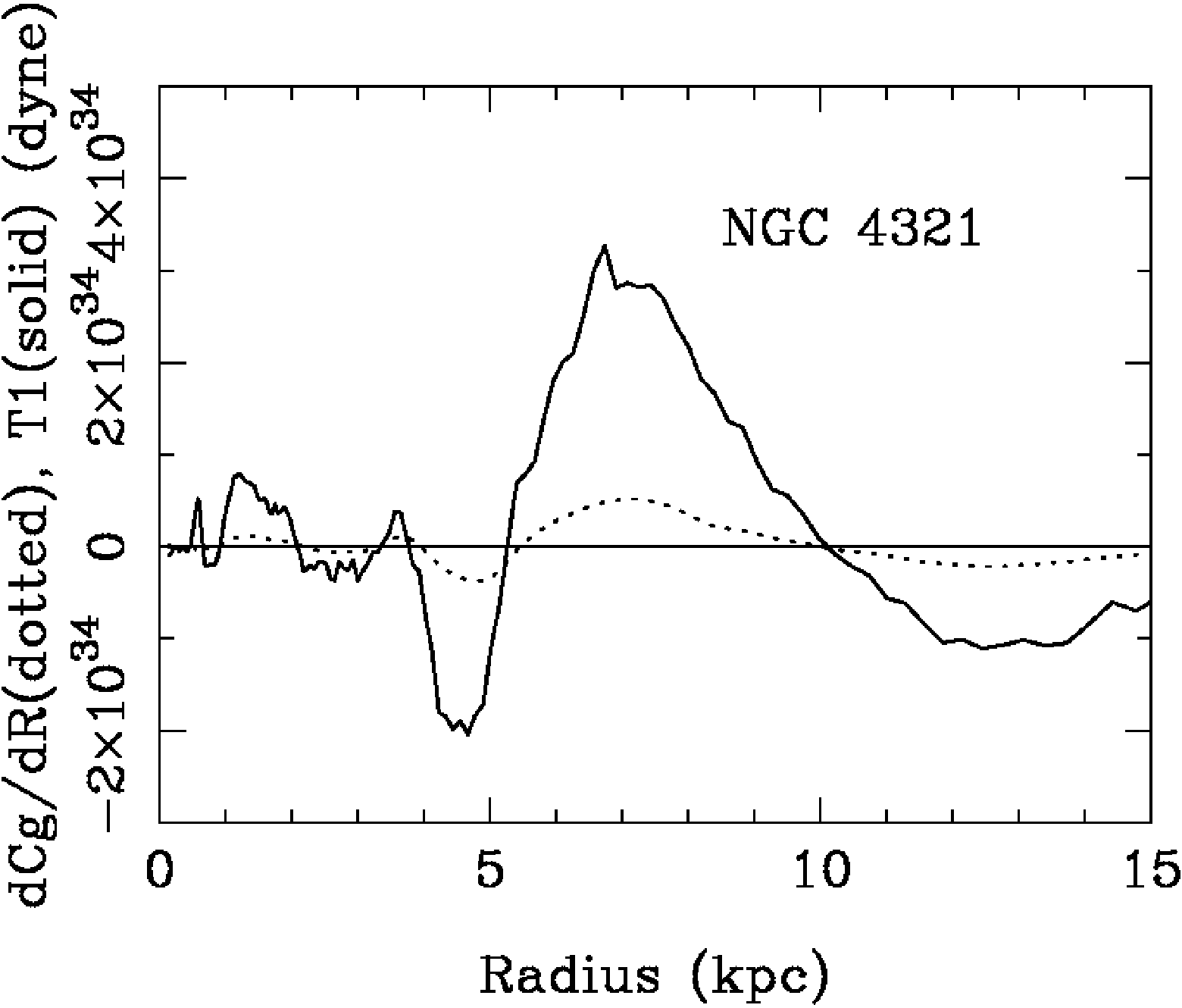}
\includegraphics{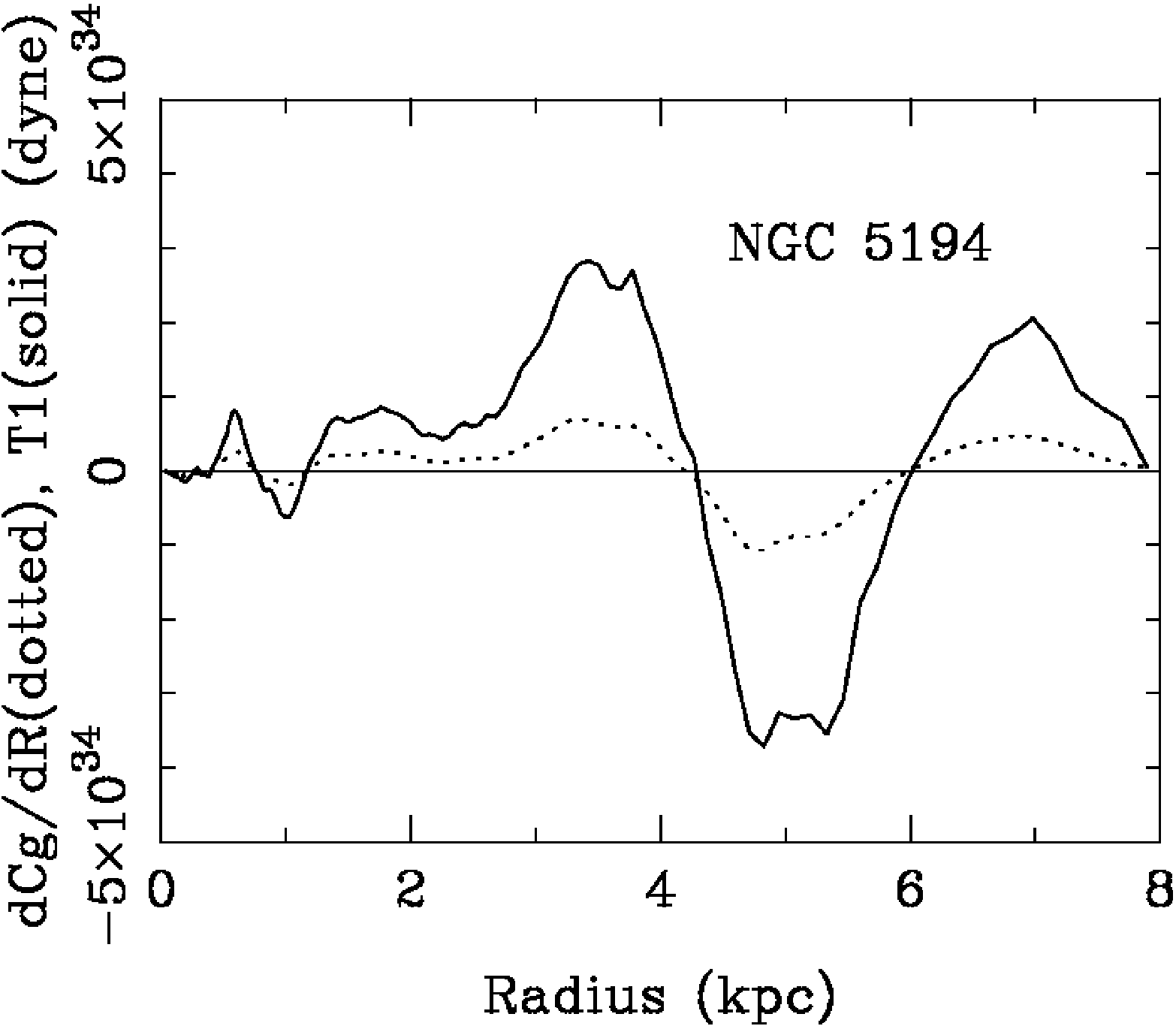}
\includegraphics{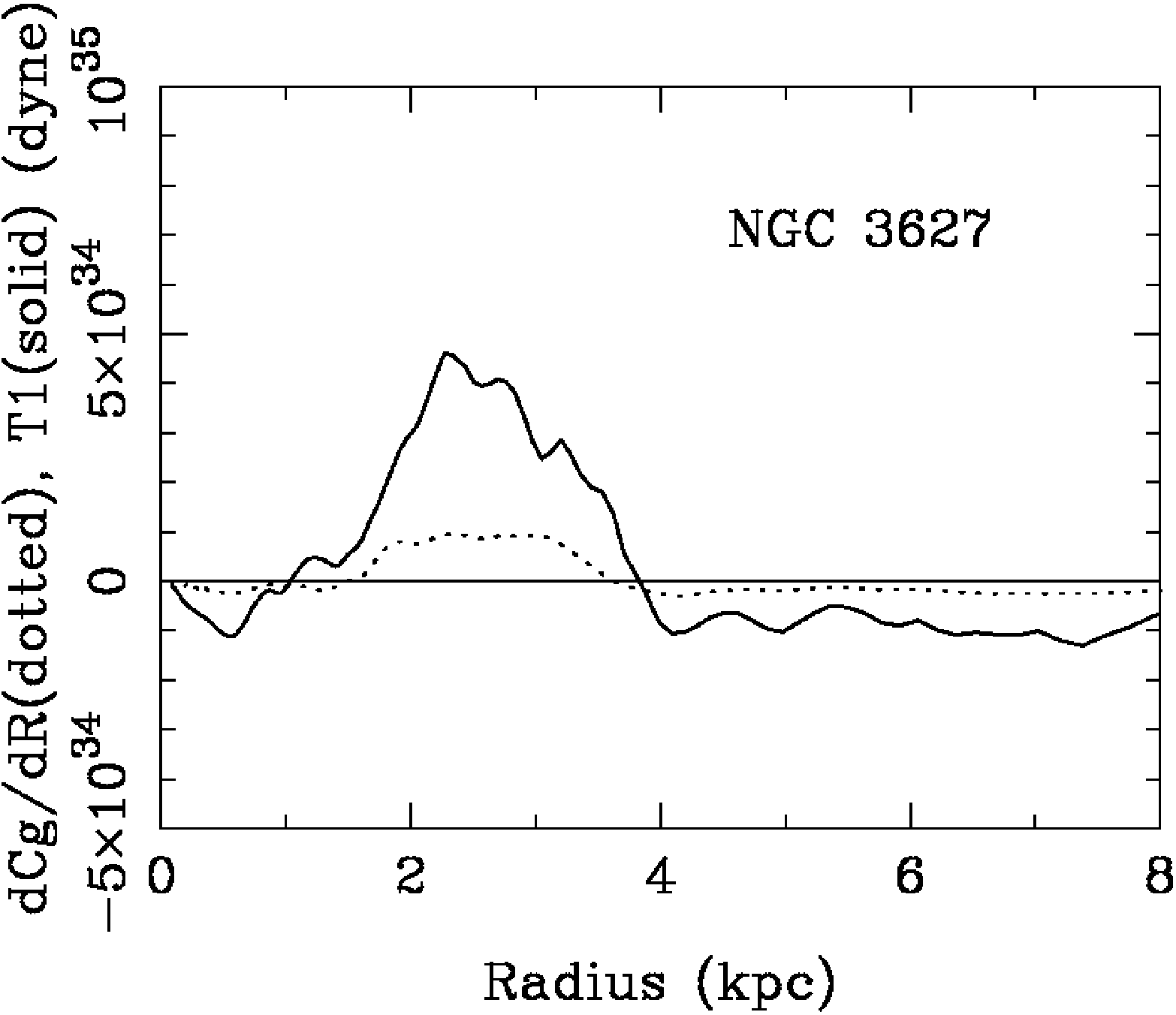}
\includegraphics{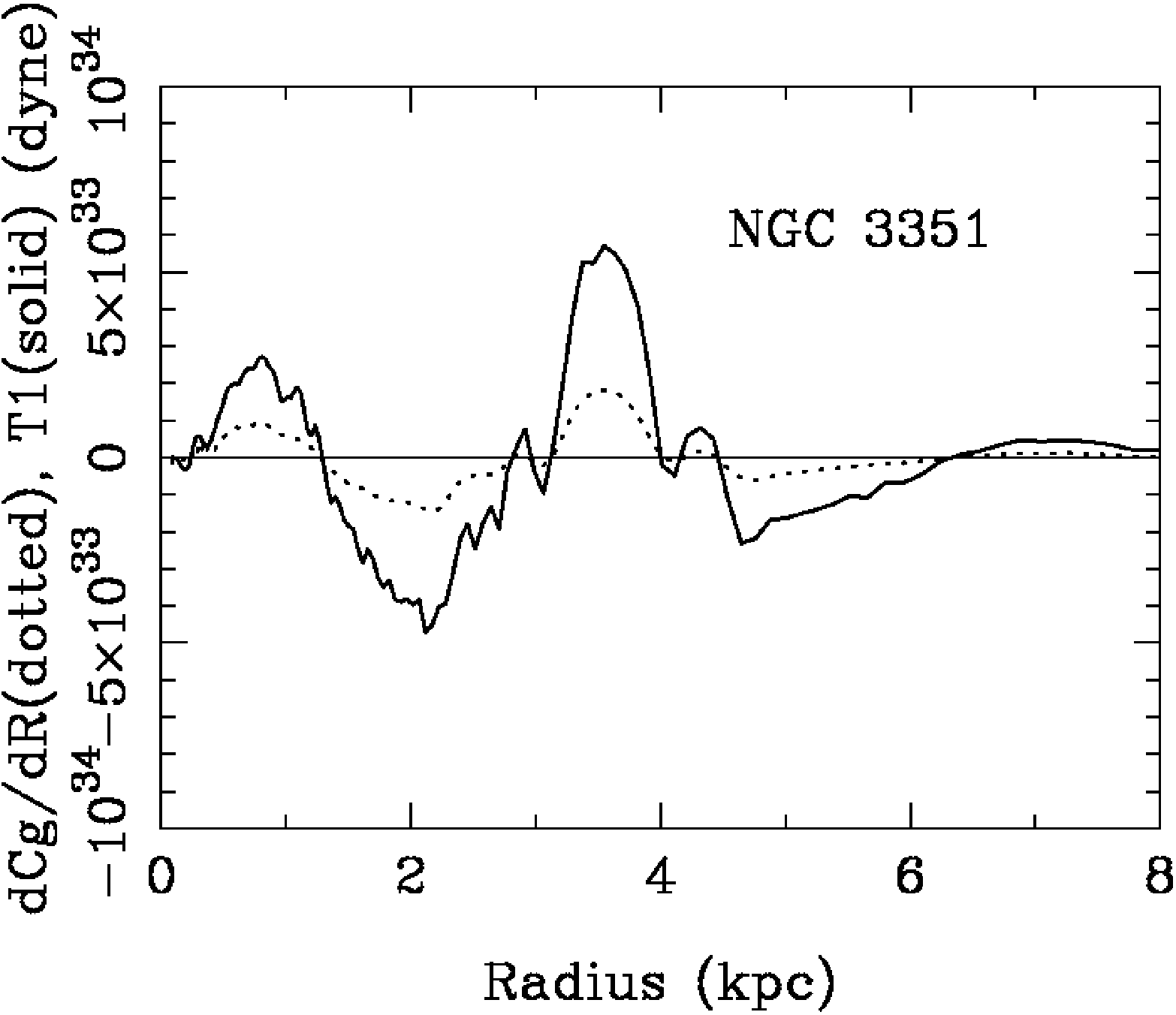}
\includegraphics{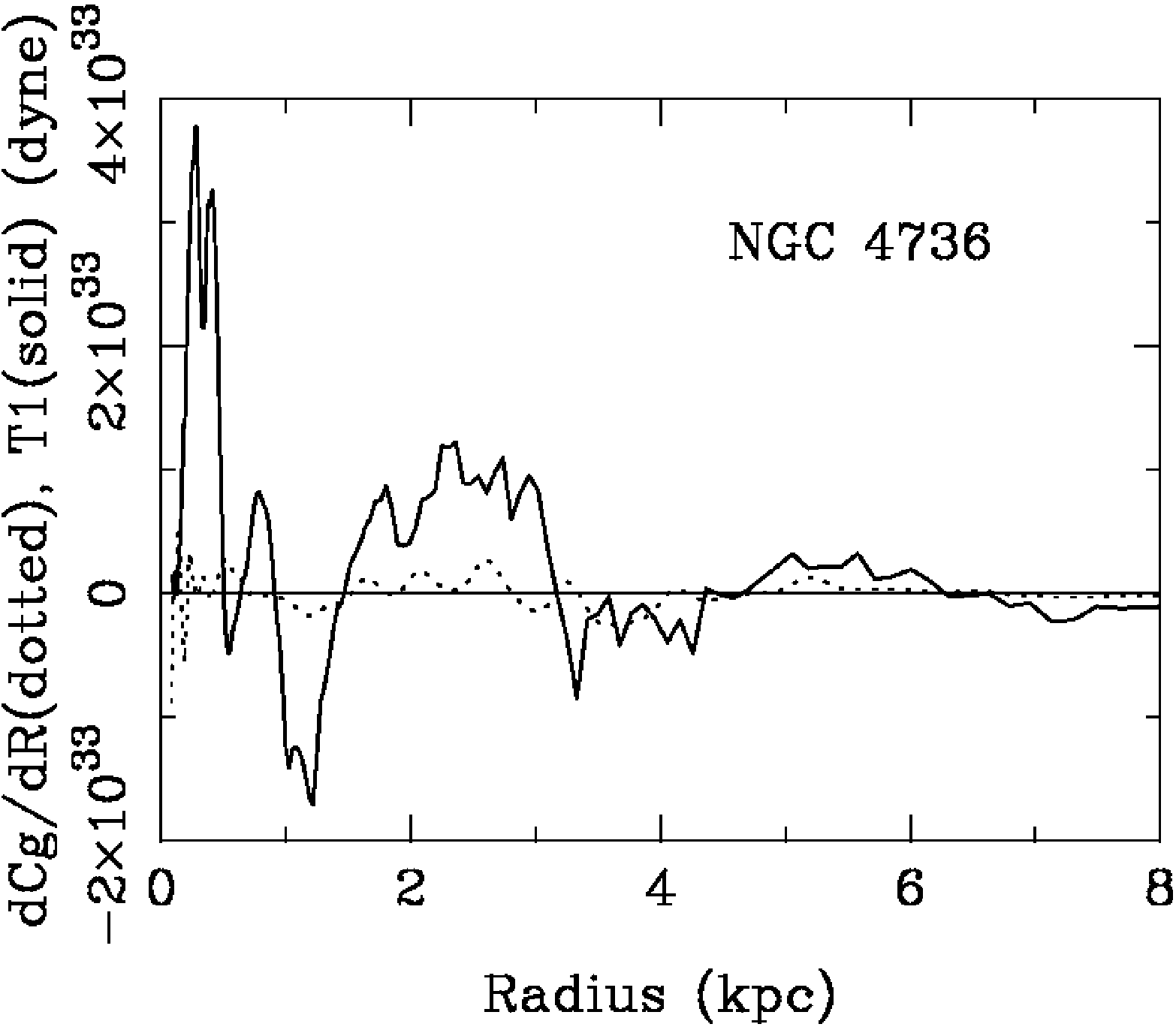}
\caption{Radial gradient of gravitational torque coupling integral compared
with the volume torque integral for the six sample galaxies.  For
NGC 628, used IRAC 3.6 $\mu$m data; for NGC 4321, 3351, 3627, 5194,
used the average of IRAC 3.6 $\mu$m and SDSS i-band data; for NGC 4736,
used SDSS i-band data; to these the usual gas maps were added.}
\label{fg:Fig17}
\end{figure}

Figure \ref{fg:Fig17} shows the calculated radial gradient of the
gravitational torque coupling integral $dC_g(R)/dR$ (dotted curve) as
compared to the volume torque integral $T_1(R)$ (solid curve). There is
a factor of 3-4 difference between the $T_1(R)$ and $d C_g(R)/dR$ for
most of the galaxies in the sample (for NGC 4736, this ratio in the
inner region is as high as 8), indicating that the remainder, which is
contributed by the gradient of the advective torque coupling, $d
C_a(R)/dR$, is in the same sense but much greater in value than
$dC_g(R)/dR$. Note that this difference between the volume torque
integral and the gradient of the (surface) gravitational torque couple
is only expected in the new theory: in the earlier theory of LBK these
two are supposed to be equal to each other (e.g., see Appendix A2 of
Z98).

Furthermore, if the LBK theory is used literally, one should not expect
any mass flow rate at all over most of galactic radii in the
quasi-steady state, except at isolated resonance radii (since the total
angular momentum flux, $C_a(R)+C_g(R)$, is expected to be constant
independent of radius in the LBK theory).  The existence of a mass flux
across the entire galactic disk, or the multiple peaked-distribution of
total torque coupling/angular momentum flux, is thus also contrary to
the LBK's original assumption of wave trains in galaxy disks, but is
consistent with the conjecture that the patterns in these galaxies are
quasi-steady modes. This conjecture is further supported by the fact
that we had successfully used the potential-density phase shifts to
locate the resonances for the grand-design galaxies in ZB07 and BZ09,
which should not have been possible if most of these patterns were
transient wave trains. Without the modal nature of the patterns one
should not expect to be able to use the mass distribution alone to
determine a quasi-kinematic quantity like the corotation radius.

We have also experimented with using the $m$=2 component only to
calculate phase shift and mass flow rates, and have compared the
results of these calculations with those based on all Fourier modes
(i.e., using the full torque expression of equation 2).  For most of the
grand-design galaxies in our sample, the $m$=2 component makes a
dominant contribution in these calculations, which further confirms the
modal nature of these observed grand-design patterns (in N-body
simulations, whenever a transient feature appears, the $m$=2 torque
calculation and the full torque calculation would differ drastically).

We have repeated the calculations for $dC_g(R)/dR$ and $T_1(R)$ using
different scale height values.  Even though both of these quantities
are affected by a particular choice of the scale height, the ratio
between the two appears little affected.  This is likely due to the
fact that the potential that is being affected by the scale height
choice enters into both the $C_g$ and the $T_1(R)$ expressions, so
their ratio is less sensitive to the choice of scale height. Therefore,
our conclusion that there is a significant difference between these two
quantities at the nonlinear regime of the wave mode is robust.

\subsection{The Relative Contributions of Stellar and Gaseous Mass Flows}

In the past few decades, related subjects of secular evolution, bulge
growth, and the morphological evolution of galaxies along the Hubble
sequence were studied mainly within the framework of the secular
redistribution of the gas (interstellar medium) components under the
influence of a barred potential, and the resulting growth of the
so-called ``pseudobulges'' (Kormendy 2012 and references therein =
K12).  Through gas accretion alone, a galaxy could at most build
late-type pseudobulges, but not intermediate and early type bulges.
This is because the observed gas inflow rates and nuclear
star-formation rates are both insufficient to account for the formation
of earlier-type bulges through the secular inflow of gas, and the
subsequent conversion of gas to stars during a Hubble time (K12).  Yet
observationally, early type galaxies (including disky-Es) are known to
form a continuum with the intermediate and late-type disky galaxies, in
terms of their structural parameters, kinematics, and stellar
populations (Jablonka, Gorgas, \& Goudfrooij 2002; Franx 1993;
Cappellari et al. 2013), and thus their formation and evolution
mechanisms are also expected to form a continuous trend.

With the recognition of the role of collective effects in the secular
mass redistribution process, the study of secular morphological
transformation of galaxies should now give equal emphasis to the roles
of stellar and gaseous mass redistribution.  Due to their different
intrinsic characteristics (radial surface density distribution,
compressibility, star-formation correlation, and dissipation
capability), stars and gas do play somewhat different roles in the
secular evolution process. In this subsection we illustrate with our
sample galaxies some of the specifics of these roles.

\begin{figure}
\vspace{370pt}
\includegraphics{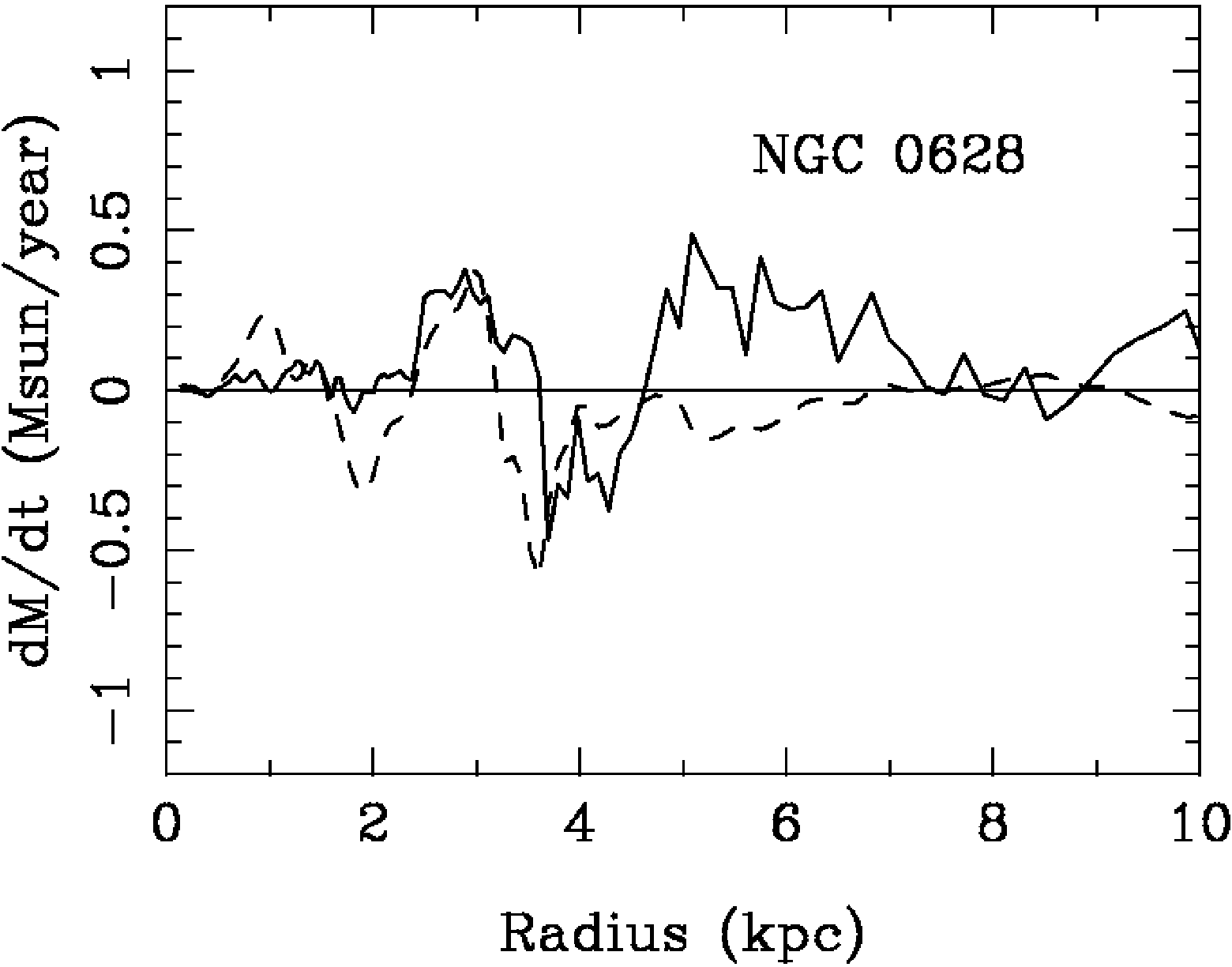}
\includegraphics{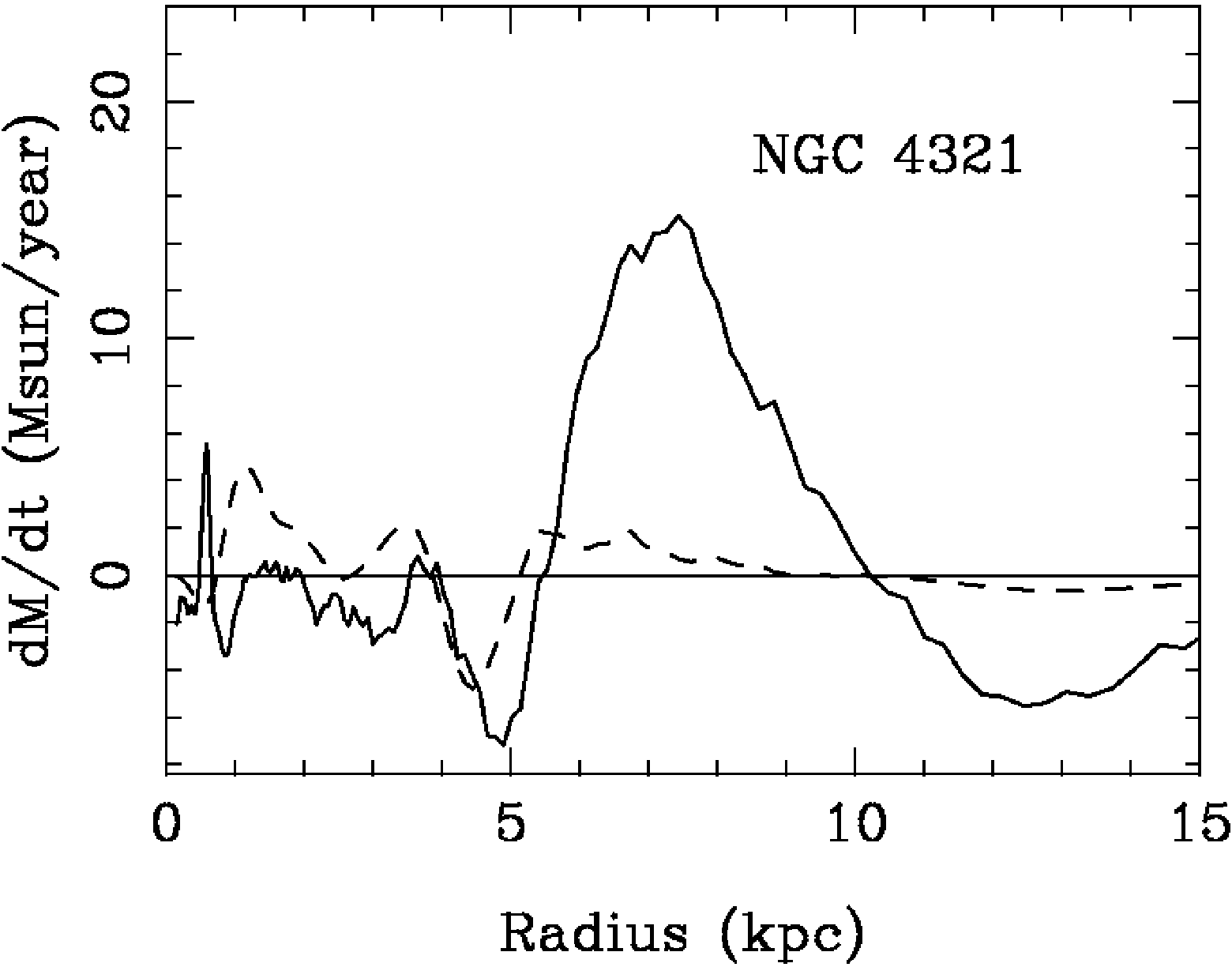}
\includegraphics{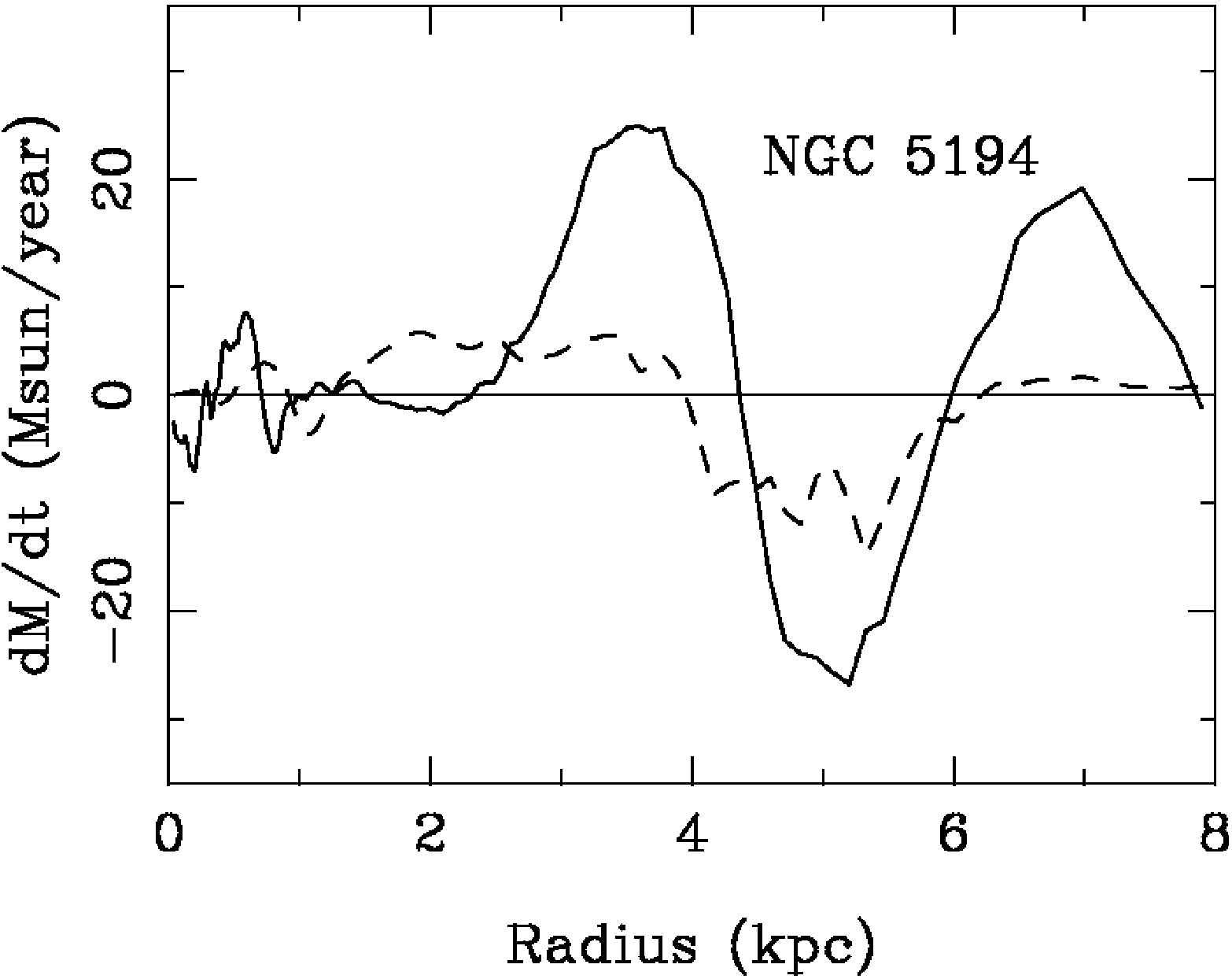}
\includegraphics{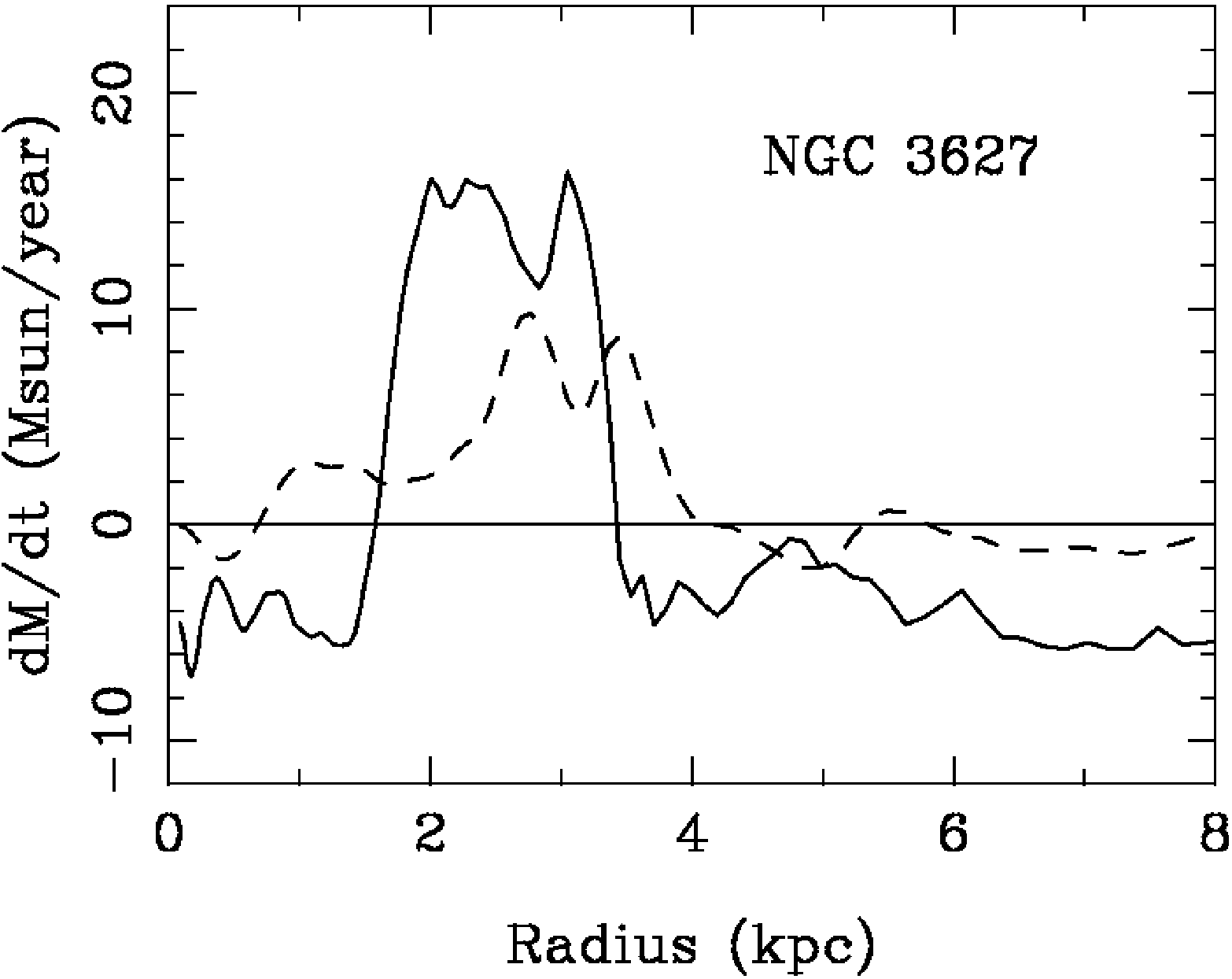}
\includegraphics{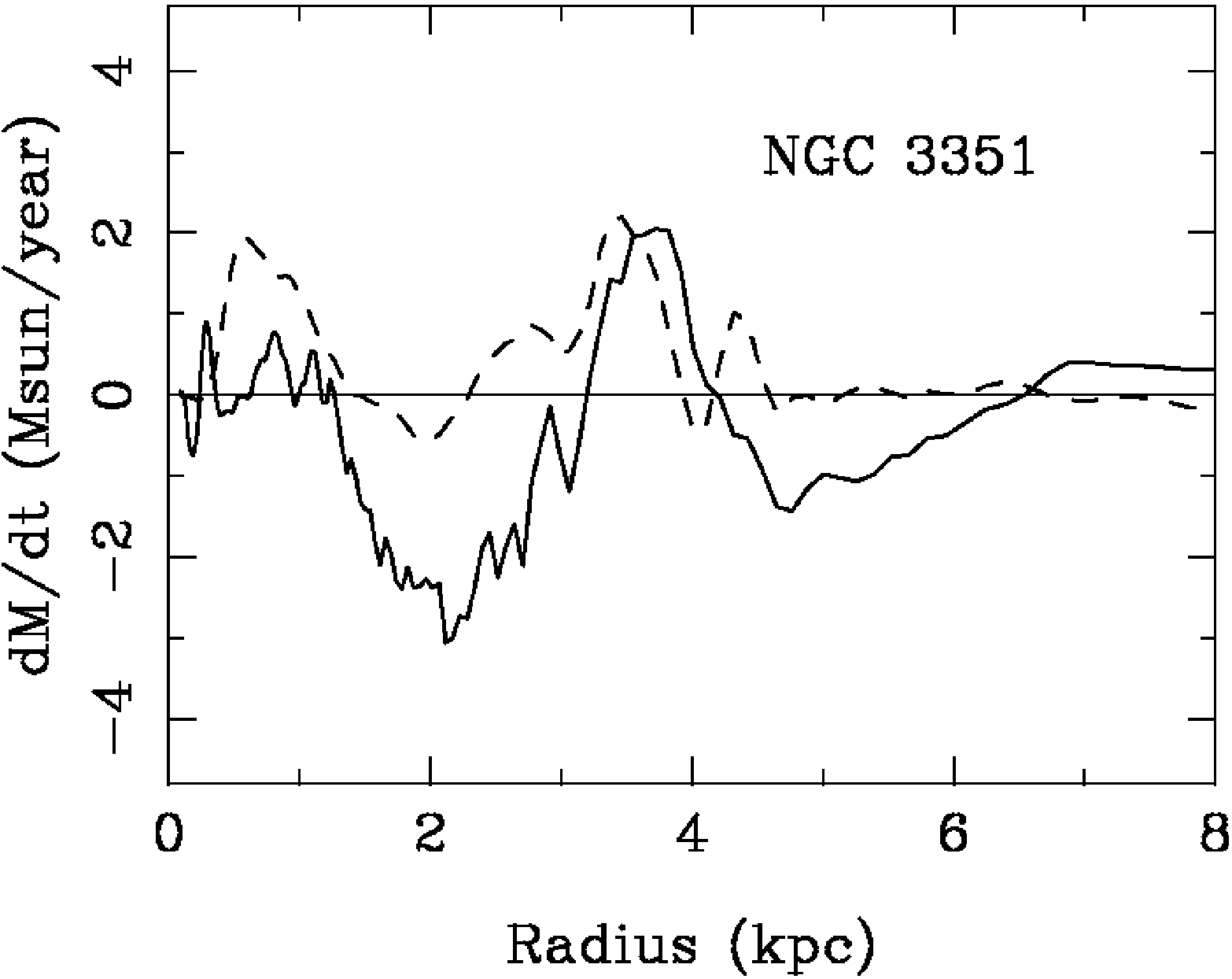}
\includegraphics{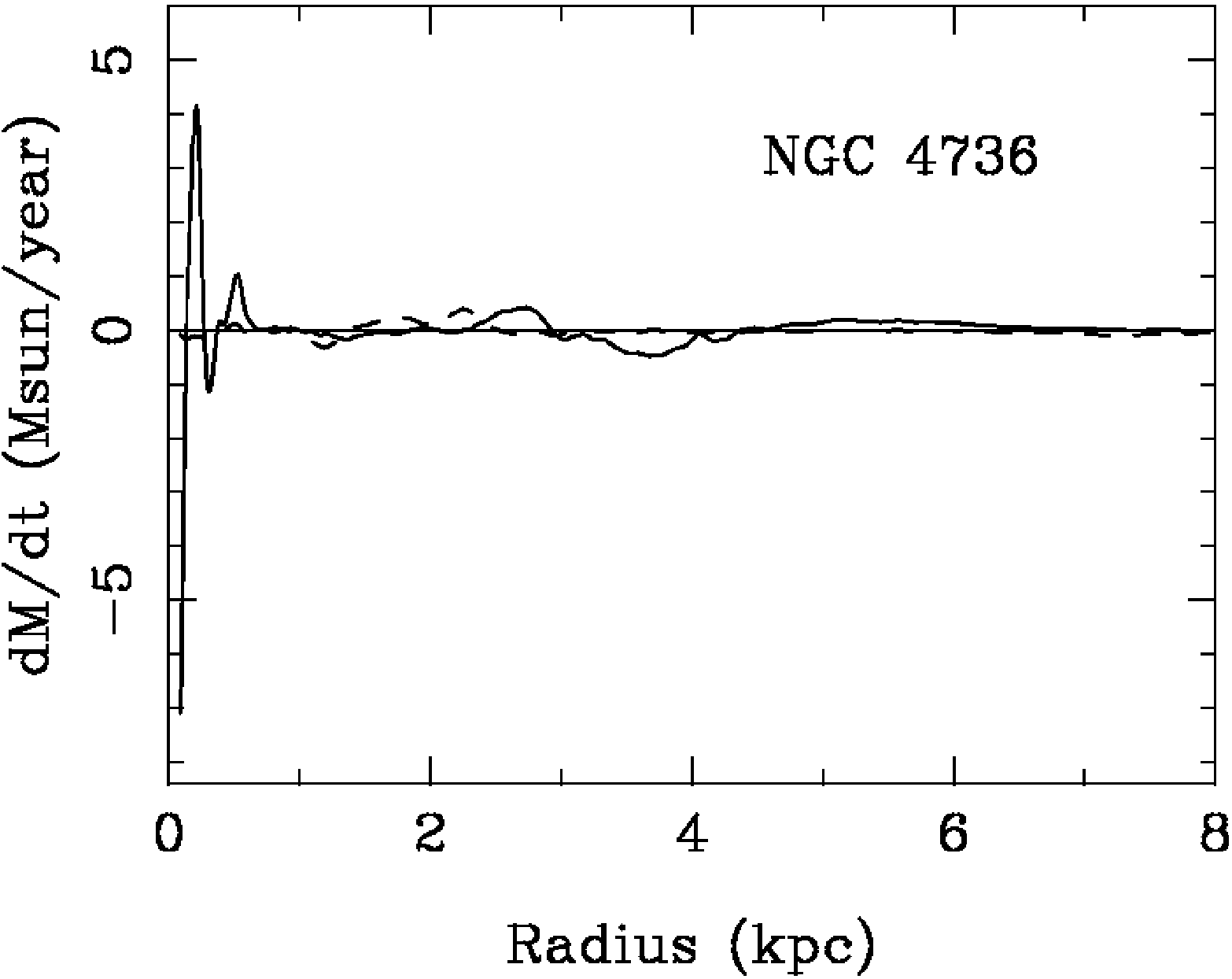}
\caption{Stellar and gaseous mass flow rates for the six sample galaxies,
calculated from the IRAC 3.6 $\mu$m for NGC 628,
an average of IRAC 3.6 $\mu$m and SDSS i-band for NGC 4321, 5194, 3627,
3351, and SDSS i-band for NGC 4736, plus VIVA, THINGS and BIMA SONG data.
The total potentials used for these
calculations were the same as previously derived using IRAC
and/or SDSS for the stellar contributions, with appropriate averaging,
plus the gas contributions.}
\label{fg:Fig18}
\end{figure}

\begin{figure}
\vspace{370pt}
\includegraphics{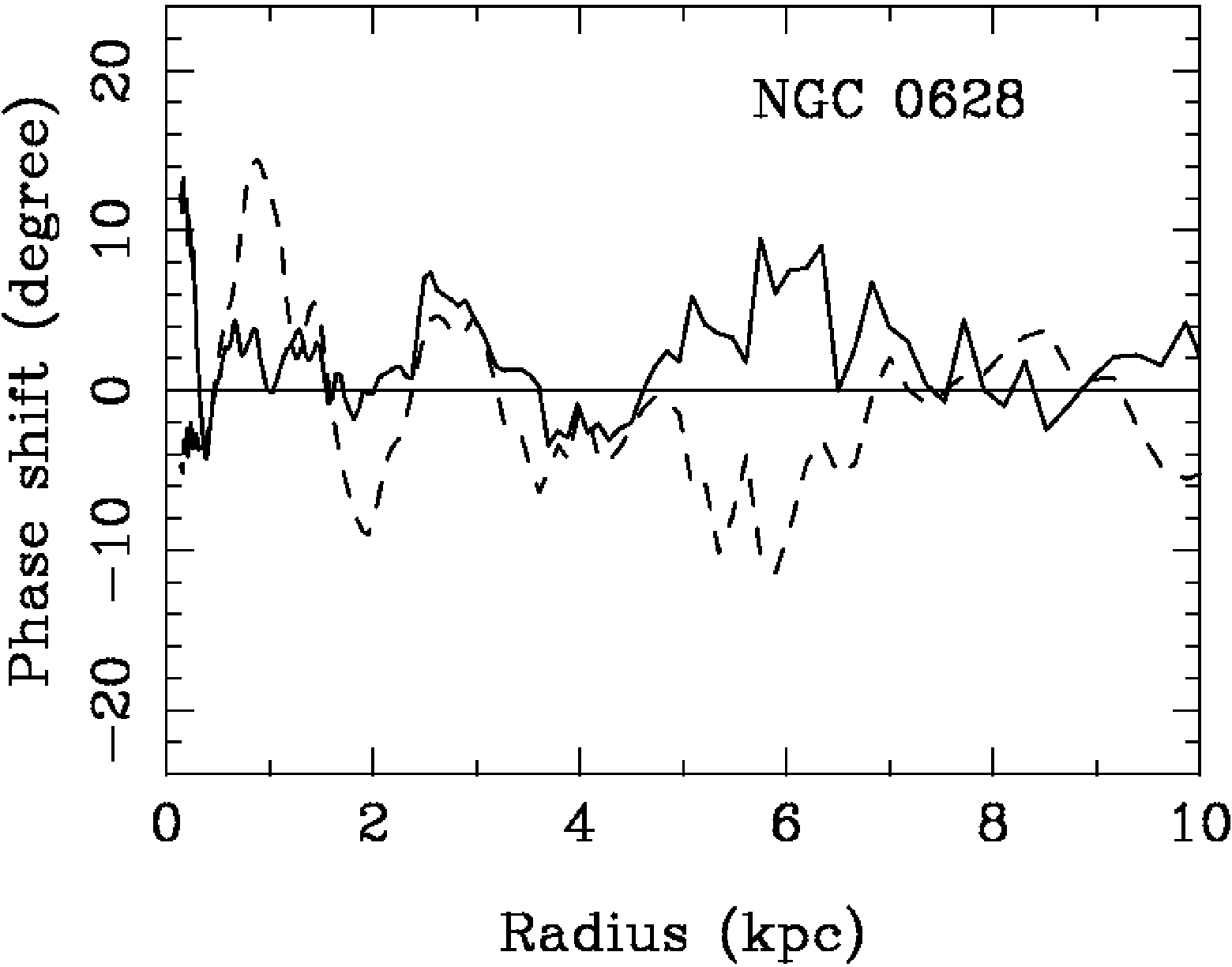}
\includegraphics{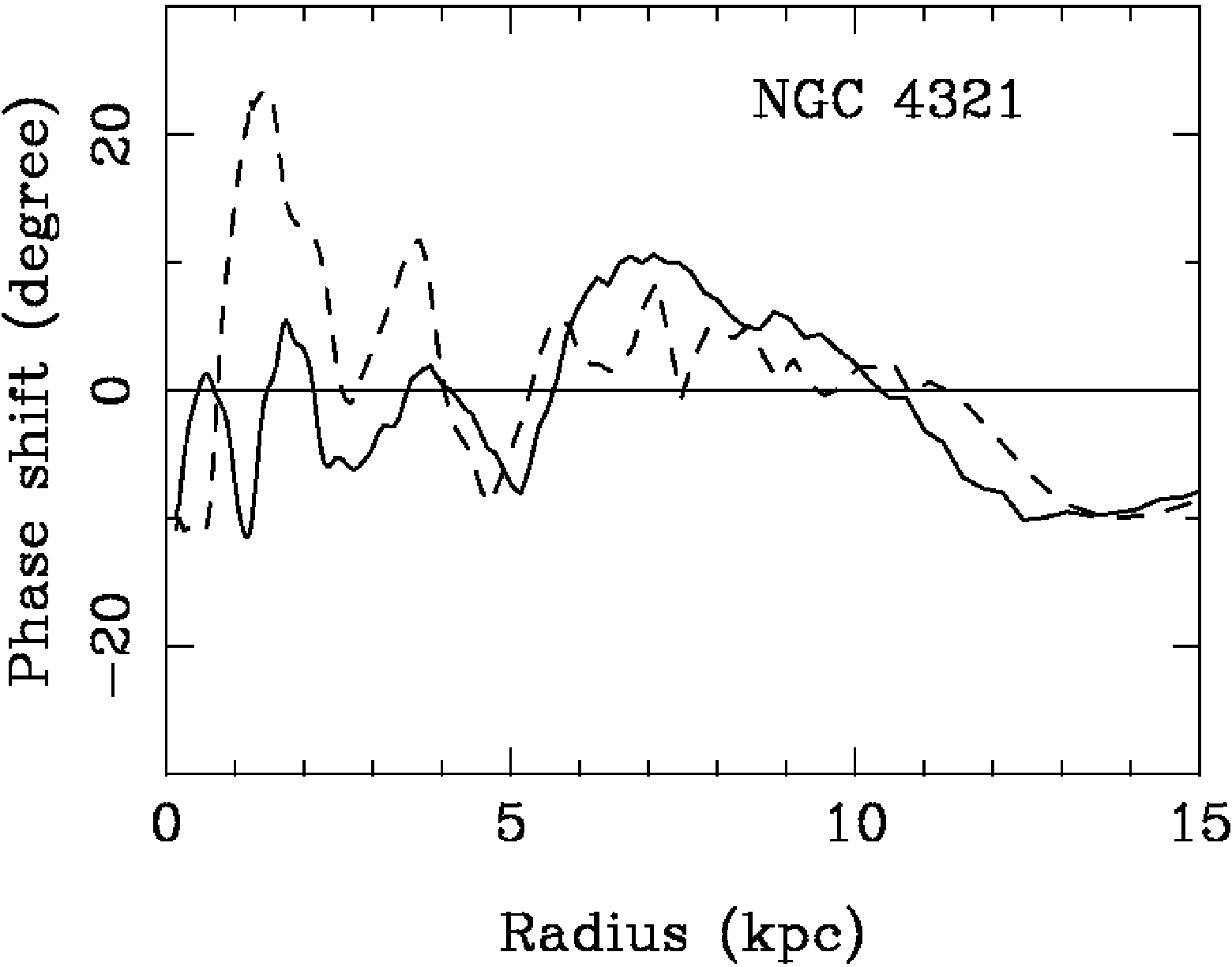}
\includegraphics{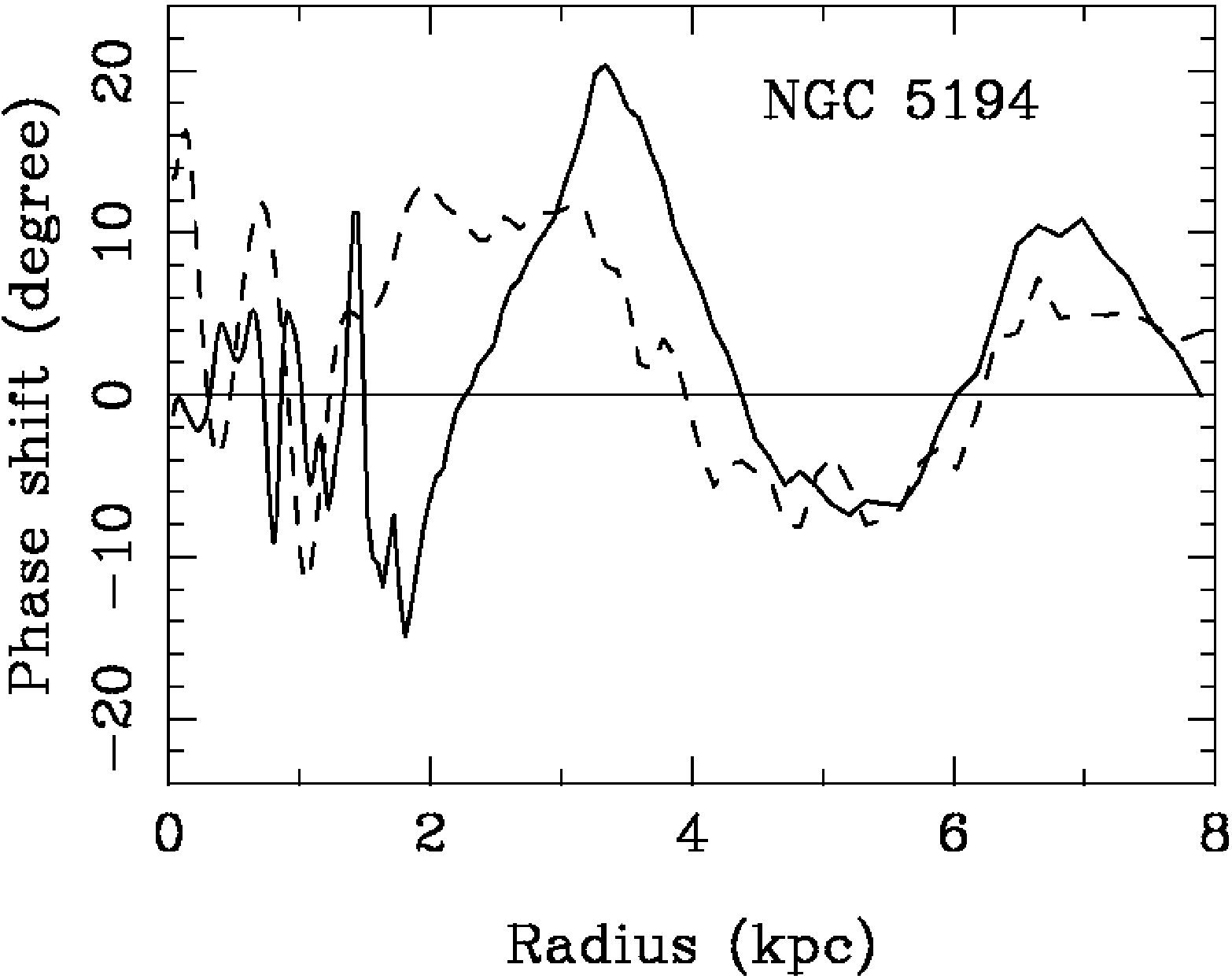}
\includegraphics{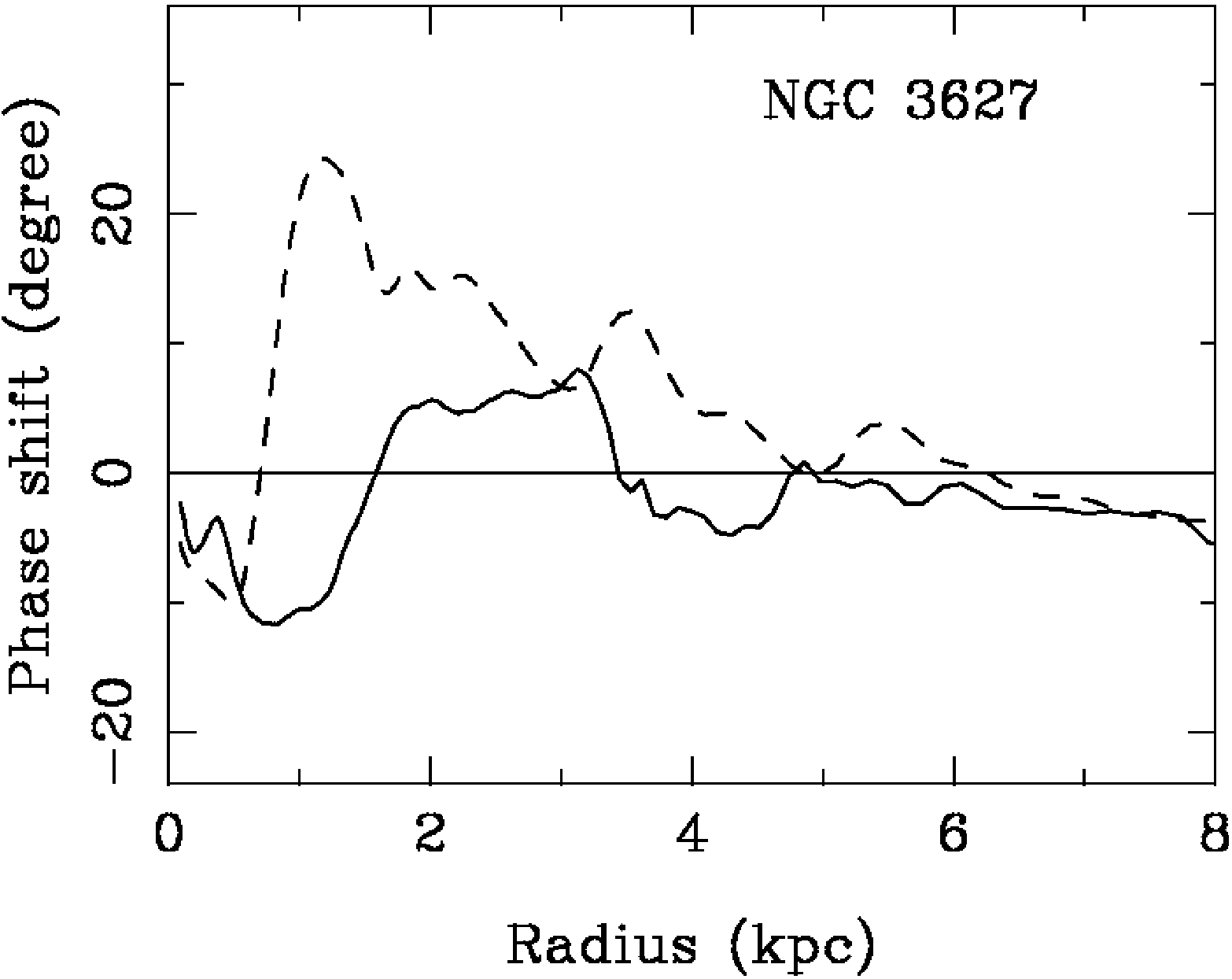}
\includegraphics{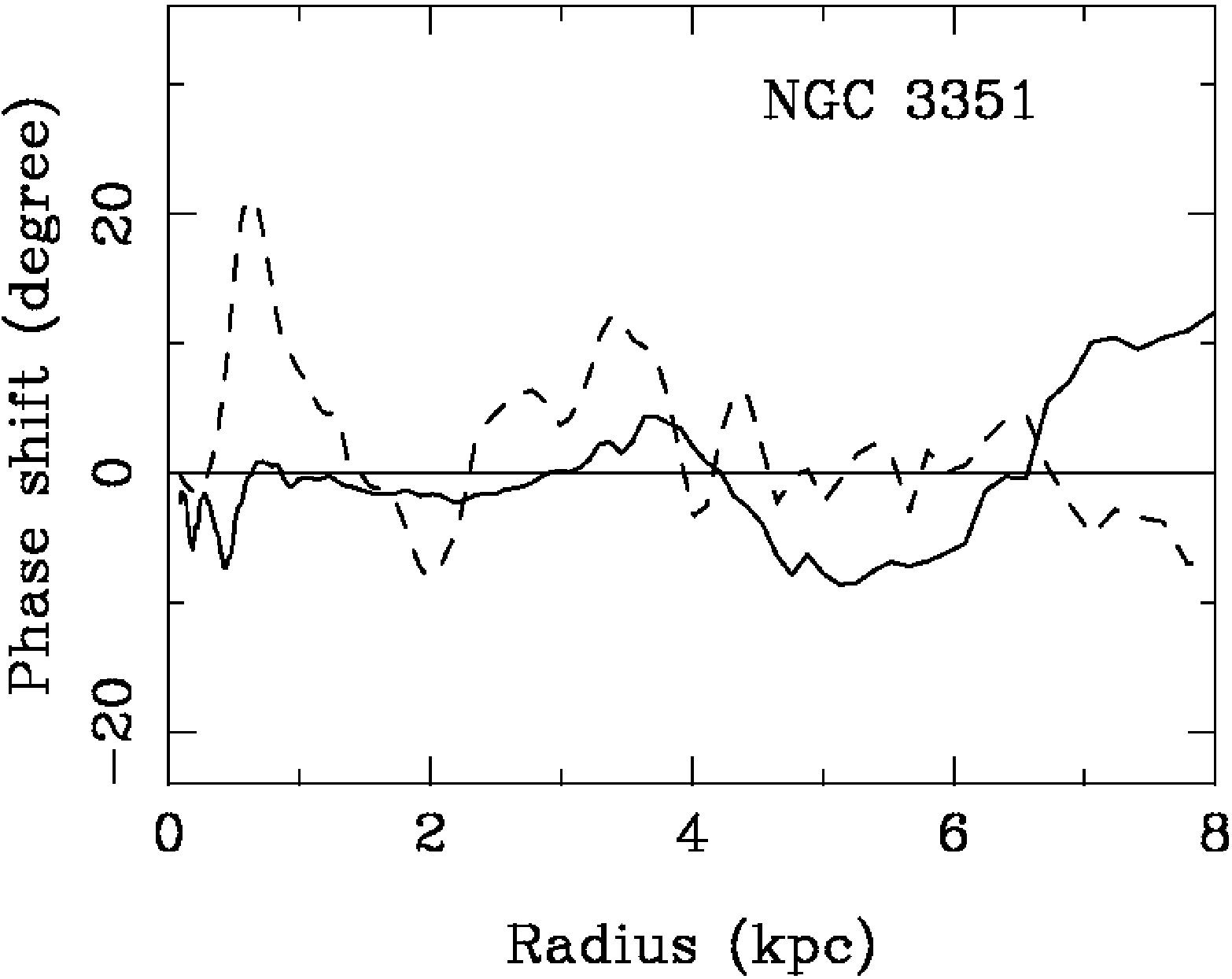}
\includegraphics{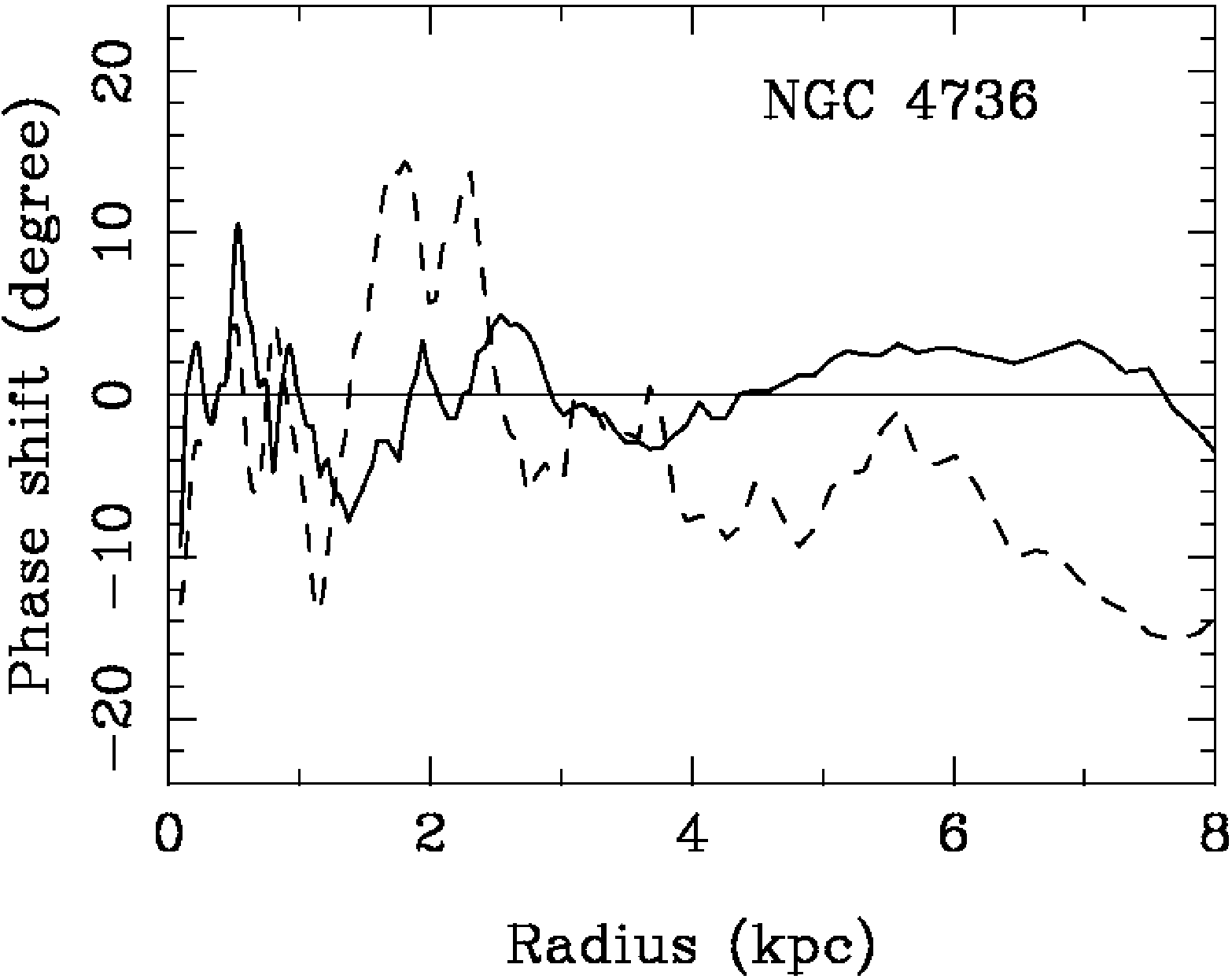}
\caption{Stellar and gaseous phase shift with respect to total
potential for the six galaxies. {\it Solid lines:} Stellar phase shifts.
{\it Dashed lines:} Gaseous phase shifts.
The stellar maps were derived using the IRAC 3.6 $\mu$m data
for NGC 628, using an average of IRAC and SDSS data for NGC 4321, 3351, 3627,
5194, and using SDSS i-band data for NGC 4637. 
The maps were from VIVA, THINGS and BIMA SONG observations.
The total potentials used for these
calculations were the same as previously derived using IRAC
and/or SDSS for the stellar contributions, with appropriate averaging,
plus the gas contributions.}
\label{fg:Fig19}
\end{figure}

In Figure~\ref{fg:Fig18} and \ref{fg:Fig19}, we present the comparison
of stellar and gaseous (HI plus H$_2$) mass flow rates, as well as
their respective phase shifts with respect to the total (star plus gas)
potential. The phase shift plots tell us the relative efficiency of the
stellar and gaseous mass accretion processes, since
the angular momentum exchange rate between the disk matter and
the density wave, per unit area, can be written as (Z98, equation 25)

\begin{equation}
{\overline{ {{dL} \over {dt}}}} (R)
= {1 \over 2} F^2 v_c^2 \tan i \sin(m \phi_0) \Sigma_0
\label{eq:e}
,
\end{equation}
where $F^2 \equiv F_{\Sigma} F_{\cal{V}}$ is the product
of the fractional density and potential wave amplitudes,
and m is the number of spiral arms (usually taken to be 2).
Therefore, using the previously derived mass flow rate
equation (\ref{eq:eq2}) and volume torque expression
equation (5), which applied for the unit-width annulus
rather than per unit area, we have

\begin{equation}
\sin(m \phi_0)_i
={{1} \over {\pi v_c R F_i^2}}{ {{dM_i } \over {dt}} \over {(\Sigma_0)_i}}
\end{equation}
where the subscript i represents stars or gas, respectively.
So for similar wave amplitudes between stars and gas, the mass component
that has higher phase shift will have higher mass flow rate per
unit surface density.

For galaxy NGC 628, the stellar and gaseous accretion efficiencies are
similar, as are the respective total mass accretion rates, since this
is a late-type spiral galaxy and is gas rich.  The alignment of the
stellar and gaseous phase shifts is poor, however, especially for the
outer disk, indicating that the galaxy is yet to evolve into a state of
dynamical equilibrium.

NGC 4321 is relatively quiescent and of intermediate Hubble type. From
Figure~\ref{fg:Fig19}, we see that for much of the central region
(except the very center) the gaseous phase shift with respect to their
common potential is much larger than the stellar phase shift,
indicating that gas leads in phase in this region compared to stars,
revealing a higher dissipation rate and mass redistribution efficiency
of the gas compared to the stars for the central region of this galaxy.
The values of the phase shifts of the stars and gas are comparable for
the outer region, indicating that the two mass components have similar
mass-redistribution efficiency there. The shapes of the positive and
negative humps of phase shifts for stars and gas have more similar
radial distributions for this galaxy compared to NGC 628, especially
for the outer region, indicating a higher degree of dynamical
equilibrium. The overall contribution of the stars to mass
redistribution, however, is much higher than for the gas in the
outer region (Figure~\ref{fg:Fig18}), because of the higher overall
stellar surface density there.

For NGC 5194 (M51), the stellar and gaseous phase shifts have
significantly different radial distributions compared to NGC 4321, even
though both galaxies are of intermediate Hubble type.  This is likely
because of the non-dynamical-equilibrium state of M51 due to the tidal
pull of the companion, and the inevitable evolution towards forming a
new set of nested resonances, with the gas playing a leading role in
seeking the new dynamical equilibrium because of its more dissipative
nature, and the stellar component lagging somewhat behind in this
action. But at every moment of this re-establishment of the dynamical
equilibrium, the overall density (i.e. the sum total of stellar and
gaseous) still has a much more coherent phase shift distribution with
respect to the total potential than each component considered
separately (i.e. compared to Figure~\ref{fg:Fig7}).  The overall mass
flow rate of stars is much higher than gas for this galaxy.

For NGC 3627, even though the phase shift in the central region shows
that gas has a higher accretion efficiency than stars, the overall
accretion rate of stars much exceeds that of the gas. The second CR
location is not shown here because of the more limited radial range
plotted, but is present in the total gas phase shift curve when
examined further outward.

For NGC 3351, we see that the central region gas-star relative phase
shift is even more prominent than for NGC 4321.  This is likely due to
the fact that the straight bar potential in the central region of this
galaxy is mostly contributed by stars, which has small phase shift with
respect to the total potential, and gas thus contributes a much larger
phase shift (through its dissipation in the bar potential and the phase
offset of its density peak from the stellar density peak) to the
overall potential-density phase shift. The overall mass flow rates are
contributed similarly by stars and gas for this galaxy.  Note that the
somewhat chaotic appearance of phase shifts in the outer regions of this
galaxy is due to the low surface density there, and thus noise begins
to dominate.

For NGC 4736, gas leads stars in parts of the radial range, but the
overall contribution to the mass flow is mostly due to stars,
especially for the central region, due to the fact that in this
early-type galaxy the stellar surface density much exceeds that of the
gas. Once again the low surface density in the outer disk of this
early type galaxy leads to the more chaotic phase shift distribution
there.

We see from this set of plots that one of the old myths of galactic
studies, that the gas always torques stars inside and outside CR in the
right sense (i.e. that gas should lead stars in azimuth inside CR, and
vice versa outside CR), is only true in a small number of instances.
To obtain a reliable CR estimate, one really needs to use the total
density (stars plus gas) and total potential, and calculate the phase
shift zero crossings between these two components to determine the CR
locations. In the absence of the gas surface densities, stellar surface
density alone and the potential calculated from it (as is done in
ZB07), come as the next best compromise.  Both of these approaches are
much more reliable than using the phase shift between the stellar
density and gas density distributions to determine CR.

We also see that in the majority of the galaxies in the local universe,
secular mass flow is dominated by the stellar mass redistribution
rather than by gas redistribution, unlike what has been emphasized by
many of the earlier works on secular evolution.  Furthermore, even for
the gas accretion in disk galaxies, the mechanism responsible for its
viscosity is still the collective gravitational instabilities (through
the scattering of gas cloud in the gravitational potential, which has
encounter mean-free-path on the order of a kpc, and the net effect
of which manifests as the phase shift between the gas density and the total
potential), rather than the microscopic gaseous viscosity due to the
collision of gas particles, which has molecular mean-free-path that was
long known to be inadequate both for the accretion phenomenon needed to
form young stars, and for the accretion phenomenon observed in the
gaseous disks of galaxies -- thus the well-known need for anomalous
viscosity in generalized accretion disks (Lin \& Pringle 1987).  Z99
showed that the large-scale density wave-induced gravitational
viscosity is likely to be the source of anomalous viscosity in both the
stellar and gaseous viscous accretion disks of galactic and stellar
types.

\subsection{The Relative Contributions of Atomic and Molecular Mass Flows}

\begin{figure}
\vspace{370pt}
\includegraphics{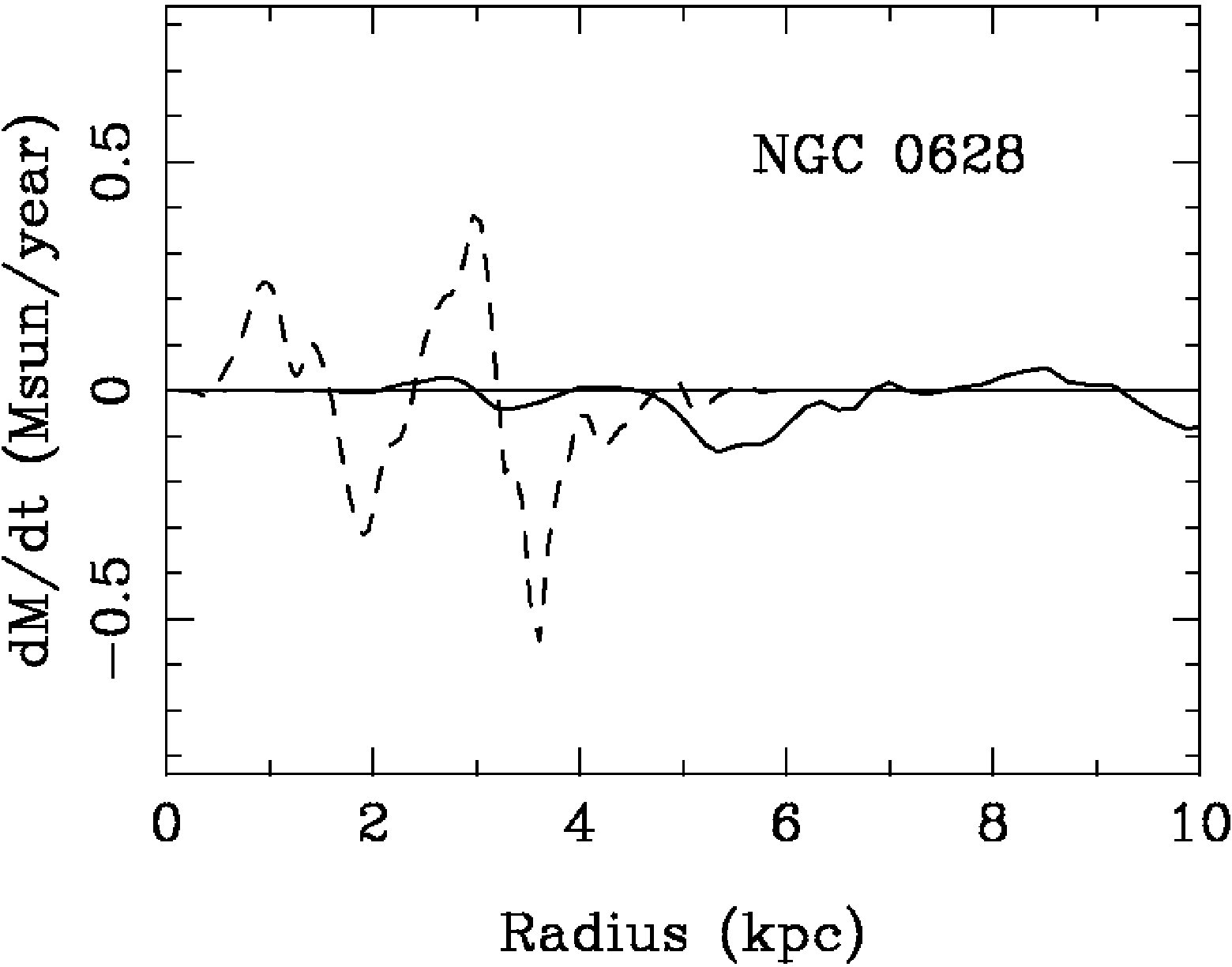}
\includegraphics{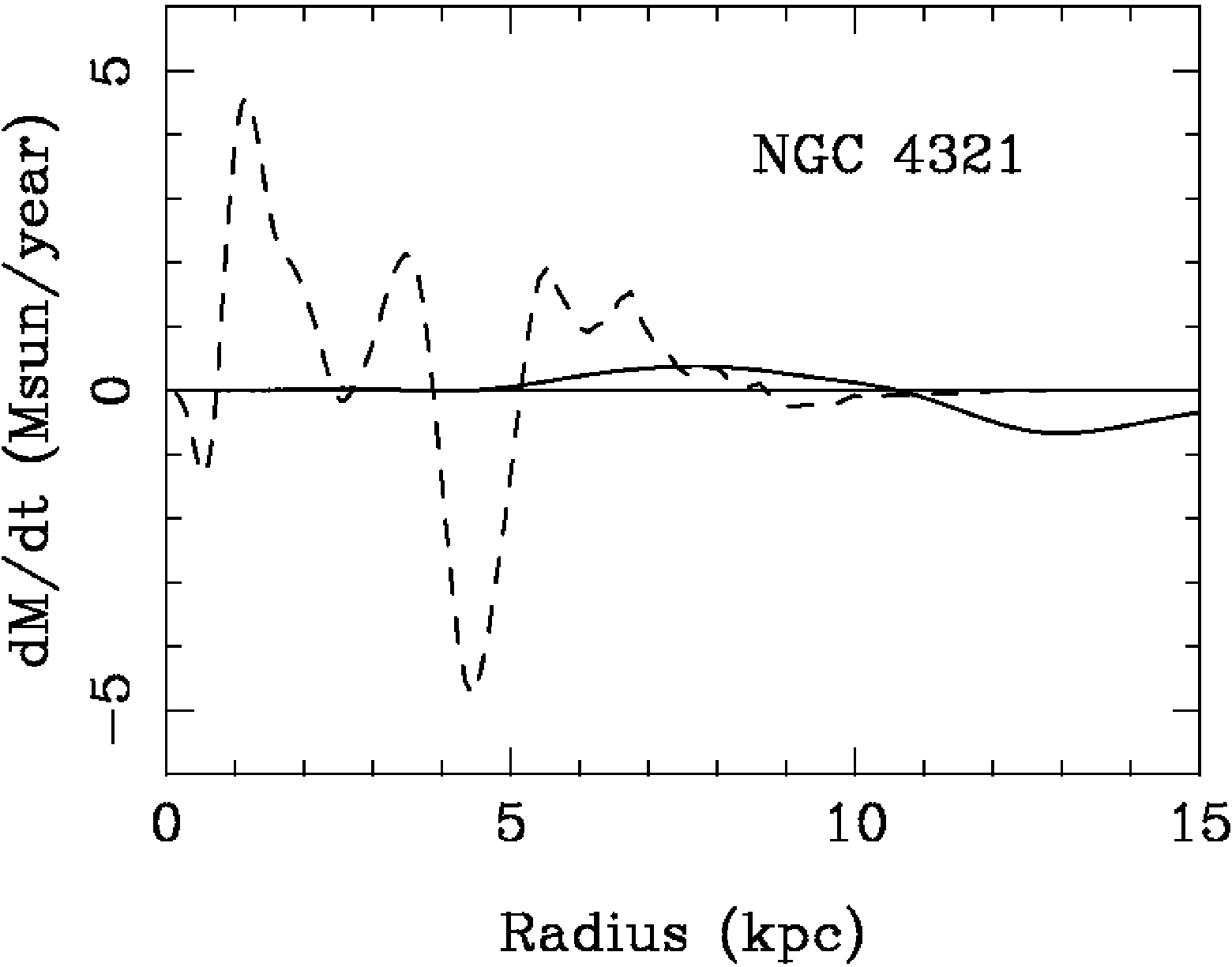}
\includegraphics{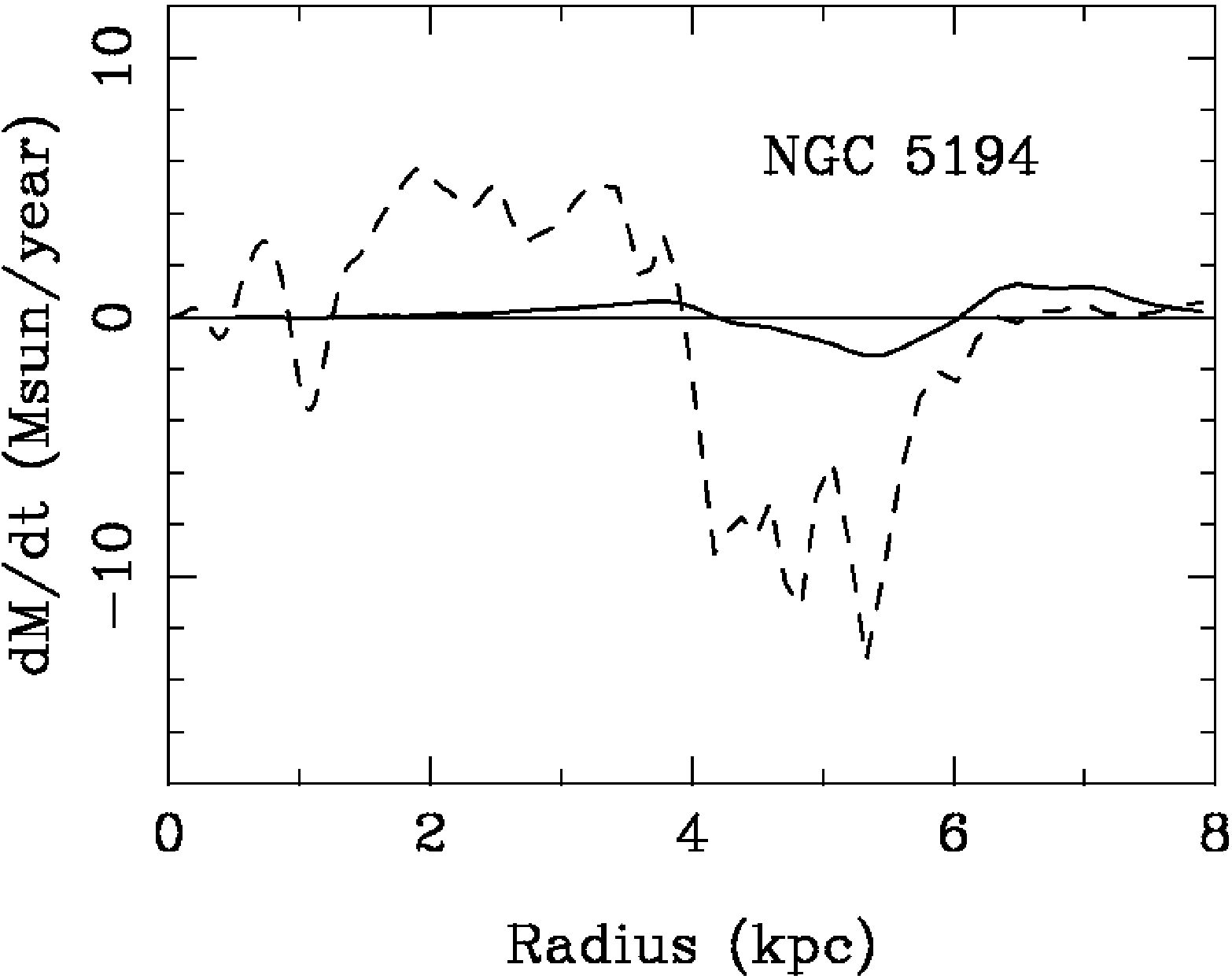}
\includegraphics{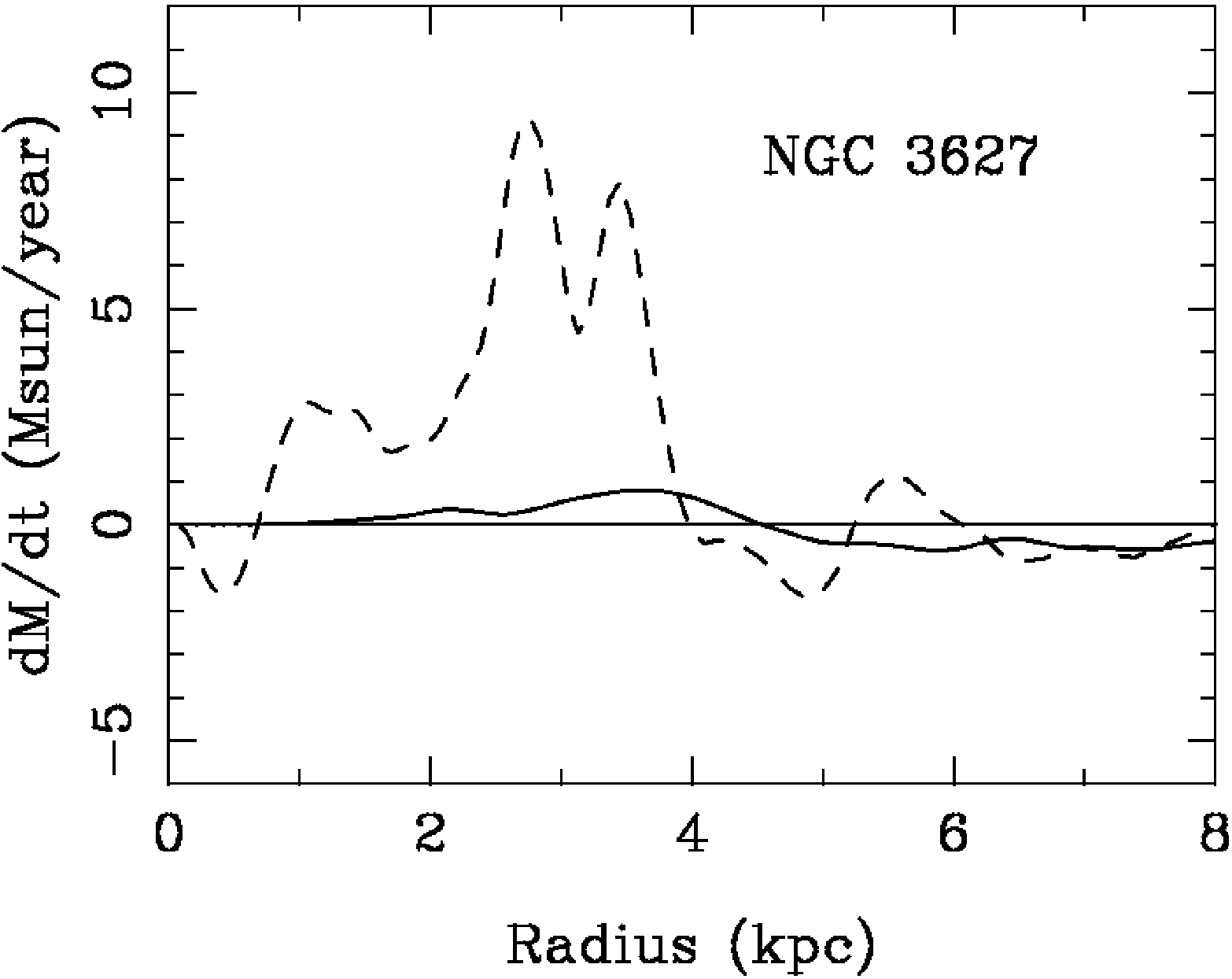}
\includegraphics{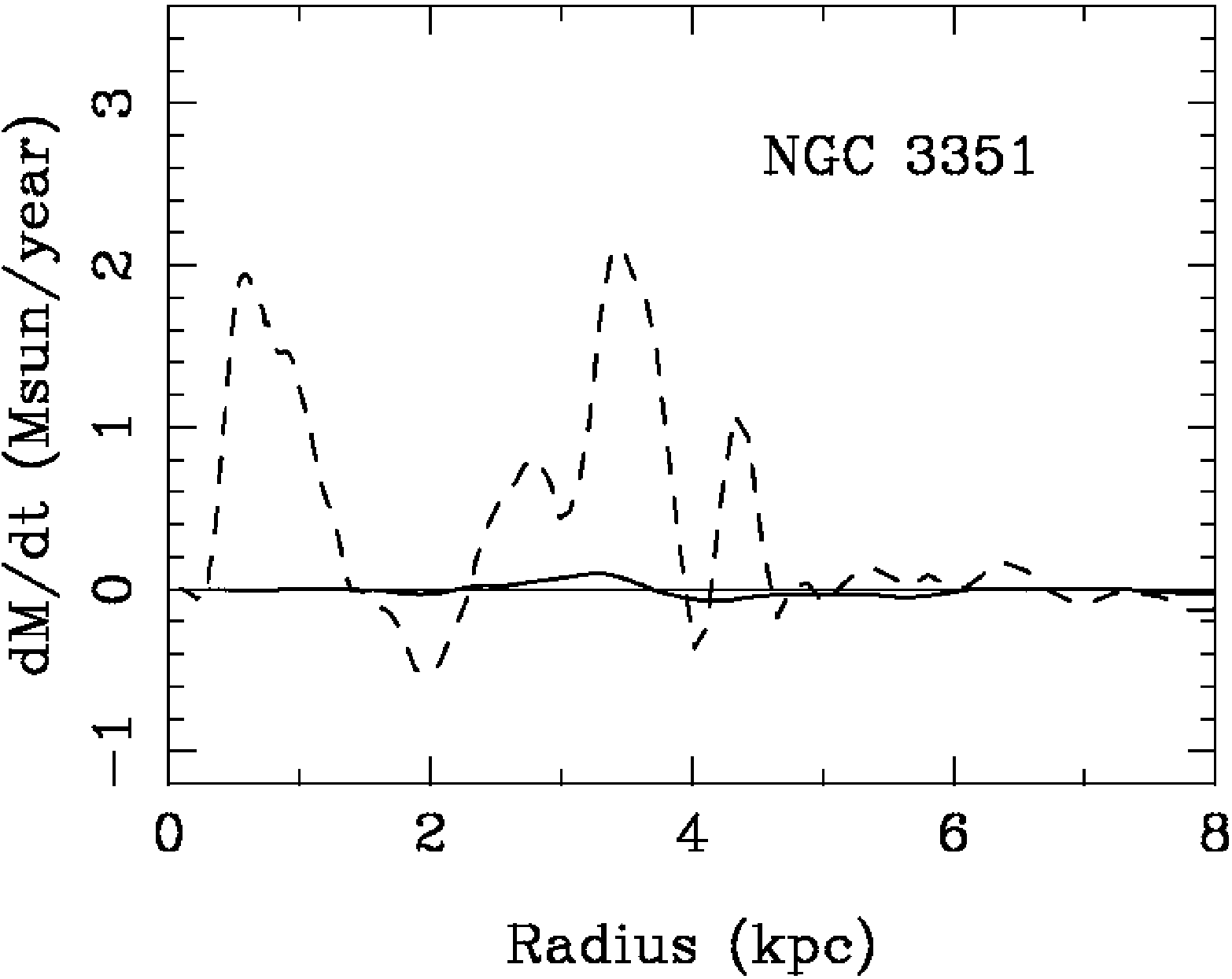}
\includegraphics{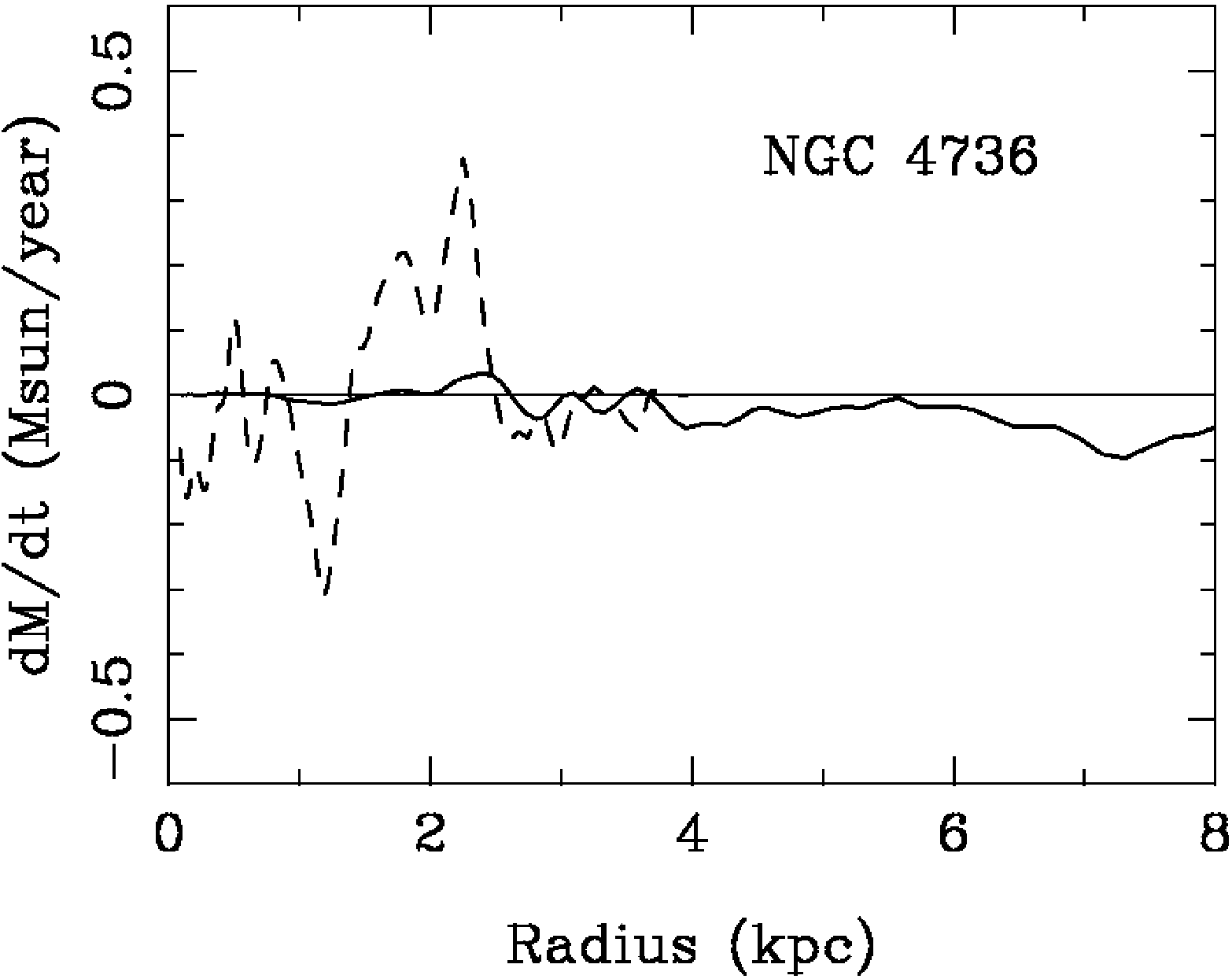}
\caption{HI and $H_2$ mass flow rates for the six sample galaxies. 
{\it Solid lines:} HI mass flow rates. 
{\it Dashed lines:} H$_2$ mass flow rates.
The HI mass flow rates were derived using mass maps from VIVA and THINGS
observations, and H$_2$ flow rates were derived using mass maps
from BIMA SONG observations. 
The total potentials used for these
calculations were the same as previously derived using IRAC
and/or SDSS for the stellar contributions, with appropriate averaging,
plus the gas contributions.}
\label{fg:Fig20}
\end{figure}

In Figure~\ref{fg:Fig20},
we present the comparison of $HI$ and $H_2$ mass flow rates for our
six galaxies.  It is seen that for all six galaxies the $H_2$
mass flow follows more closely the stellar mass flow distributions,
whereas the HI has a more smooth and gradual distribution.  This
is likely to be a result of the process of the formation of molecular
gas (as well as the formation of molecular clouds and complexes)
at the spiral arms due to the density wave shock, and their subsequent
dissociation.  The HI gas distribution, on the other hand, follows a more
quiescent dynamics, even though a mild correspondence to the density
wave patterns can be discerned.

We emphasize that these plots should not be read as that the HI gas is
less responsive to the gravitational perturbation of the density wave
than H$_2$, but rather that where the response of the gas happens in
the spiral shock is also where HI is converted to H$_2$, so the
responsiveness will show up more prominently in $H_2$ than in HI.

\section{DISCUSSION: IMPLICATIONS ON THE SECULAR MORPHOLOGICAL
TRANSFORMATION OF GALAXIES}

In the results presented so far, a picture emerges of the secular
morphological transformation of galaxies driven mostly by collective
effects mediated by large-scale density wave modes.  This evolution
will slowly change the basic state mass distribution towards that of an
ever increasing central (bulge) mass together with the build-up of an
extended outer envelope. These changes are accompanied by a corresponding
change in the morphology, kinematics, and other physical properties of
the density wave modes (including the formation of successive nested
resonances), so at every stage of the transformation from the late to
the early Hubble types, there should be a good correspondence between
the basic state properties and the density wave modes. Once the
galaxy has settled onto an initial quasi-steady state, as it transforms
its morphology from late to intermediate Hubble types, only during
brief periods in its subsequent life, when it encounters
gravitational perturbation from a companion galaxy, such as the case
for M51 or NGC 3627, will the coherence of its phase shift curve and
its kinematic and dynamical equilibrium be temporarily disturbed. A new
equilibrium state is expected to be restored once the perturbation
ceases and the galaxy adjusts its mass distribution and modal pattern
to be once again mutually compatible. The reason these new equilibrium
states are always possible is because for every commonly-occurring basic
state configuration there is almost always a set of unstable modes
corresponding to it (Bertin et al.  1989).

As a result of this coordinated co-evolution of the basic state of the
galaxy disk and the density-wave modes it supports, over the major
span of a galaxy's lifetime the effective evolution rate depends 
on the stage of life a galaxy is in.  Late- or intermediate-type galaxies, 
having larger equilibrium amplitude and open-spiral/skewed-bar
modes in their outer disks, would lead to larger mass flow rates and thus 
higher secular evolution rates in the outer disk.  In contrast, for early type
galaxies, the secular evolution activity is shifted to the central region
of a galaxy, and in its outer region the tightly-wrapped spirals or
more straight bars both lead to small potential-density phase shifts
and low mass flow rates. 

The secular evolution rate also depends on the interaction state of a
galaxy, with a galaxy in group or cluster environments generally having
larger evolution rates due to the large-amplitude, open spiral and bar
patterns excited (Zhang 2008).  Interaction-enhanced, density-wave
mediated evolution appears to underlie the so-called morphological
Butcher-Oemler effect (Butcher \& Oemler 1978) in rich clusters.
Furthermore, there is the well-known morphology-density relation
(Hubble \& Humason 1931; Dressler 1980), which has been shown by
various teams to hold over more than 4 orders of magnitude in mean
density spanning environments from small field groups, to poor
clusters, to rich-cluster outskirts, all the way to dense cluster
central regions.  The universality of such a correlation is in fact a
powerful illustration of the unifying role played by
``nurture-assisted-nature'' type of processes during galaxy evolution.
On the other hand, as we see in the current work, even for strongly
interacting galaxies such as M51 and NGC 3627, the environment exerts
its effect through the innate mechanism already present in individual
galaxies:  the density wave patterns that are excited during the
interaction are still the intrinsic modes, the effect of interaction
having only enhanced the amplitudes of these modes.

The flip side of this correlation of evolution rate with environment is
that some disk galaxies in isolated environments could have a very
small secular evolution rate throughout the span of a Hubble time,
which explains the observational fact that many disk galaxies in the
field are found to have evolved little during the past few Gyr
(essentially, this is why we still observe disk-dominated 
galaxies in the nearby universe
at all, instead of all galaxies having evolved into bulge-dominated
configurations).  Still, an isolated environment cannot be automatically
equated to a slow evolution rate.  One of the most impressive examples
of a strongly excited density wave pattern we have encountered is in
the SB(s)b galaxy NGC 1530 (ZB07), which lies in a surprisingly
pristine environment, i.e., nearly totally isolated.  Yet NGC 1530 has
by far the largest mass flow rate of all the galaxies we have applied
the PDPS method to so far, on the order of more than 100 $M_{\odot}$ yr$^{-1}$,
due to its large surface density and the presence of a strong 
bar-spiral modal structure.  This shows that the initial conditions
that galaxies inherited at birth are likely to have played an important
role as well in determining the subsequent morphological evolution
rate.

The secular morphological transformation of galaxies along
the Hubble sequence also implies that the required external gas
accretion rate to sustain the current level of star formation 
in disk galaxies can be much reduced from the amount previously sought:
As a galaxy's Hubble type evolves from late to early, more and more of
its store of primordial gas will be exhausted, and its star-formation
rate will naturally decline.  But this is precisely the observed
trend of star-formation in galaxies along the Hubble sequence: i.e., the
early-type disk galaxies do not have nearly as much star
formation activity as late-type galaxies.  Galaxies thus do not
have to sustain their ``current'' level of star formation over cosmic time,
since what is current today will be history by the next phase of
their morphological evolution.

A further inference is that as galaxies evolve from late to early
Hubble types, the central potential well will gradually become deeper,
thus more and more nested resonances form in the nuclear region as a
result (Zhang et al. 1993).  These successive resonances form a
continued chain of mass accretion into the central region of a galaxy
while the bulge itself grows, and this process could naturally account
for the observed correlations between the galaxy bulge mass and central
black-hole mass.

Finally, we note that the continuous evolution across the S0
boundary into disky Es may erase the distinction between pseudobulges
and classical bulges, i.e., many galaxies of the Milky Way type or earlier
were previously thought to have classical bulges because of their
$r^{1/4}$ central density profile, yet the building up of these intermediate-
to early-type galactic bulges is most likely through the secular mass 
accretion process over the past Hubble time -- provided only that
the disky bulges further relaxed into the $r^{1/4}$ shaped bulges with time.  
In the end, as has
already been pointed out by Franx (1993), there might not be a clear distinction
between early type disks and disky Es, only a gradual variation
of the bulge-to-disk ratio.  The recent results from the ATLAS$^{3D}$
project of a significant disk component in all low-mass ellipticals,
as well as the structural and population continuity between the
early-type disk galaxies and disky ellipticals 
(Cappellari et al. 2013) also support this continuous evolution 
trend from late type disk galaxies all the way to disky Es.

\section{CONCLUSIONS}

We have presented a study of a sample of six galaxies with a broad
range of Hubble types in order to obtain an initial estimate of their
radial mass accretion/excretion rates, and thus to gauge the relevance
of secular mass redistribution processes to the morphological
transformation of galaxies along the Hubble sequence.  Our results show
that the mass flow rates obtained in typical disk galaxies are able to
produce significant evolution of their Hubble types over cosmic
time, especially if such disk galaxies have undergone external 
perturbation due to tidal interactions with companion galaxies in a
dense environment. We have found that the reasons past studies have
often concluded that secular evolution is only important for building
up late-type pseudo-bulges are, first of all, due to the neglect of the
important role of stellar mass accretion, and secondly, the neglect of
the dominant role played by collective effects enabled by
self-organized density wave modes. 

\section*{REFERENCES}

\noindent
Bell, E. F. \& de Jong, R. S. 2001, ApJ, 550, 212

\noindent
Bell, E.F., McIntosh, D.H., Katz, N., \& Weinberg, M.D. 2003, ApJS, 149, 289

\noindent
Bertin, G., Lin, C.C., Lowe, S.A., \& Thurstans, R.P. 1989,
ApJ, 338, 78

\noindent
Binney, J., \& Tremaine, S. 2008, Galactic Dynamics, second ed. (Princeton:
Princeton Univ. Press) (BT08)

\noindent
Buta, R. 1988, ApJS, 66, 233

\noindent
Buta, R. \& Combes, F. 1996, Fundamentals of Cosmic Physics, 17, 95

\noindent
Buta, R., Corwin, H.G.,Jr., de Vaucouleurs, G., de Vaucouleurs, A., \& Longo, G. 1995, AJ, 109, 517

\noindent
Buta, R. \& Williams, K. L. 1995, AJ, 109, 517

\noindent
Buta, R.J., \& Zhang, X. 2009, ApJS, 182, 559 (BZ09)

\noindent
Buta, R. et al. 2010, ApJS, 190, 147

\noindent
Butcher, H., \& Oemler, A., Jr. 1978, ApJ, 219,18; 226,559; 

\noindent
Cappellari, M. et al. 2013, MNRAS, 432, 1862

\noindent
Chemin, L., Cayatte, V., Balkowski, C., Marcelin, M., Amram, P., van Driel, W.,
\& Flores, H. 2003, A\&A, 405, 89

\noindent
Chung, A., van Gorkom, J. H.; Kenney, J.D.P., Crowl, H., Vollmer, B. 2009
AJ, 138, 1741

\noindent
Contopoulos, G. 1980, A\&A, 81, 198

\noindent
Devereux, N.A., Kenney, J.D., \& Young, J.S. 1992, ApJ., 103, 784

\noindent
de Grijs, R. 1998, MNRAS, 100, 595

\noindent
Donner, K.J., \& Thomasson, M. 1994, A\&A, 290, 785

\noindent
Dressler, A. 1980, ApJ, 236, 351

\noindent
Eskridge, P.B. et al. 2002, ApJS, 143, 7

\noindent
Flagey, N., Boulaner, F., Verstraete, L., Miville Desch\^enes, M. A., Noriega Crespo, A., \&
Reach, W. T. 2006, A\&A, 453, 969

\noindent
Foyle, K., Rix, H.-W., \& Zibetti, S. 2010, MNRAS, 407, 163

\noindent
Franx, M. 1993, in Proc. IAUS 153, 
Galactic Bulges, eds. H. Dejonghe \& H.J. Having (Dordrecht:
Kluwer), 243

\noindent
Gnedin, O., Goodman, J., \& Frei, Z. 1995, AJ, 110, 1105

\noindent
Gonz\'alez, R.A., \& Graham, J.R. 1996, ApJ, 460, 651

\noindent
Gunn, J.E. et al. 1998, AJ, 116, 3040

\noindent
Haan, S., Schinnerer, E., Emsellem, E., Garcia-Burillo, S.,
Combes, F., Mundell, C.G., \& Rix, H. 2009, ApJ, 692, 1623

\noindent
Helfer, T. 2003, 
Thornley, M. D., Regan, M.W., Wong, T., Sheth, K., 
Vogel, S. N., Blitz, L.; Bock, D.C.-J.
ApJS, 145, 259

\noindent
Helou G. et al., 2004, ApJS, 154, 253

\noindent
Hernandez. O., Wozniak, H., Carignan, C., Amram, P.,
Chemin, L., \& Daigle, O. 2005, ApJ, 632, 253

\noindent
Hubble, E. \& Humason, M. 1931, ApJ, 74, 43

\noindent
Jablonka, P., Gorgas, J., \& Goudfrooij, P. 2002,
Ap\&SS, 281, 367

\noindent
Jalocha, J.,  Bratek, L., Kutschera, M. 2008, ApJ, 679, 373

\noindent
Kendall, S., Kennicutt, R. C., Clarke, C., \& Thornley, M. 2008, MNRAS, 387, 1007

\noindent
Kennicutt, R.C. et al. 2003, PASP, 115, 928

\noindent
Kormendy, J. 2012, in Secular Evolution of Galaxies, XXIII Canary Islands Winter
School of Astrophysics, J. Falc\'on-Barroso \& J. H. Knapen, Cambridge, Cambridge
University Press, p. 1 

\noindent
Lin, D.N.C., \& Pringle, J.E., 1987, MNRAS,
225, 607

\noindent
Lynden-Bell, D., \& Kalnajs, A.J. 1972, MNRAS, 157, 1

\noindent
Mart\'inez-Garc\'ia, E.E., Gonz\'alez-L\'opezlira, R.A., Bruzual, A. G.
2009, ApJ, 694, 512

\noindent
Mart\'inez-Garc\'ia, E.E., Gonz\'alez-L\'opezlira, R.A., Bruzual, A. G.
2011, ApJ, 734, 122

\noindent
Meidt, S., et al. 2012, ApJ, 744, 17

\noindent
Nicol, M.-H. 2006, MPIA Student Workshop, http://www.
\newline mpia-hd.mpg.de/70CM/3rdworkshop/presentations/Marie-Helene$\_$Nicol$\_$dark$\_$matter$\_$SF$\_$rate.pdf

\noindent
Oh, S., de Blok, W.J.G., Walter, F., Brinks, E., \& Kennicutt, R.C.,Jr. 2008, AJ, 136, 2761

\noindent
Quillen, A. C., Frogel, J. A., \& Gonz\'alez, R. 1994, ApJ, 437, 162

\noindent
Reach, W. T. et al. 2005, PASP, 117, 978

\noindent
Schwarz, M. P. 1984, MNRAS, 209, 93

\noindent
Sellwood, J.A. 2011, MNRAS, 410, 1673

\noindent
Shostak, G.S., van Gorkom, J.H., Ekers, R.D., Sanders, R.H.,
Goss, W.M., \& Cornwell, F.J. 1983, A\&A, 119, L3

\noindent
Shu, F.S. 1992, The Physics of Astrophysics, vol. II.  Gas Dynamics,
(Mill Valley, Univ. Sci. Books)

\noindent
Sofue, Y. 1996, ApJ, 458, 120

\noindent
Sofue, Y., Tutui, Y., Honma, A., Tomita, A., Takamiya, T., Koda, J.,
\& Takeda, Y. 1999, ApJ, 523, 136

\noindent
Tully, R.B. 1974, ApJS, 27, 415

\noindent
Walter, F., Brinks, E., de Blok, W.J.G., Bigiel, F.,
Kennicutt, R.C.Jr., Thornley, M.D., Leroy, A. 2008, AJ, 136, 2563

\noindent
Worthey, G. 1994, ApJS, 95, 107

\noindent
York, D.G. et al. 2000, AJ, 120, 1579

\noindent
Zhang, X. 1996, ApJ, 457, 125 (Z96)

\noindent
Zhang, X. 1998, ApJ, 499, 93 (Z98)

\noindent
Zhang, X. 1999, ApJ, 518, 613 (Z99)

\noindent 
Zhang, X. 2008, PASP, 120, 121

\noindent
Zhang, X. \& Buta, R. 2007, AJ, 133, 2584 (ZB07)

\noindent
Zhang, X., Wright, M.C.H., \& Alexandria, P. 1993, ApJ,
418,100

\section*{APPENDIX A. DESCRIPTIONS OF THE PROCEDURES FOR OBTAINING
SURFACE MASS DENSITY MAPS}

Stellar surface mass density maps can be made from two-dimensional
images using surface colors as indicators of stellar mass-to-light
ratio (Bell \& de Jong 2001). Calibrated surface brightness maps can be
converted to units of $L_{\odot}$ pc$^{-2}$, and then multiplied by
color-inferred $M/L$ values in solar units to give the surface mass
density $\Sigma$ $(i,j)$ in units of $M_{\odot}$ pc$^{-2}$ at pixel
coordinate $(i,j)$. Thus our approach is two-dimensional and not
based on azimuthal averages of the luminosity distribution.

It is widely regarded that the best images to use for mapping stellar
mass distributions are infrared images, because these penetrate
interstellar dust more effectively than optical images and also because
such images are more sensitive to the light of the old stellar
population that defines the backbone of the stellar mass distribution.
For our study here, we used two principal types of images: (1) an
Infrared Array Camera (IRAC) image taken at 3.6$\mu$m for the {\it
Spitzer Infrared Nearby Galaxies Survey} (SINGS, Kennicutt et al.
2003); and (2) a 0.8$\mu$m $i$-band image obtained from the Sloan
Digital Sky Survey (SDSS; Gunn et al.  1998; York et al. 2000). The
pixel scales are 0\rlap{.}$^{\prime\prime}$75 for the 3.6$\mu$m images
and 0\rlap{.}$^{\prime\prime}$396 for the $i$-band images.
For NGC 3351 and 3627, the
$i$-band images were rescaled to the scale of the 3.6$\mu$m images.
For NGC 5194, the 3.6$\mu$m mass map was rescaled to the scale of the
$i$-band to insure that the same area is covered on the two images.

Bell \& de Jong (2001) give linear relationships between the log of the
$M/L$ ratio in a given passband and a variety of color indices in the
Johnson-Cousins systems. Bell et al. (2003) give the same kinds of
relations for SDSS filters.  For the 3.6$\mu$m images of NGC 628, 3351,
3627, and 5194, we used $B-V$ as our $M/L$ calibration color index,
while for M100 we used $B-R$. For the SDSS $i$-band, we used $g-i$ as
our main color. Using different color indices to scale the two
base images from array units to solar masses per square parsec
means that independent photometric calibrations are used, allowing
us to examine effects that might be due to $M/L$ uncertainties or
systematics.

The base images we have used have different advantages and
disadvantages for mass map calculations. For example, IRAC 3.6$\mu$m
images have the advantages of much greater depth of exposure than most
ground-based near-IR images and also they give the most extinction-free
view of the old stellar background. Nevertheless, 3.6$\mu$m images are
affected by hot dust connected with star-forming regions and by a
prominent 3.3$\mu$m emission feature due to a polycyclic aromatic
hydrocarbon compound that also is associated with star-forming regions
(see Meidt et al. 2012).  These star-forming regions appear as
conspicuous ``knots" lining spiral arms in 3.6$\mu$m images, such that
the appearance of a galaxy at this mid-IR wavelength is astonishingly
similar to its appearance in the $B$-band, minus the effects of
extinction (e. g., Buta et al. 2010).

The advantages of the SDSS $i$-band are the high quality of the SDSS
images in general (especially with regard to uniformity of background),
the reduced effect of star-forming regions compared to the $B$-band and
the 3.6$\mu$m band, and the pixel scale which is almost a factor of two
better than for the 3.6$\mu$m IRAC band. Nevertheless, extinction at
0.8$\mu$m is more than 40\% of that in the $V$-band, so $i$-band images
are considerably more affected by extinction than are 3.6$\mu$m images.

We use the 3.6$\mu$m and $i$-band mass maps as consistency checks on
our results since neither waveband is perfect for the purpose intended.
In practice, the star-forming region problems in the 3.6$\mu$m
band can be reduced using an 8.0$\mu$m image if available (Kendall et al.
2008). These ``contaminants" can also be eliminated using Independent
Component Analysis (Meidt et al. 2012) if no 8.0$\mu$m image is
available. 

\subsection*{A1. Procedure}

{\it SDSS images} - Images were downloaded from the SDSS archive using
the on-line DAS Coordinate Submission Form. For the large galaxies in
our sample, it was necessary to download multiple images per filter to cover
the whole object (ranging from 3 for NGC 3351 to 9 for NGC 4736). These
were mosaiced, cleaned of foreground and background objects, and then
background-subtracted using routines in the Image Reduction and
Analysis Facility (IRAF). SDSS images were available for five of the six
sample galaxies, excluding NGC 628.

The zero points for the $g$ and $i$-band images were obtained using
information given on the SDSS DR6 field pages. The airmasses $x_g$ and
$x_i$, calibration zero points $aa_g$ and $aa_i$, and extinction
coefficients $kk_g$ and $kk_i$, were extracted from this page and the
zero points appropriate to the main galaxy fields were derived as

$$zp_g = -(aa_g+kk_g x_g)+2.5log(a_{pix}t)$$
$$zp_i = -(aa_i+kk_i x_i)+2.5log(a_{pix}t)$$

\noindent
where $a_{pix}$ is the pixel area equal to (0.396)$^2$ or (0.75)$^2$
if rescaled to the 3.6$\mu$m image and $t$ is the integration time of
53.907456s.  For the galaxies that were rescaled, we
matched the coordinates of the SDSS $g$- and $i$-band images to the
system of the 3.6$\mu$m image, which made it necessary to modify $zp_g$
and $zp_i$ to account for the new pixel size. This was done using IRAF
routines GEOMAP and GEOTRAN. Foreground stars were selected that were
well-defined on all three of the images. GEOMAP gave the rotation,
shifting, and scale parameters, while GEOTRAN performed the actual
transformations. Before the final transformations, the point spread
function (PSF) of the images was checked. If the PSFs of the $g$- and
$i$-band images were significantly different, the image with the best
seeing was matched to the other using IRAF routine GAUSS. For M100, the
higher resolution of the SDSS images relative to the 3.6$\mu$m image
was retained.

The next step was to deproject these images in flux-conserving mode.
For this purpose, IRAF routine IMLINTRAN was used with the adopted
orientation parameters (see Table 1). When possible, we used kinematic
orientation parameters for the deprojections. In the case of NGC 3351,
we also used isophotal ellipse fits to deduce these parameters.
No photometric decomposition was used for the deprojections; the bulges
were assumed to be as flat as the disk. 

The SDSS $i$-band mass map was derived as follows. The absolute magnitude of
the Sun was taken to be $M_i$ = 4.48 (CNA Willmer), with which the zero
point needed to convert $i$-band surface brightnesses to solar $i$-band
luminosities per square parsec is $zp_{\odot i}$ = 26.052.  To convert
from these units to solar masses per square parsec, Bell et al. (2003)
give a simple relationship:

$$log{M\over L_i} = -0.152 + 0.518(\mu_g - \mu_i)_o$$

\noindent
where $(\mu_g - \mu_i)_o$ is the reddening-corrected $g-i$ color index.
The only correction made was for Galactic reddening. 
This was judged using information from the NASA/IPAC Extragalactic Database
(NED).\footnote{This research has made use of the NASA/IPAC Extragalactic
Database (NED), which is operated by the Jet Propulsion Laboratory, California
Institute of Technology, under contract with the National Aeronautics
and Space Administration.} 
Although NED lists extinction in the broad-band Johnson filters like B 
and V, the extinctions in g and i were not listed at the time we did our
study. For these, we used the York Extinction Solver (YES; McCall 2004) on 
the NED website.  The actual use of the
above equation requires that some account be made of noise, because
SDSS images are not as deep as 3.6$\mu$m images. For our analysis, it
is important to use the two-dimensional color index distribution, not
an azimuthal average except in the outer parts of the disk where there
is little azimuthal structure. Our procedure was to derive
azimuthally-averaged surface brightness and color index profiles, and
to interpolate colors from these profiles in the outer regions where
noise made the actual pixel color too uncertain to be used in the above
equation. Smoothing was also used even in intermediate regions.
Usually, the colors in the inner regions were used without smoothing,
and then annular zones of increasing radius used an $n$$\times$$n$
median box smoothing.

The median-smoothed raw $i$-band counts, $C_i (i,j)$, at array position
$(i,j)$, were then converted to surface mass density $\Sigma (M_{\odot}
pc^{-2})$ through

$$\Sigma(i,j) = C_i(i,j) \times 10^{-0.4(zp{_i}
- A{_i} - zp_{\odot i})} $$
$$\times 10^{-0.152 + 0.518[\mu_g(i,j) - \mu_i(i,j)
- A_g + A_i)]}$$

The final SDSS mass maps usually would have left-over foreground stars
that had not been removed, or dark spots in areas of star-formation
where the color index was affected by too much blue light. These spots
were removed using IRAF routine IMEDIT.

{\it 3.6$\mu$m Images} - The procedure for these is in many respects
similar to the procedure used for the SDSS images, except that
different color index maps are used and these are not directly linked
to the $M/L$ ratio at 3.6$\mu$m. For 3.6$\mu$m maps, we used $B-V$ or
$B-R$ colors where the individual images are calibrated using
photoelectric multi-aperture photometry. The sources and error analysis
of such photometry, originally used to derive total magnitudes and
color indices for RC3, are described in Buta et al. (1995) and Buta \&
Williams (1995).\footnote{A full catalogue of $UBV$ measurements may be
found at http://kudzu.astr.ua.edu/devatlas/revUBV.ecat.txt.} The
typical uncertainty in the $B$-band zero point from this approach is
0.017 mag while for $V$ it is 0.015 mag, based on 8-27 measurements.
Fewer measurements are available for $R$-band calibrations from this
approach; the uncertainties in these are described by Buta \& Williams
(1995). The $B$ and $V$ images used were downloaded from the SINGS
database webpage. $B-R$ was used only for M100 and is based on images
due to B. Canzian from observations with the USNO 1.0m telescope.

The surface brightnesses in the 3.6$\mu$m images were derived using a
common zero point of 17.6935, based on the calibration from Reach et
al. (2005). IRAC images are in units of MegaJanskys per steradian.

For our study here, we have made a ``hot dust correction" to the
3.6$\mu$m images using a procedure similar to that outlined by Kendall
et al. (2008) since all of our galaxies have an 8.0$\mu$m image
available. The first step is to match the coordinate systems of the 3.6
and 8.0$\mu$m images and then subtract a fraction (0.232) of the
3.6$\mu$m flux from the 8.0$\mu$m image to correct the latter for
continuum emission (Helou et al. 2004). Then, a fraction $R_{3.6/8.0}$
of the net dust map is subtracted from the 3.6$\mu$m map to give an
image corrected for the hot dust emission. In general this method did
improve our mass maps. The factor we used, $R_{3.6/8.0}$ = 0.059, is at
the low end of the flux ratios found by Flagey et al. (2006) for the
Galactic diffuse interstellar medium.

For converting 3.6$\mu$m surface brightnesses to solar luminosities per
square parsec, the absolute magnitude of the Sun was taken to be 
$M_{3.6}$ = $M_L$ = 3.27 for the $L$-band, which is close to the same
wavelength (Worthey 1994). This corresponds to $zp_{\odot,3.6}$ =
24.842.  To convert from these units to solar masses per square parsec,
two steps are used. The first is to derive the $K$-band $M/L$ ratio
from the corrected broadband colors. For $B-R$, the
relation used is (Bell et al. 2003):

$$log{M\over L_K} = -0.264 + 0.138(\mu_B - \mu_R)_o$$

\noindent
where $(\mu_B - \mu_R)_o$ is the Galactic reddening-corrected color
index based on extinctions listed in NED. This relation differs substantially
from that for the same color and near-IR band listed in Table 1
of Bell \& de Jong (2001), which Bell et al. (2003) suggest is due
to a larger metallicity scatter than accounted for in the earlier paper.

The second step is to convert ${M\over L_K}$ into ${M\over L_{3.6}}$. We
used a simple relation due to Oh et al. (2008), based on stellar population
synthesis models with a range of metallicities and star formation histories;

$${M\over L_{3.6}} = 0.92{M\over L_K} - 0.05$$

\noindent
Because of the higher signal-to-noise in the 3.6$\mu$m image, it was
not necessary to use the staggered median smoothing approach used for
the SDSS images. The surface mass densities were then derived from the
array values $C_{3.6}$ using

$$\Sigma(i,j) = C_{3.6}(i,j) \times
10^{-0.4(zp_{3.6} - A_{3.6} - zp_{\odot,3.6})} \times
$$
$$
0.92(10^{-0.264 + 0.138[\mu_B(i,j) - \mu_R(i,j) - A_B + A_R)]}) - 0.05$$

\noindent 
For those cases where $B-V$ was used instead (NGC 628, 3351, 3627, and
5194), the $M/L$ relation applied was (Bell et al. 2003)

$$log{M\over L_K} = -0.206 + 0.138(\mu_B - \mu_V)_o$$

\noindent
where again the Galactic extinction corrections were taken from NED.

{\it Addition of gas} - Our analysis requires total mass maps, and thus
it is essential to add in the contributions of atomic and molecular
gas. Five of our sample galaxies were observed in The HI Nearby Galaxy
Survey (THINGS; Walter et al.  2008), while M100 was observed as part
of the VLA Imaging of Virgo Spirals in Atomic Gas (VIVA) program (Chung
et al. 2009). All 6 are included in the BIMA Survey of Nearby Galaxies
(BIMA SONG, Helfer et al. 2003). VIVA provides an HI image with
resolution 31\arc\ $\times$ 28\arc\ and pixels 10\arc\ $\times$ 10\arc.
THINGS provides HI maps with a resolution of
$\approx$6$^{\prime\prime}$ and pixels 1\rlap{.}$^{\prime\prime}$5 in
size. The maps in all cases are publicly available, and the procedure
for adding both HI and CO into the mass maps was the same. For the HI
map, the total flux in the image was integrated to a radius consistent
with the HI size of the object using IRAF routine PHOT. The map was
then scaled to the measured total flux $S_{HI}$ given in Table 5 of
Walter et al. (2008) and in Table 3 of Chung et al. (2009). With this
scaling, each pixel in the image then has the same units, Jy km s$^{-1}$, 
and can be converted to mass using $ M_{HI}(i,j) = 2.36\times
10^5 D^2 S_{HI}(i,j)$, where $D$ is the distance in Mpc. Dividing each
value by the number of square parsecs in a pixel, this gives the
distance-independent surface mass density of HI gas in units of
$M_{\odot}$ pc$^{-2}$.

For the CO map, Table 4 of Helfer et al. (2003) gives the global CO
flux, $S_{CO}$, for each galaxy. The same procedure as for the HI map
gives the scaling of each pixel, such that the mass in each pixel is
$M_{H_2}(i,j) = 7845.0 D^2 S_{CO}(i,j)$, where a conversion factor of
$X$ = 2$\times$10$^{20}$ has been used (Helfer et al. 2003).  The scale of
the publicly available images is 1\arcs 0 per pixel.

The pixel sizes of the two gas maps were different from the pixel sizes
of the 3.6$\mu$m and $i$-band images. IRAF routine IMLINTRAN was used
to create scaled maps with the same pixel sizes, outputted to an
appropriate center of the galaxy. Each paper gave the right ascension
and declination of the pointing center, which was compared with the
coordinates in the RC3 to judge where we should set the centers in our
mass maps. Each scaled map had its own flux-scale
factor to keep the total masses the same as published by Helfer et al.
(2003), Walter et al. (2008), and Chung et al. (2009).

\noindent
{\it Gravitational Potentials} - The potentials were calculated using the
2D Cartesian approach outlined by Binney \& Tremaine (2008) as described
in ZB07 (and similar to the approach used by Quillen et al. 1994).
An important parameter needed in this calculation is the vertical
scale height, assuming an exponential vertical density distribution.
We used the approximate radial scalelengths listed in Table 1 and
information from de Grijs (1998) to judge scale heights. Being bright
galaxies, there are other sources of radial scale-length determinations
for our sample. A literature search showed good agreement between
our estimated values and other sources except for NGC 4736, whose
complex structure causes a large spread in values, and for NGC 5194,
which is complicated by its companion.

\noindent
{\it Uncertainties in mass map determinations and derived results} -
The uncertainties in our mass maps come from a variety of sources. In
general, photometric calibration uncertainties are small, and less than
0.05 mag. The principal uncertainties come from the $M/L$ calibrations,
effects of dust, and from deprojection and orientation parameter
uncertainties. According to Bell et al. (2003), typical uncertainties
in a color-dependent $M/L$ involving a near-IR band (such as $M/L_K$)
is $\pm$0.1 dex for redder $B-V$ and $B-R$ colors, and $\pm$0.2 dex for
bluer colors. We have shown that the Bell et al. (2003) $B-V$/$B-R$
calibrations with 3.6$\mu$m as the base stellar mass image give
azimuthally-averaged radial surface mass density profiles very similar
to those given by the the Bell et al. (2003) $g-i$ calibration with
the $i$-band as the base stellar mass image. From comparisons between
the two base images and using also the Bell \& de Jong (2001)
$B-V$/$B-R$ calibrations, we found that phase shift distributions are
more robust to $M/L$ uncertainties than are mass flow rates, although
even the latter agree fairly well between the two base images.

On the issue of orientation parameters, we did not experiment with different
values but simply point to section 5.2 of BZ09 where tests were made of
the impact of such uncertainties, including problems of bulge deprojection.
The CR radii we list in Table 2 are for the assumed orientation parameters
in Table 1.

Uncertainties in the assumed vertical scale-heights were examined for M100.
Reducing $h_z$ from 12\arcs 6 in Table 1 to 3\arcs 8 had almost no effect
on the phase shift distribution. In the outer disk near 7 kpc, the flow
rate is increased by about 10\% but at 1 kpc the difference is about 30\%. 
Thus even a drastic difference in $h_z$ has only a relatively small effect
on our results.

The issue of dust enters in the uncertainties in two ways: through the
significant impact of dust on the optical $M/L$ calibrating colors, and
through the ``hot dust correction" to the 3.6$\mu$m image. The former
is less of a problem than might be thought. As noted by Bell et al.
(2003), the effects of dust approximately cancel out to 0.1-0.2 dex
when estimating color-derived $M/L$ values because, in most passbands,
stellar populations and dust predict about the same amounts of
reddening per unit fading. That is, while dust reddens the starlight,
redder colors imply higher $M/L$, which effectively can reduce the
impact of dust lanes. Our results here basically verify this point. 
The uncertainty in the hot dust correction lies mainly in the
factor R$_{3.6/8.0}$. We chose a low end value from Flagey et al.
(2006) for this correction, but higher values may be appropriate
for some galaxies (e.g., Kendall et al. 2008).

\begin{figure}
\vspace{160pt}
\centerline{
\includegraphics{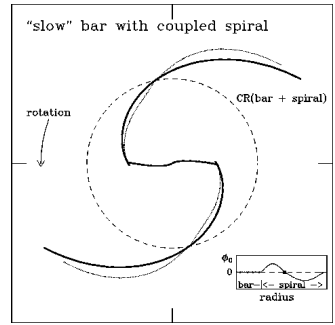}
}
\caption{Schematic of the phase shift distribution for the slow bar.}
\label{fg:Fig21}
\end{figure}

\begin{figure}
\vspace{120pt}
\centerline{
\includegraphics{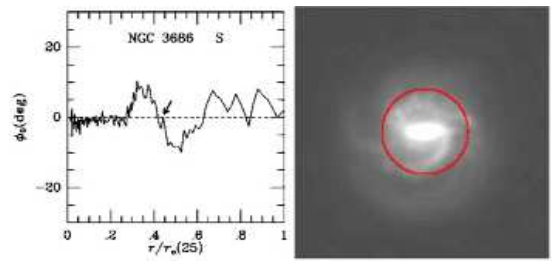}
}
\caption{Examples of the phase shift distribution and morphology
for the slow bar}
\label{fg:Fig22}
\end{figure}

\section*{APPENDIX B. CATEGORIES OF BAR-SPIRAL MODAL MORPHOLOGY}

In this appendix, we give schematics of the phase shift
distributions of the various types of bar-spiral morphology, followed
by examples of real galaxies that we have analyzed in BZ09.  The
sequence we present with these examples agrees roughly with the order
of Hubble sequence from late to early, in order to show possible
evolutionary connections between the various morphological patterns.

Figures \ref{fg:Fig21} and \ref{fg:Fig22} show a schematic and an
example of a slow bar. This kind of bar is characterized by the fact
that the bar ends well within its CR radius, with possible spiral
structure emanating from the bar ends.  The hosts are predominantly
late-type galaxies, with a significant flocculent pattern in the outer
regions.  Closer inspection of the phase shift plot for NGC 3686 shows
that around the location of the bar end, a new phase shift P/N
transition is in the process of forming. Thus the slow bar appears to
be a short-lived phase in the process of evolving into a bar-driven
spiral, which we will analyze next.

\begin{figure}
\vspace{160pt}
\centerline{
\includegraphics{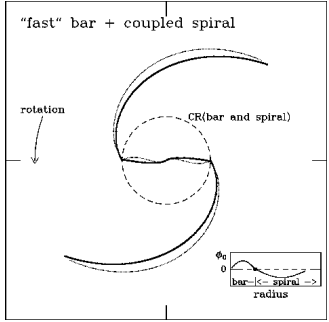}
}
\caption{Schematic of the phase shift distribution for the
fast bar with coupled spiral pattern.}
\label{fg:Fig23}
\end{figure}

\begin{figure}
\vspace{120pt}
\centerline{
\includegraphics{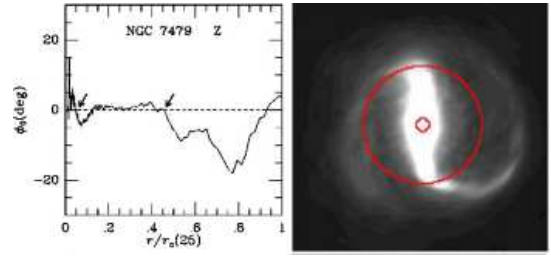}
}
\caption{Examples of the phase shift distribution and morphology for the
fast bar with coupled spiral pattern.}
\label{fg:Fig24}
\end{figure}

Figures \ref{fg:Fig23} and \ref{fg:Fig24} show a schematic and an
example of a fast bar with a coupled spiral, commonly referred to as a
bar-driven spiral.  This kind of pattern most often appears in
intermediate-type galaxies, and appears to have evolved through stages
of either a slow bar or else a skewed long bar.

\begin{figure}
\vspace{160pt}
\centerline{
\includegraphics{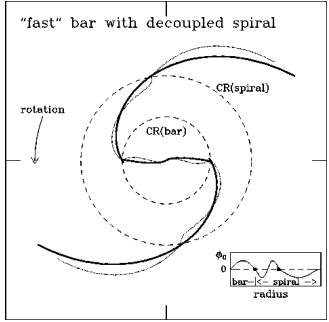}
}
\caption{Schematic of the phase shift distribution for the
fast bars with decoupled spiral pattern.}
\label{fg:Fig25}
\end{figure}

\begin{figure}
\vspace{220pt}
\centerline{
\includegraphics{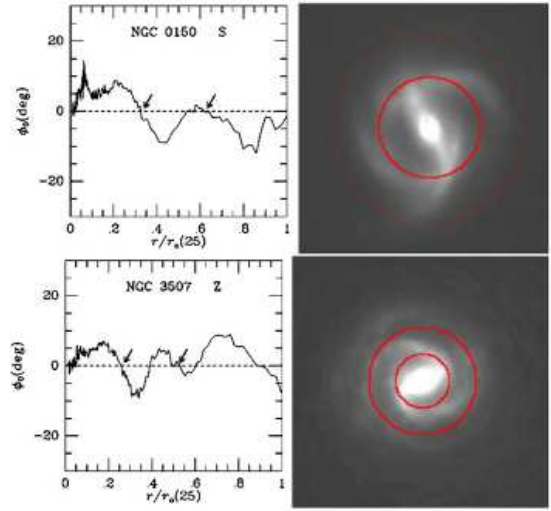}
}
\caption{Examples of the phase shift distribution and morphology for the
fast bars with decoupled spiral pattern.}
\label{fg:Fig26}
\end{figure}

Figures \ref{fg:Fig25} and \ref{fg:Fig26} show a schematic and two
examples of fast bars with decoupled spirals. On one hand, these appear
to be a further evolutionary stage of a bar-driven spiral, with the
inner bar decoupled from the outer spirals and with the bar ends
coinciding roughly with the CR radii. The signature of the decoupling
(at the N/P phase shift crossing) shows up as branching of the spiral
arms disconnected from the bar-end.  Close inspection of the images and
the phase shift plots show that the two examples we give will evolve
into somewhat different configurations later on:  NGC 3507 appears to
be evolving towards an inner bar-driven spiral followed by an outer
spiral, whereas, NGC 150 appears to be evolving towards a super-fast bar
with decoupled spiral that we will discuss next.

\begin{figure}
\vspace{160pt}
\centerline{
\includegraphics{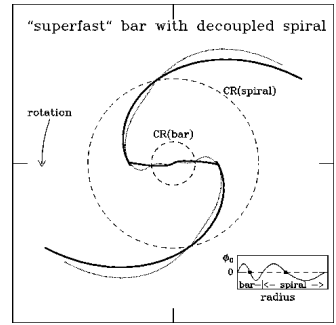}
}
\caption{Schematic of the phase shift distribution for the
super-fast bars with decoupled spiral pattern.}
\label{fg:Fig27}
\end{figure}

\begin{figure}
\vspace{120pt}
\centerline{
\includegraphics{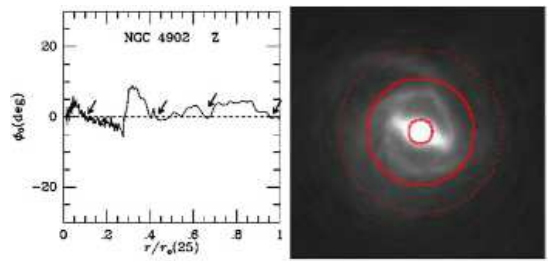}
}
\caption{Examples of the phase shift distribution and morphology for the
super-fast bars with decoupled spiral pattern.}
\label{fg:Fig28}
\end{figure}

Figures \ref{fg:Fig27} and \ref{fg:Fig28} show a schematic and an
example of a super-fast bar with a decoupled spiral. The phrase
``super-fast" means that the bars extend significantly beyond their CR
radii. In ZB07, we showed the case of a single isolated bar in the
galaxy NGC 4665, where the bar extends about 10-20\% beyond the
PDPS-implied CR radius, and argued that in this case it is reasonable
to expect the bar to be longer than the CR radius because the SWING
amplified over-reflected waves from the inner disk must penetrate CR
into the outer disk as a transmitted wave in order to have the overall
angular momentum budget balance. In our current plot, the super-fast
section of the bars are straight segments emanating from the location
of an inner oval.  The end of the bar coincides not with CR but with
the next N/P phase shift crossing after the CR.  This is reasonable
because modal growth requires that for a complete self-sustained mode
there must be a positive phase shift packet followed by a negative
phase shift packet, with the two packets joining at CR -- this in turn
is because the density wave/mode has negative energy and angular
momentum density inside CR (Shu 1992), and for its spontaneous growth
the potential must lag the density -- which leads to the positive
potential-density phase shift -- inside CR in order for the wave to
torque the disk matter in the correct sense to lead to its own
spontaneous growth by losing angular momentum to the disk matter.  This
is true vice versa for the modal content outside CR. Therefore, we see
that for the kind of twin-bars joining the central oval the mode has
little choice but to have the N/P phase shift crossing be at the end of
the bar.

Direct evidence from morphology of NGC 4665 supporting our claim that an
N/P phase shift crossing for this galaxy coincides with the location
where the modal pattern speed changes discontinuously is that there is
a pronounced ring-like structure at the radius of the N/P crossing,
obviously caused by the snow-plough effect of the interaction of the
inner and outer modes.  A similar configuration of a central oval joined
by straight super-fast bars are also observed in NGC 3351, NGC 1073,
NGC 5643, and in the central region of NGC 4321.  As a matter of fact,
since this configuration requires the central oval connecting to the
straight bars, it is always found in the central configuration of
nested modes.

Comparing the last two types of morphologies, we can clearly see that
the ``fast bars with decoupled spirals" appear to be evolving into
``super-fast bars with decoupled spirals" I.e., in the phase shift
plot for NGC 150 we see that a new P/N crossing at the end of inner
oval is in the process of forming, or dropping down to zero.  When it
is fully formed this will become a super-fast bar.

It is no coincidence that the super-fast bars, which often appear in
early type galaxies, have a more rounded nuclear pattern with two very
straight segments connecting to it: both of these patterns correspond
to very small phase shift. The largest phase shift occurs for patterns
with 45$^o$  pitch angle, but when the pitch angle is close to 0$^o$ or
90$^o$, the phase shift becomes zero.  Small phase shifts lead
to slow secular evolution rates as characterize early type galaxies. The
fact that bar-driven spirals mostly have skewed nuclear bar patterns
followed by trailing spiral segments that taper into narrow tails,
whereas super-fast bars most often have rounded nuclear patterns
followed by very straight bar segments which broaden into dumb-bell
shaped pile-up of material at the mode-decoupling radii shows that
super-fast bars are real, and are of completely different modal
category than bar-driven spirals.

Note also that the above four categories are the main types of bar-spiral
associations, and they do not exhaust all the morphological
types encountered in real galaxies.  For example, the above
categories did not include the cases of either pure spiral
galaxies (i.e. NGC 5247 analyzed in ZB07), or pure bar 
galaxies (i.e. NGC 4665 also analyzed in ZB07).

From the above analysis we see that distinctive phase shift patterns
seem to delineate distinctive galaxy morphology proto-types, with the
morphology of galaxies within a type category repeatable to a high
degree. These morphological features also appear to correlate with the
Hubble types of the basic state of the galactic disks (i.e., their
being early, intermediate, or late).  These correlations are naturally
explained under the modal theory of density wave patterns (Bertin et al.
1989).  The fact that the potential-density phase shift method can
consistently classify the typical resonance structures for the given
morphological types of galaxies shows that its success is not an
accident, but rather supported by the underlying modal structure of the
density wave patterns.

\end{document}